\documentclass[useAMS,usenatbib,usegraphicx,onecolumn]{mn2e}

\usepackage{amssymb,amsfonts,amsmath,amstext,amsgen,amsopn,amsxtra,indentfirst,lscape,times,rotating}

\title[Atmospheric circulation of exoplanets II]{Atmospheric circulation of tidally locked exoplanets: \\ II. Dual-band radiative transfer and convective adjustment}
\author[Heng, Frierson \& Phillipps]{Kevin Heng$^{1}$\thanks{E-mail: kheng@phys.ethz.ch, heng@ias.edu (KH)}, Dargan M.W. Frierson$^{2}$\thanks{Email: dargan@atmos.washington.edu (DMWF)} and Peter J. Phillipps$^{3}$\thanks{Email: Peter.Phillipps@noaa.gov (PJP)}\\
$^{1}$Zwicky Fellow, ETH Z\"{u}rich, Institute for Astronomy, Wolfgang-Pauli-Strasse 27, CH-8093, Z\"{u}rich, Switzerland\\
$^{2}$Department of Atmospheric Sciences, University of Washington, ATG Building, Box 351640, Seattle, WA 98195, U.S.A.\\
$^{3}$Geophysical Fluid Dynamics Laboratory, 201 Forrestal Road, Princeton, NJ 08540, U.S.A.}

\begin{document}

\date{Submitted 2011 May 20.  Re-submitted 2011 July 15 and 2011 August 16.  Accepted 2011 August 18.}

\pagerange{\pageref{firstpage}--\pageref{lastpage}} \pubyear{2011}

\maketitle

\label{firstpage}

\begin{abstract}
Improving upon our purely dynamical work, we present three-dimensional simulations of the atmospheric circulation on Earth-like (exo)planets and hot Jupiters using the GFDL-Princeton \textit{Flexible Modeling System} (\texttt{FMS}).  As the first steps away from the dynamical benchmarks of Heng, Menou \& Phillipps (2011), we add dual-band radiative transfer and dry convective adjustment schemes to our computational setup.  Our treatment of radiative transfer assumes stellar irradiation to peak at a wavelength shorter than and distinct from that at which the exoplanet re-emits radiation (``shortwave" versus ``longwave"), and also uses a two-stream approximation.  Convection is mimicked by adjusting unstable lapse rates to the dry adiabat.  The bottom of the atmosphere is bounded by a uniform slab with a finite thermal inertia.  For our models of hot Jupiters, we include an analytical formalism for calculating temperature-pressure profiles, in radiative equilibrium, which accounts for the effect of collision-induced absorption via a single parameter.  We discuss our results within the context of: the predicted temperature-pressure profiles and the absence/presence of a temperature inversion; the possible maintenance, via atmospheric circulation, of the putative high-altitude, shortwave absorber expected to produce these inversions; the angular/temporal offset of the hot spot from the substellar point, its robustness to our ignorance of hyperviscosity and hence its utility in distinguishing between different hot Jovian atmospheres; and various zonal-mean flow quantities.  Our work bridges the gap between three-dimensional simulations which are purely dynamical and those which incorporate multi-band radiative transfer, thus contributing to the construction of a required hierarchy of three-dimensional theoretical models.
\end{abstract}


\section{Introduction}
\label{sect:intro}

Much of the previous research on three-dimensional simulations of exoplanetary atmospheric circulation has focused on the atmospheric dynamics (e.g., \citealt{cs05,cs06,koskinen07,showman08,dd10,rm10,tc10,hmp11}), while others \citep{showman09,lewis10,selsis11} have developed sophisticated schemes which combine dynamics with multi-band radiative transfer.  Despite making progress, our understanding of atmospheric circulation on extrasolar planets (or ``exoplanets") remains incomplete.  To move forward, we need to construct a hierarchy of three-dimensional simulations of varying sophistication, so as to elucidate the isolated effects of various pieces of physics as well as the complex interplay between them \citep{po84,held05,p10}.  The purpose of the present study is to bridge the gap between these two bodies of work by elucidating the details of a computational setup of intermediate sophistication, so as to serve as a foundation for future work.  Our work may be partially regarded as the three-dimensional generalizations of \cite{hubeny03}, \cite{hansen08} and \cite{guillot10}.

As natural extensions to purely dynamical simulation work, the effects of radiative transfer and convection need to be considered.  As a first step in adding radiative transfer, we simplify the treatment by assuming that most of the stellar emission peaks at a wavelength which is shorter than and distinct from that at which the exoplanet re-radiates the absorbed heat \citep{hansen08,ll08,guillot10,p10}.  For example, the Sun-like host star of an exoplanet has an emission spectrum which peaks in the optical, since
\begin{equation}
\lambda_{\star,{\rm max}} \approx 0.5 ~\mu\mbox{m} ~\left( \frac{T_\star}{6000 \mbox{ K}} \right)^{-1}
\end{equation}
using Wien's law, where $T_\star$ is the effective temperature of the star.  For M stars (red dwarfs), the stellar spectrum instead peaks at $\lambda_{\star,{\rm max}} \gtrsim 0.8 ~\mu\mbox{m} ~(3700 \mbox{ K}/T_\star)$.  The equilibrium or effective temperature of the exoplanet is\footnote{This expression is considered to be order-of-magnitude because there is a factor of order unity associated with the exoplanetary albedo and the efficiency of heat redistribution from the day to the night side.}
\begin{equation}
T_{\rm eq} \sim 1300 \mbox{ K} ~\left( \frac{T_\star}{6000 \mbox{ K}} \right) \left( \frac{R_\star}{R_\odot} \right)^{1/2} \left( \frac{a}{0.05\mbox{ AU}} \right)^{-1/2},
\end{equation}
where $R_\star$ is the stellar radius and $a$ denotes the spatial separation between the star and the exoplanet.  As the stellar photons travel downward into the atmosphere, they are eventually absorbed and re-emitted.  The emission spectrum of an exoplanet is expected to peak in the infrared, since
\begin{equation}
\lambda_{p,{\rm max}} \sim 2.2 ~\mu\mbox{m} ~\left( \frac{T_\star}{6000 \mbox{ K}} \right)^{-1} \left( \frac{R_\star}{R_\odot} \right)^{-1/2} \left( \frac{a}{0.05\mbox{ AU}} \right)^{1/2}.
\end{equation}
Thus, to a first approximation one can regard the star as emitting ``shortwave" radiation onto the exoplanet, which then re-emits in the infrared (``longwave").  The conversion of a shortwave photon into multiple longwave photons serves to increase entropy.  Our treatment of the radiative transfer then requires the specification of the shortwave ($\tau_{\rm S}$) and longwave ($\tau_{\rm L}$) optical depths, as well as the flux associated with stellar irradiation (${\cal F}_{\downarrow_{\rm S}}$).  Also as a first step, we mimic convection by adjusting unstable atmospheric lapse rates to the stable dry adiabat \citep{manabe}.  The key advantages of our simpler setup, compared to simulations with multi-band radiative transfer, are that it allows for a more efficient sampling of parameter space and clean comparisons to analytical models.

In \S\ref{sect:fms}, we describe our computational setup, including our schemes for stellar irradiation, shortwave/longwave absorption, radiative transfer and convection.  Prior to simulating the atmospheric circulation on hot Jupiters, we test our computational setup using Earth-like models (\S\ref{sect:earthlike}).  In \S\ref{sect:hj}, we describe our method for simulating hot Jovian atmospheres --- including an analytical formalism for calculating temperature-pressure profiles which includes the effect of collision-induced absorption --- and examine results from several models based on the values of parameters appropriate to HD 209458b.  The implications of our results, as well as a concise summary of the present study, are discussed in \S\ref{sect:discussion}.  Table \ref{tab:params} summarizes the list of adopted parameter values as well as the commonly-used symbols.  Appendix \ref{append:pbl} contains the technical details of the atmospheric boundary layer scheme used only for our Earth-like models.

\begin{table*}
\centering
\caption{Table of parameters, symbols and their values}
\label{tab:params}
\begin{tabular}{lcccc}
\hline\hline
\multicolumn{1}{c}{Symbol} & \multicolumn{1}{c}{Description}  & \multicolumn{1}{c}{Units} & \multicolumn{1}{c}{Earth} & \multicolumn{1}{c}{Hot Jupiter}\\
\hline
\vspace{2pt}
${\cal F}_0$ & stellar irradiation constant &  W m$^{-2}$ & 938.4 & $9.5 \times 10^5$ \\
$\Delta_{\rm T}$ & meridional temperature gradient parameter & --- & 1.4 & --- \\
$n_{\rm S}$ & power law index for shortwave optical depth & --- & 2 & 1 \\
$n_{\rm L}$ & power law index for longwave optical depth & --- & 4 & 2 \\
$\tau_{\rm S_0}$ & surface optical depth of shortwave absorbers & --- & 0.2 & 1401 \\
$\tau_{\rm L_{eq}}$ & surface optical depth of longwave absorbers at equator & --- & 6 & $4.67 \times 10^6$ \\
$\tau_{\rm L_{pole}}$ &surface optical depth of longwave absorbers at poles & --- & 1.5 & $4.67 \times 10^6$ \\
$f_l$ & strength of well-mixed longwave absorbers & --- & 0.1 &  1/2000 \\
\hline
$c_P$ & specific heat capacity at constant pressure (of the atmosphere) & J K$^{-1}$ kg$^{-1}$ & 1004.64 & 14308.4 \\
${\cal R}$ & ideal gas constant (of the atmosphere) & J K$^{-1}$ kg$^{-1}$ & 287.04 & 4593 \\
$\kappa \equiv {\cal R}/c_P$ & adiabatic coefficient$^\ddagger$ & J K$^{-1}$ kg$^{-1}$ & 2/7 & 0.321 \\
$P_0$ & reference pressure at bottom of simulation domain & bar & 1 & 220 \\
$C_{\rm int}$ & areal heat capacity of lower atmospheric boundary & J K$^{-1}$ m$^{-2}$ & $10^7$ & $10^5$ \\
$g_p$ & acceleration due to gravity & m s$^{-2}$ & 9.8 & 9.42 \\
$R_p$ & radius of (exo)planet & km & 6371 & $9.44 \times 10^4$ \\
$\Omega_p$ & rate of rotation of (exo)planet & s$^{-1}$ & $7.292 \times 10^{-5}$ & $2.06 \times 10^{-5}$ \\
\hline
${\cal R}_{i,{\rm crit}}$ & critical bulk Richardson number & --- & 1 & --- \\
$z_{\rm rough}$ & roughness length & m & $3.21 \times 10^{-5}$ & ---\\
$\kappa_{\rm vK}$ & von K\'{a}rm\'{a}n constant & --- & 0.4 & ---\\
$f_b$ & surface layer fraction & --- & 0.1 & ---\\
\hline
$t_\nu$ & hyperviscous time & day$^\dagger$ & 0.1 & $10^{-5}$ \\
$\Delta t$ & time step & s & 1200 & 120 \\
$T_{\rm init}$ & initial temperature & K & 264 & 1824 \\
$N_{\rm v}$ & vertical resolution & --- & 20 & 33 \\
\hline
$\Phi$ & latitude & degrees & -$90^\circ$--$90^\circ$ & -$90^\circ$--$90^\circ$\\
$\Theta$ & longitude & degrees & 0---$360^\circ$ & 0---$360^\circ$\\
$\Gamma_0 \equiv g_p/c_P$ & dry adiabatic lapse rate & K km$^{-1}$ & $\approx 9.8$ & $\approx 0.7$ \\
$\theta_{\rm T}$ & potential temperature & K & $\approx 300$--500 & $\sim 10^3$--$10^5$ \\
$\Psi$ & Eulerian mean streamfunction & kg s$^{-1}$ & $\sim 10^8$--$10^{10}$ & $\sim 10^{13}$--$10^{15}$ \\
\hline
\hline
\end{tabular}\\
$\dagger$: expressed in terms of a (exo)planetary day (i.e., rotational period).\\
$\ddagger$: we term $\kappa$ the ``adiabatic coefficient", following \cite{p10}, and avoid calling it \\
the ``adiabatic index" in order to not confuse it with the ratio of specific heat capacities.\\
Note: one ``Earth day" is exactly 86400 s (as used in the text and simulation/analysis codes).
\end{table*}

\section{The GFDL-Princeton \textit{Flexible Modeling System}}
\label{sect:fms}

In this section, we describe our computational setup, which is based on the \texttt{Lima} release of the \texttt{FMS}, developed by the Geophysical Fluid Dynamics Laboratory (GFDL) at Princeton University \citep{gs82}.  As the \texttt{FMS} is set up to work in MKS (metres, kilogrammes and seconds) units, most of the discussion will follow suit.  Unless specified otherwise, the term ``day" refers to an Earth day (exactly 86400 s).

\subsection{Dynamics}

We implement the spectral dynamical core, which solves the three-dimensional primitive equations of meteorology --- these equations assume hydrostatic equilibrium and other consistent approximations such as a small aspect ratio for the atmosphere (e.g., see \citealt{hmp11} and references therein).  \cite{hs94} originally proposed the comparison of purely dynamical simulations with simplified thermal forcing and performed via different methods of solutions, known as the ``Held-Suarez benchmark" for Earth.  \cite{hmp11} extended the Held-Suarez benchmark to tidally-locked exoplanets by examining three additional cases: a hypothetical exo-Earth, a shallow model for hot Jupiters and a deep model of HD 209458b.\footnote{The adjectives ``shallow" and ``deep" refer to the fact that one and about 5 orders of magnitudes in vertical pressure are simulated, respectively.}  The spectral and finite difference (B-grid) dynamical cores of the \texttt{FMS} were both subjected to these tests.  In the case of the deep model of HD 209458b, the predictions for the wind speeds are found to depend upon the chosen magnitude of the hyperviscosity, although the qualitative features of the simulated wind and temperature maps remain largely invariant.

Within the spectral dynamical core, the (linear) fluid dynamical quantities are expressed as a truncated sum $N_{\rm h}$ of spherical harmonics --- the larger the value of $N_{\rm h}$, the higher the horizontal resolution.  The fiducial resolution we will adopt in the present study is T63L$N_{\rm v}$ ($N_{\rm h}=63$), which corresponds to a horizontal grid of 192 by 96 points in longitude versus latitude.  (See Table 2 of \citealt{hmp11} for a list of $N_{\rm h}$ and their corresponding horizontal resolutions.)  A finite difference solver with $N_{\rm v}$ grid points is used for the vertical coordinate, which is cast in terms of the ``$\sigma$-coordinate" \citep{p57},
\begin{equation}
\sigma \equiv \frac{P}{P_s},
\end{equation}
where $P$ denotes the vertical pressure and $P_s$ is the time-dependent pressure at the surface.  By contrast, the reference pressure at the bottom of the simulation domain, $P_0$, is a constant.  Analogous to $\sigma$, we define $\sigma_0 \equiv P/P_0$.  For Earth-like simulations, the reference pressure is usually set to $P_0 = 1$ bar = $10^5$ Pa = $10^6$ dyn cm$^{-2}$.

Numerical noise accumulates at the grid scale and has to be damped via a ``hyperviscous" term with a damping order of $n_\nu = 4$, such that the operator $\nabla^{2n_\nu}$ acts on the relative vorticity (see \S3.1 of \citealt{hmp11}).  The pragmatic aim is to apply horizontal dissipation on a time scale $t_\nu$ that is a small fraction of an exoplanetary day.  For example, in the Held-Suarez benchmark for Earth, the adopted time scale is $t_\nu \approx 0.1$ Earth day.  For our hot Jupiter simulations, we use $t_\nu \approx 10^{-5}$ HD 209458b day, consistent with the value used in \cite{hmp11}; this is also, to within an order of magnitude, the highest value needed.  It is important to keep in mind that the application of hyperviscosity and the specification of $t_\nu$ is unsupported by any fundamental theory --- horizontal dissipation is solely a numerical tool meant to prevent a simulation from failing due to spectral blocking.

As in \cite{hmp11}, the simulations are started from an initial state of windless isothermality with $T_{\rm init}$ denoting the initial temperature, in contrast to using initial temperature-pressure profiles obtained from one-dimensional radiative-convective calculations (e.g., \citealt{showman09,dd10,lewis10}).  The possible dependence of our results on more exotic initial conditions \citep{tc10} is deferred to a future study.  There is an initialization period for each simulation which we disregard, during which the temperatures and velocities, averaged over the entire (exo)planet, are decreasing/increasing from some initial values to their quasi-steady values.  For Earth-like simulations, we find the initialization period to be 200 days.  For hot Jupiters, a longer initialization period of 500 days is needed.  After quasi-equilibrium is attained, we execute the simulations for a further 1000 days and temporally average the ouput.  Minor asymmetries in the physical quantities between the northern and southern hemispheres are artifacts of averaging over a finite period of time (i.e., in this case, 1000 days).

\subsection{Stellar irradiation and shortwave absorption}
\label{subsect:irradiation}

At the top of the atmosphere, stellar irradiation is specified via the function,
\begin{equation}
{\cal F}_{\rm TOA} = {\cal F}_0 ~{\cal G},
\label{eq:flux_toa}
\end{equation}
where ${\cal F}_0$ is the stellar irradiation constant,\footnote{The Solar constant is typically taken to be about 1367 W m$^{-2}$.}
\begin{equation}
{\cal F}_0 = \left( \frac{R_\star}{a} \right)^2 \sigma_{\rm SB} T^4_\star \approx
\begin{cases}
1370 \mbox{ W m}^{-2} ~\left( \frac{R_\star}{R_\odot} \right)^2 \left( \frac{a}{1 \mbox{ AU}} \right)^{-2} \left( \frac{T_\star}{5780 \mbox{ K}} \right)^4 & \mbox{ (Earth),}\\
9.5 \times 10^5 \mbox{ W m}^{-2} ~\left( \frac{R_\star}{1.146 ~R_\odot} \right)^2 \left( \frac{a}{0.0468 \mbox{ AU}} \right)^{-2} \left( \frac{T_\star}{6000 \mbox{ K}} \right)^4 & \mbox{ (HD 209458b)},
\end{cases}
\label{eq:stellar_constant}
\end{equation}
where $\sigma_{\rm SB}$ denotes the Stefan-Boltzmann constant and the dimensionless stellar irradiation function ${\cal G}={\cal G}(\Phi, \Theta)$ is in general a function of both the latitude $\Phi$ and the longitude $\Theta$.  Since hot Jupiters orbit at distances $a \sim 10~R_\star$ from their host stars, we have ${\cal F}_0 \approx 6.4 \times 10^5$ W m$^{-2}$ for $R_\star = R_\odot$.  By contrast, we have ${\cal F}_0 \approx 51$ W m$^{-2}$ for Jupiter, which is a factor $\sim 10^4$ less than for hot Jupiters.  In the case of HD 209458b, we use the values of $R_\star$, $a$ and $T_\star$ measured or inferred by \cite{mazeh00} and \cite{brown01} to obtain the value shown in equation (\ref{eq:stellar_constant}); we note that $a/R_\star \approx 9$.  In general, the influence of the stellar irradiation on a (exo)planet and its dependence on geometry (through ${\cal G}$) is known as the ``thermal forcing".

The shortwave optical depth $\tau_{\rm S}$ determines how the stellar irradiation is absorbed as it travels downward into the atmosphere \citep{f07,os08,ms10},
\begin{equation}
\tau_{\rm S} = \tau_{\rm S_0} \sigma_0^{n_{\rm S}}.
\label{eq:tau_sw}
\end{equation}
Specifying the power law index as $n_{\rm S}=1$ is equivalent to having a uniformly mixed shortwave absorber, since \citep{guillot10,hhps11}
\begin{equation}
\tau_{\rm S} = \frac{\kappa_{\rm S} P}{g_p} = 10^3 ~\left( \frac{\kappa_{\rm S}}{0.01 \mbox{ cm}^2 \mbox{ g}^{-1}} \frac{P}{100 \mbox{ bar}} \right) \left( \frac{g_p}{10 \mbox{ m s}^{-1}} \right)^{-1}
\label{eq:tau_to_pressure}
\end{equation}
where $\kappa_{\rm S}$ is the shortwave opacity and $g_p$ is the surface gravity.  When $\kappa_{\rm S}$ is constant, we have $\tau_{\rm S} \propto P$.  When $n_{\rm S} > 1$, the shortwave absorber is less well-mixed and resides \emph{lower} down in the atmosphere.  The received flux as a function of vertical pressure is then
\begin{equation}
{\cal F}_{\downarrow_{\rm S}} = {\cal F}_{\rm TOA} \exp\left(-\tau_{\rm S}\right).
\label{eq:flux_s}
\end{equation}

Finally, we note that it is possible to specify a mean (exo)planetary albedo, but this only serves to diminish the value of ${\cal F}_{\rm TOA}$.  As such, one only needs to explore the effects of varying the stellar irradiation constant.

\subsection{Radiative transfer and longwave absorption}
\label{subsect:rt}

The radiative transfer scheme is based on the two-stream approximation \citep{mihalas,hansen08}, which assumes that the radiation can be separated into longwave, upward- (${\cal F}_{\uparrow_{\rm L}}$) and downward-propagating (${\cal F}_{\downarrow_{\rm L}}$) fluxes (see \S4.2 of \citealt{p10}),
\begin{equation}
\begin{split}
&\frac{d{\cal F}_{\uparrow_{\rm L}}}{d \tau_{\rm L}} = {\cal F}_{\uparrow_{\rm L}} - {\cal F}_{\rm bb}, \\
&\frac{d{\cal F}_{\downarrow_{\rm L}}}{d \tau_{\rm L}} = -{\cal F}_{\downarrow_{\rm L}} + {\cal F}_{\rm bb}, \\
\end{split}
\label{eq:rt}
\end{equation}
where ${\cal F}_{\rm bb} = \sigma_{\rm SB} T^4$ is the blackbody flux.  At the bottom of the simulated atmosphere, the longwave boundary condition is ${\cal F}^{\rm (bottom)}_{\uparrow_{\rm L}} = \sigma_{\rm SB} T^4_s$ where $T_s$ is the surface temperature; at the top, it is ${\cal F}^{\rm (top)}_{\downarrow_{\rm L}} = 0$.  The Newtonian relaxation term in the temperature equation is replaced by a radiative source term \citep{fhz06},
\begin{equation}
Q_{\rm rad} = - \frac{1}{c_P \rho} ~\frac{\partial}{\partial z} \left( {\cal F}_{\uparrow_{\rm L}} - {\cal F}_{\downarrow_{\rm L}} - {\cal F}_{\downarrow_{\rm S}}\right),
\end{equation}
where $c_P$ is the specific heat capacity at constant pressure, $\rho$ is the mass density and $z$ is the vertical coordinate.  The shortwave, upward-propagating flux is assumed to be zero, i.e., ${\cal F}_{\uparrow_{\rm S}}=0$.  Furthermore, substituting the Newtonian relaxation scheme for the dual-band radiative transfer scheme alleviates the need to rely on one-dimensional radiative-convective models (e.g., \citealt{iro05}; see also \citealt{rc78}) for calculations of the equilibrium temperature-pressure profile, thereby allowing self-consistency to be achieved.  From the perspective of equation (\ref{eq:rt}) alone, the radiative transfer scheme is strictly speaking grey (as already noted by \citealt{fhz06}) --- the shortwave, downward-propagating flux is merely diluted on its way down into the atmosphere via equation (\ref{eq:flux_s}) --- but we term it ``dual-band" to emphasize the dual-wavelength nature of our treatment.  The effects of scattering are neglected.

The bottom of the atmosphere (or the ``surface") is an idealized slab with a specifiable heat capacity, described by a single temperature $T_s$ governed by the equation,
\begin{equation}
C_{\rm int} \frac{\partial T_s}{\partial t} = {\cal F}_{\downarrow_{\rm S}} + {\cal F}_{\downarrow_{\rm L}} - {\cal F}_{\uparrow_{\rm L}},
\end{equation}
where $C_{\rm int}$ is the (areal) heat capacity of the slab.  Horizontal transport of fluid within the slab is not modelled.  In the (terrestrial) atmospheric sciences literature, this idealized slab mimics the ``mixed layer ocean", which is the $\sim 100$ m-thick layer of ocean intermediate between the atmosphere and the deep ocean, characterized by rigorous mixing induced by wave motion and turbulence.  For simplicity, we do not implement an internal heat flux, typically parametrized by the equivalent blackbody temperature $T_{\rm int}$.  For example, \cite{iro05} and \cite{guillot10} use $T_{\rm int}=100, 300$ K in their models of HD 209458b, while \cite{ls10} adopt 5.7 W m$^{-2}$ as the internal heat flux in their simulation of Jupiter (which is equivalent to $T_{\rm int} \approx 100$ K).

As the radiation is absorbed and re-emitted, the (infrared) photons encounter a longwave optical depth near the surface, described by
\begin{equation}
\tau_{\rm L_0} = \tau_{\rm L_{eq}} + \left( \tau_{\rm L_{pole}} - \tau_{\rm L_{eq}} \right) \sin^2\Phi,
\label{eq:tau_L0}
\end{equation}
where the longwave optical depths at the equator and the poles are given by $\tau_{\rm L_{eq}}$ and $\tau_{\rm L_{poles}}$, respectively.  As already noted by \cite{ms10}, equation (\ref{eq:tau_L0}) ignores the longitudinal dependence of $\tau_{\rm L_0}$ introduced by water vapour feedback and is thus inadequate for treating tidally locked, Earth-like aquaplanets.  Throughout the vertical extent of the atmosphere, the longwave optical depth is a linear combination of well-mixed and segregated atmospheric absorbers,
\begin{equation}
\tau_{\rm L} = \tau_{\rm L_w} ~\sigma_0 + \tau_{\rm L_s} ~\sigma^{n_{\rm L}}_0,
\label{eq:tau_lw}
\end{equation}
where $\tau_{\rm L_w}$ and $\tau_{\rm L_s}$ generally depend on geometry (i.e., $\Phi$ and $\Theta$) as well as atmospheric chemistry.  The second term in equation (\ref{eq:tau_lw}) represents species of absorbers where the associated pressure scale height is $1/n_{\rm L}$ that of the well-mixed species.  For example, the linear term may represent well-mixed species like carbon dioxide, while setting $n_{\rm L}=4$ approximates the structure of water vapour in the terrestrial atmosphere \citep{fhz06}.  As another example, setting $n_{\rm L}=2$ approximates the process of collision-induced absorption \citep{herzberg52} within the atmospheres of the giant planets in our Solar System \citep{ls10}.  Following \cite{fhz06,fhz07} and \cite{f07}, we use
\begin{equation}
\begin{split}
&\tau_{\rm L_w} = \tau_{\rm L_0} f_l, \\
&\tau_{\rm L_s} = \tau_{\rm L_0} \left( 1 - f_l \right),\\
\end{split}
\label{eq:tau_lw2}
\end{equation}
with $f_l$ being a dimensionless parameter that controls the strength of the well-mixed longwave absorber.

In \cite{fhz06,fhz07}, \cite{f07} and \cite{os08}, a key assumption is that the (water) moisture content does not affect radiative transfer, such that the dynamical effects of the latent heat release associated with the condensation of water vapour can be isolated from its radiative effects.  \cite{ms10} improved upon this treatment by allowing the optical depth associated with water vapour ($\tau_{\rm L_s}; n_{\rm L}=4$) to depend on the instantaneous column density, thereby providing a simple treatment of water vapour feedback.  Since we are concerned only with dry atmospheres in the present study, we will ignore this issue.

\subsection{Convection}
\label{subsect:convection}

In simulations of atmospheric circulation, convection is typically mimicked using simplified parametrizations known as ``convective adjustment" schemes (e.g., \citealt{manabe}; \S3.6.9 of \citealt{wp05}).  For a dry atmosphere in hydrostatic equilibrium, the change in temperature of a parcel of atmosphere resulting from vertical motion is (e.g., \S2.7.2 of \citealt{holton})\footnote{Equation (\ref{eq:lapse}) may be generalized to describe an atmosphere containing moisture and with a corresponding moist adiabatic lapse rate.}
\begin{equation}
\frac{1}{\theta_{\rm T}} \frac{\partial \theta_{\rm T}}{\partial z} = \frac{1}{T} \left( \Gamma_0 - \Gamma \right),
\label{eq:lapse}
\end{equation}
where the lapse rate is defined as
\begin{equation}
\Gamma \equiv - \frac{\partial T}{\partial z},
\end{equation}
with $z$ being the vertical spatial coordinate, and the dry adiabatic lapse rate is $\Gamma_0 \equiv g_p/c_P$, which has a value of about 9.8 K km$^{-1}$ for the terrestrial atmosphere.  The potential temperature,
\begin{equation}
\theta_{\rm T} \equiv T \sigma^{-\kappa}_0,
\label{eq:potential_temp}
\end{equation}
where $\kappa \equiv {\cal R}/c_P$ and ${\cal R}$ is the ideal gas constant, is the temperature a parcel of atmosphere would have if it was compressed or expanded adiabatically from the pressure $P$ to the reference pressure $P_0$ at the bottom of the simulation domain.  The adiabatic coefficient $\kappa$ is related to the number of degrees of freedom of the atmospheric gas $n_{\rm dof}$: $\kappa = 2/(n_{\rm dof}+2)$ (see \S2.3.3 of \citealt{p10}).  For example, molecular hydrogen is experimentally found to have $n_{\rm dof} \approx 5.2$, close to the theoretical value for a diatomic gas ($n_{\rm dof}=5$).  The atmosphere is unstable if $\Gamma > \Gamma_0$ (i.e., super-adiabatic lapse rates).\footnote{The condition for stability is sometimes known as the Schwarzschild criterion.}  When the lapse rate is adjusted to a state of stability --- subjected to the constraint of total energy conservation --- convective adjustment is said to be performed.  In other words, the tendency of convection is to minimize the magnitude of the potential temperature gradient $\partial \theta_{\rm T}/\partial z$ and convective adjustment is an approximate way to mimic this process \citep{rc78}.

Within the \texttt{FMS}, the dry convective adjustment scheme follows the prescription of \cite{manabe}.  At a given time, the temperature difference $\delta T$ resulting from departures from the adiabatic lapse rate is computed,
\begin{equation}
\begin{split}
&\frac{\partial \theta_{\rm T}}{\partial P} = 0, \\
&\frac{c_P}{g_p} \int \delta T ~dP = 0,\\
\end{split}
\end{equation}
where the integral is taken over the vertical extent of the unstable, model atmosphere layer.  An implicit assumption made is that when the lapse rate of a layer is super-adiabatic, convection is rigorous enough to maintain a neutral lapse rate of potential temperature.  The kinetic energy created by convection is then dissipated and instantly converted into heat, such that the total potential energy is invariant to convective adjustment.  Within each model layer, the temperature is then adjusted by $\delta T$.

It is worth noting that there exists a body of work on moist convective adjustment schemes, largely motivated by the pioneering work of \cite{manabe}, \cite{b86} and \cite{bm86}, who proposed the convective adjustment to reference temperature and humidity profiles instead of $\Gamma_0$.  \cite{f07} adopted a simplified version of the Betts-Miller scheme, where the reference humidity profile is held at a constant threshold value.

\subsection{Atmospheric boundary layer}
\label{subsect:boundary}

\begin{figure}
\begin{center}
\includegraphics[width=0.48\columnwidth]{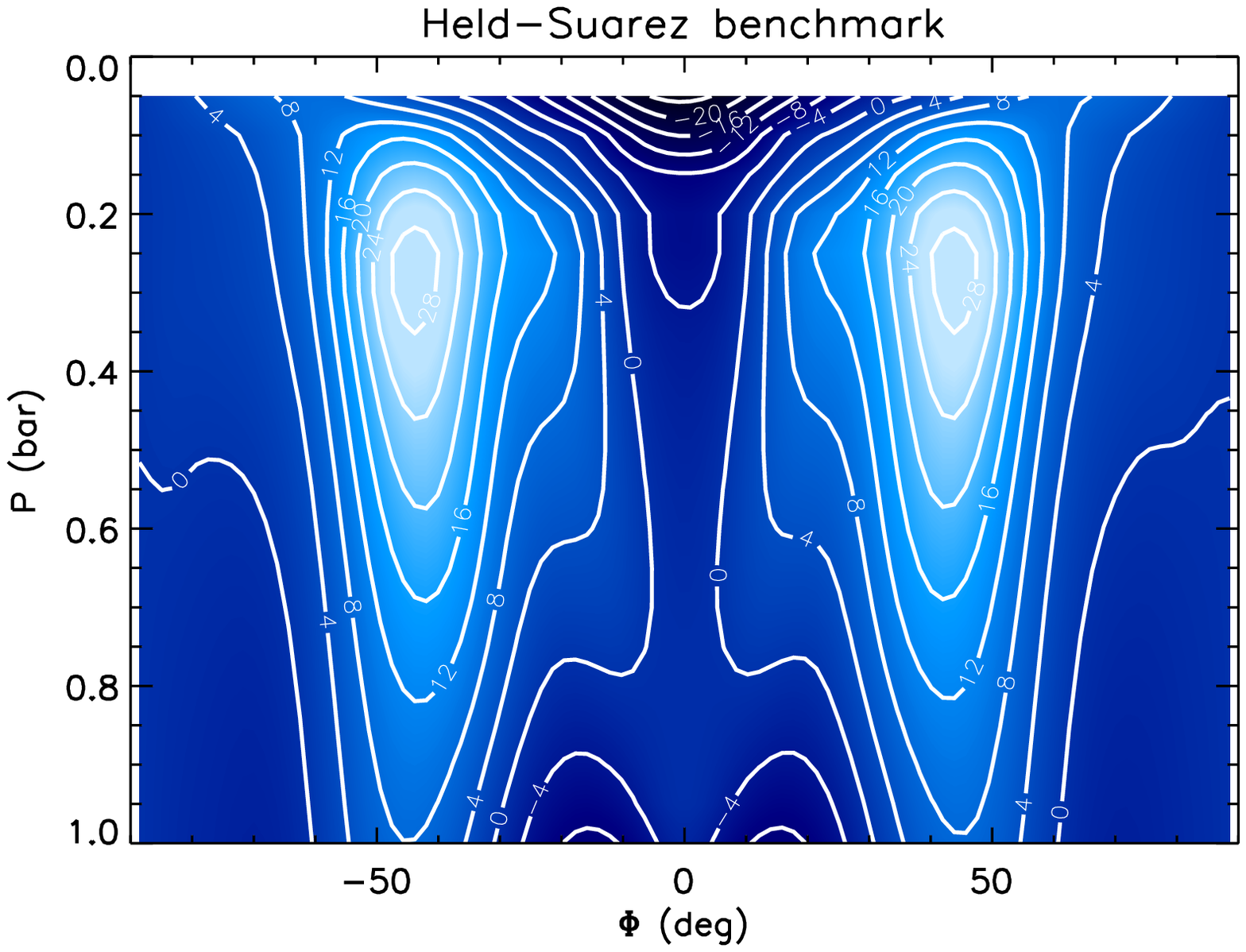}
\includegraphics[width=0.48\columnwidth]{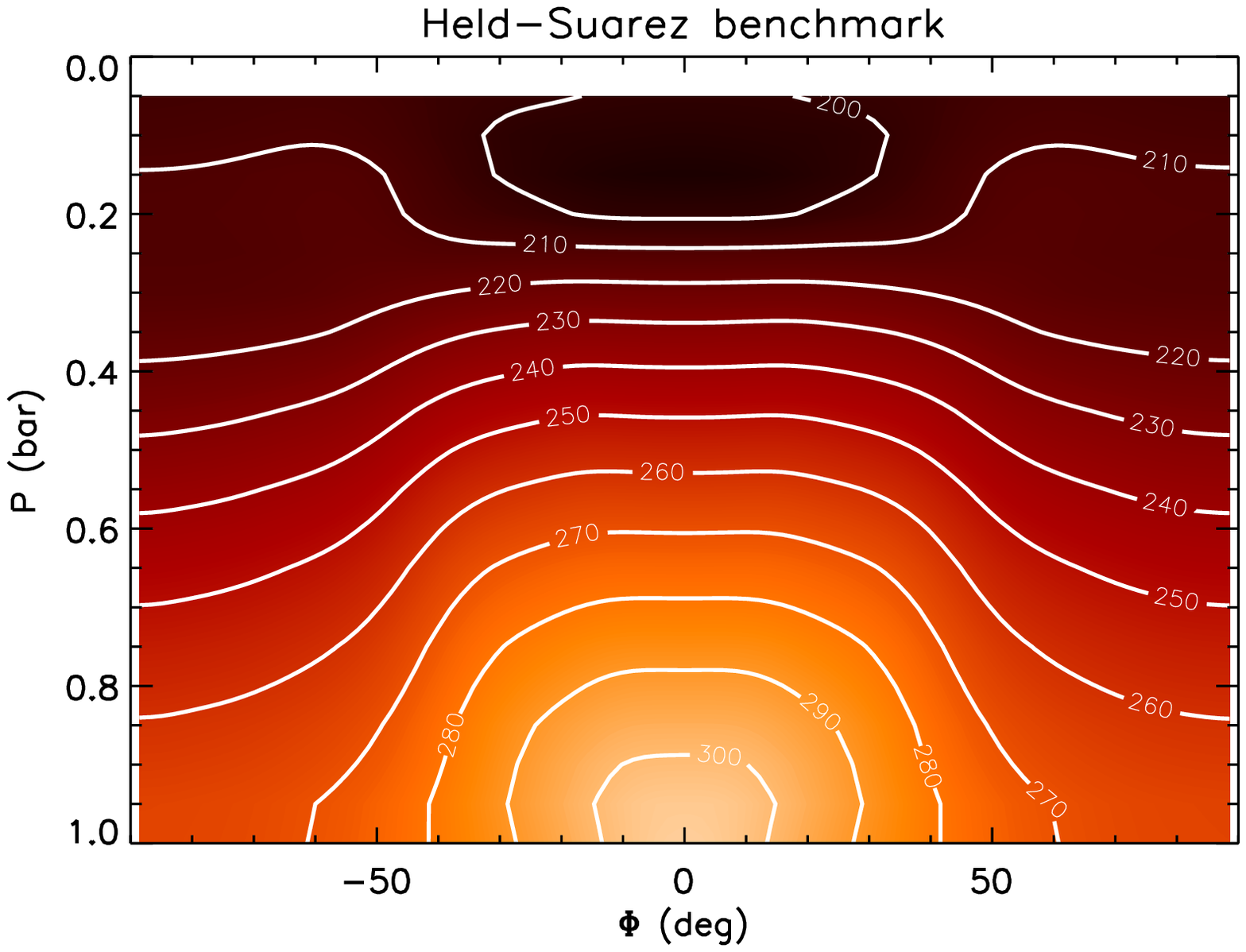}
\includegraphics[width=0.48\columnwidth]{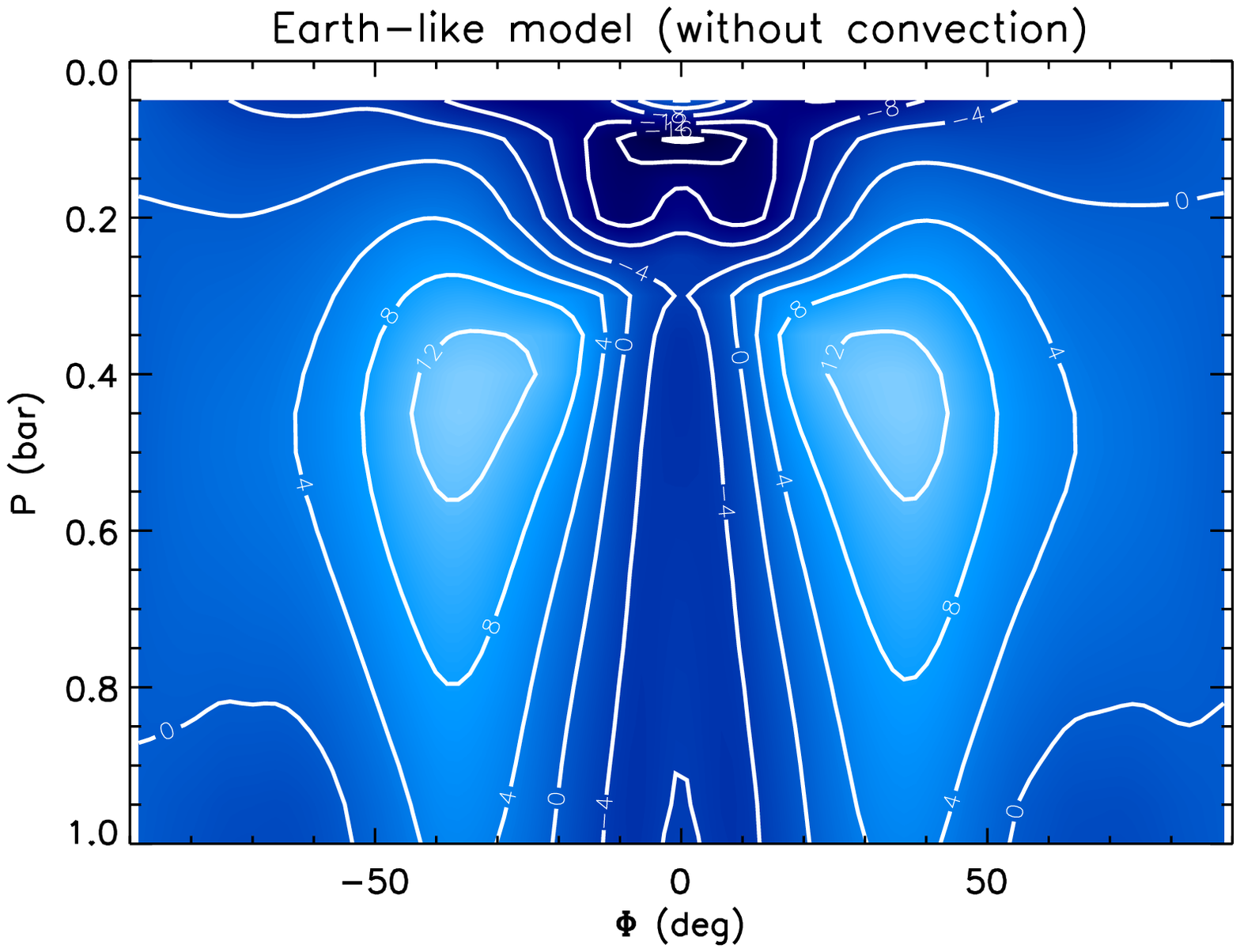}
\includegraphics[width=0.48\columnwidth]{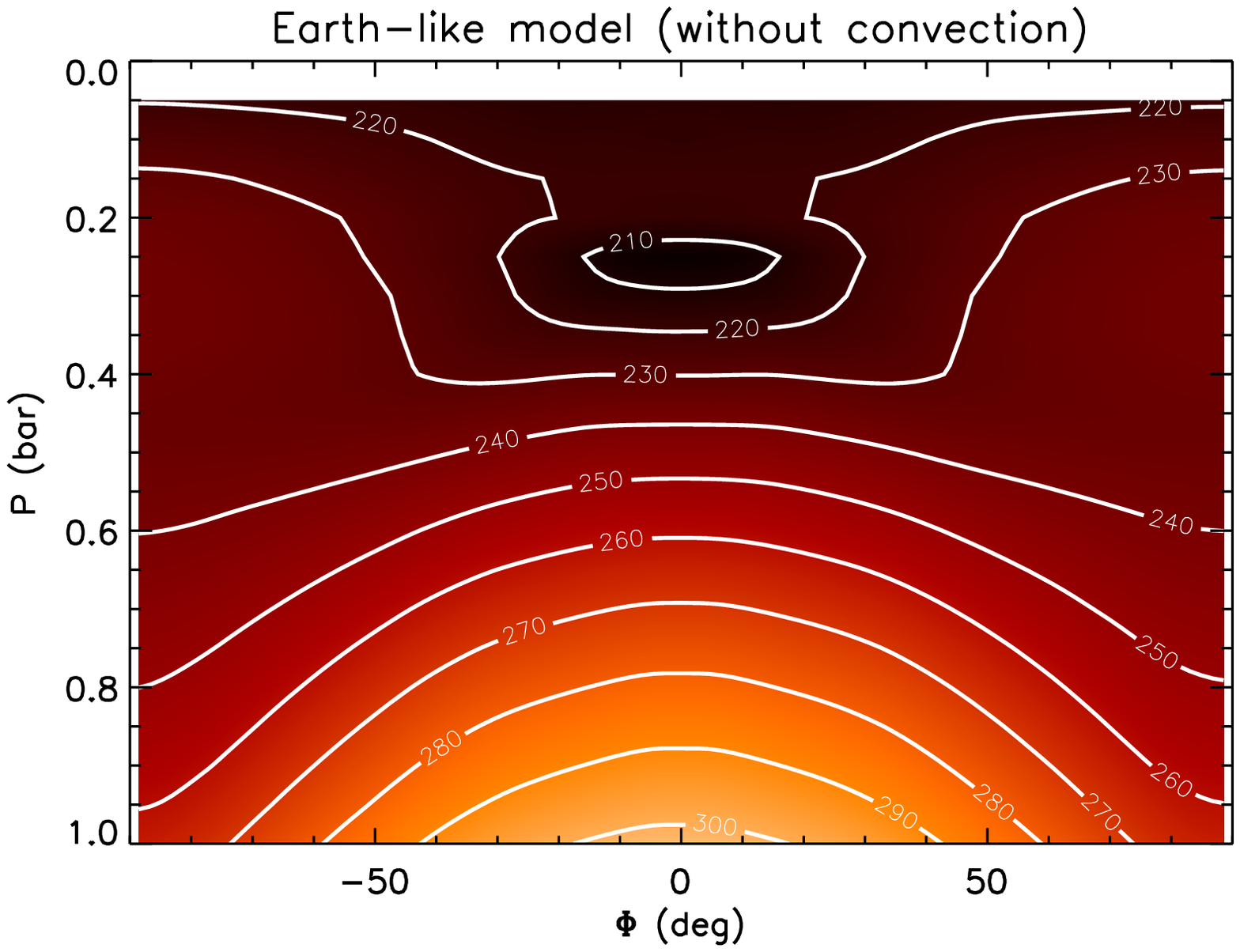}
\includegraphics[width=0.48\columnwidth]{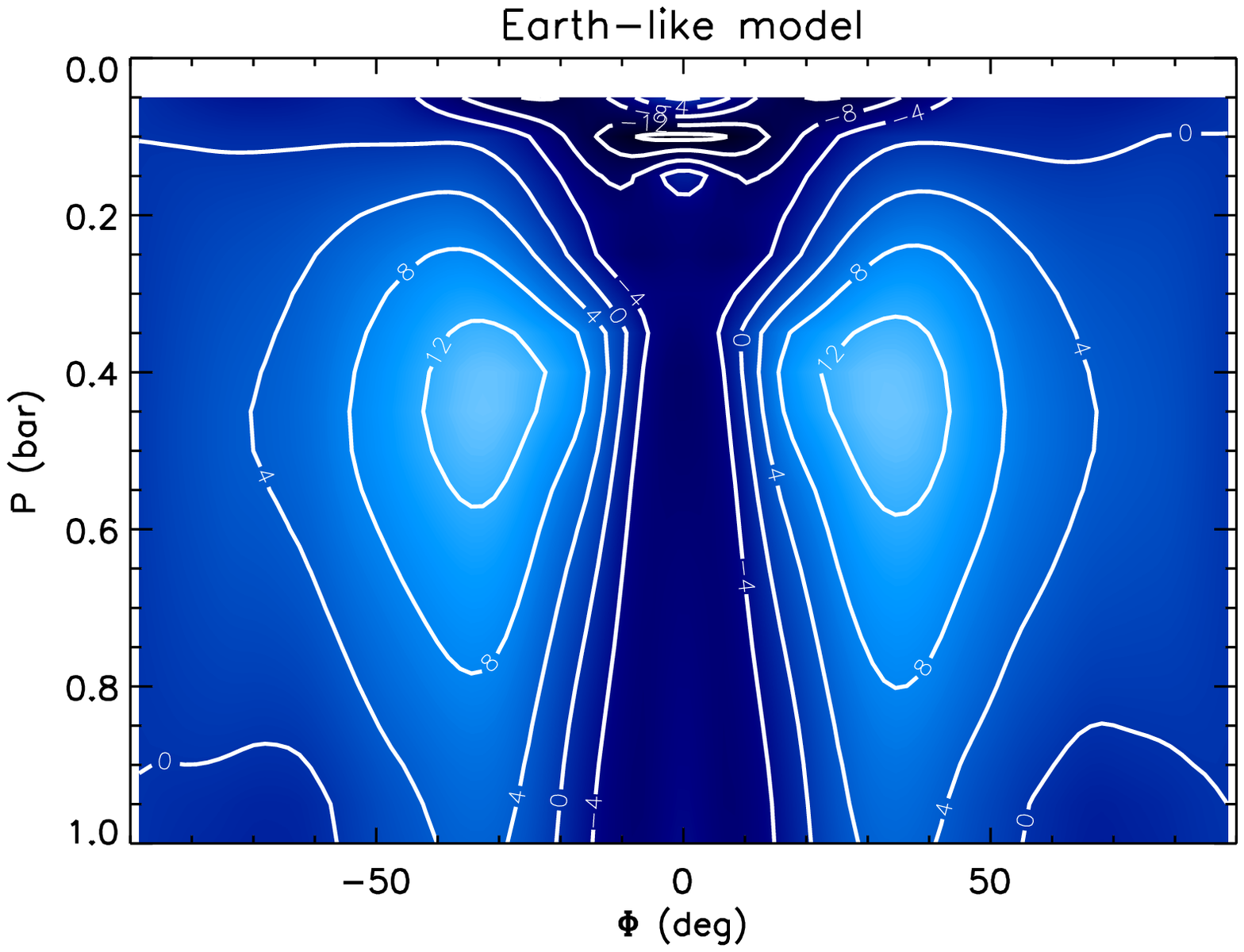}
\includegraphics[width=0.48\columnwidth]{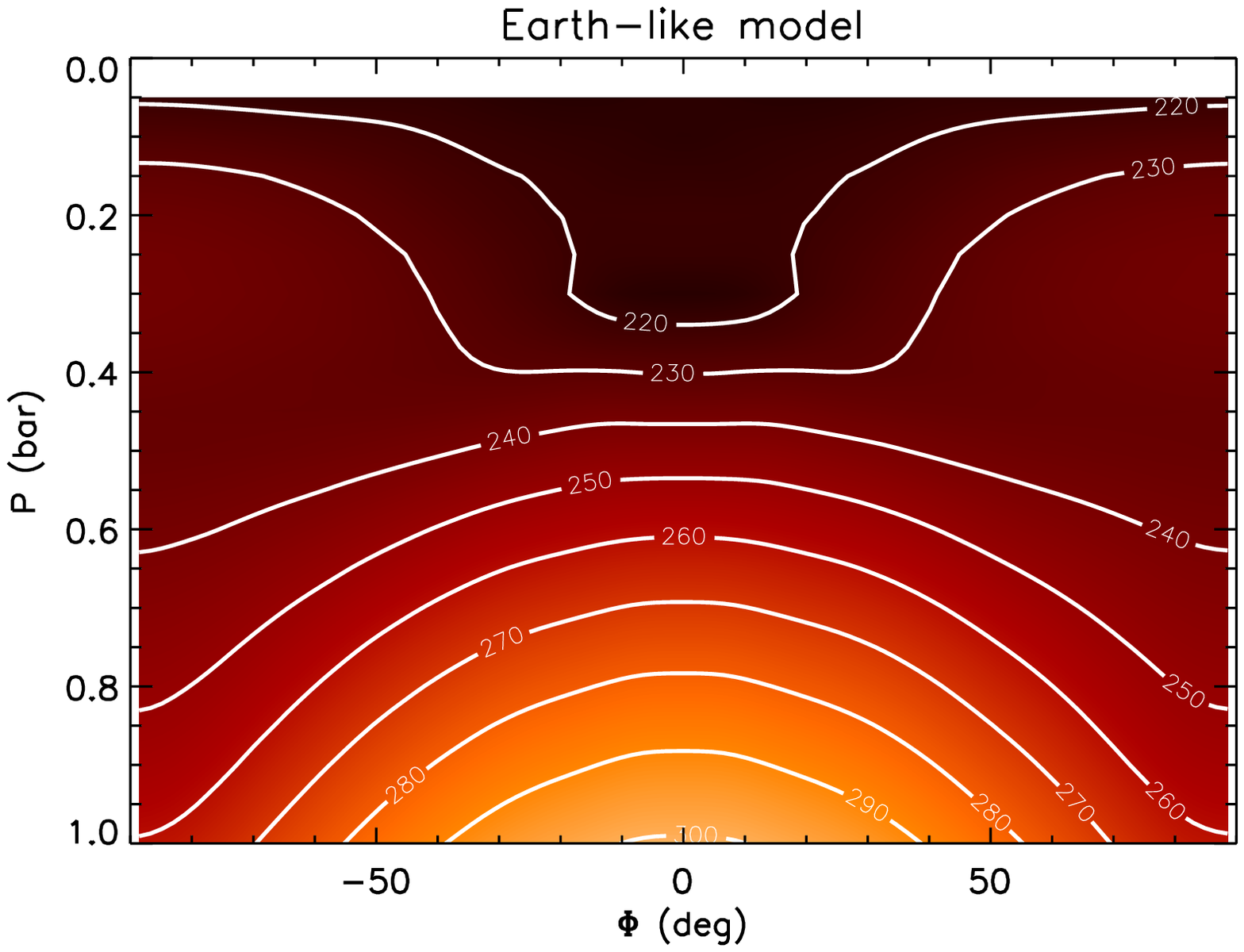}
\end{center}
\vspace{-0.2in}
\caption{Temporally-averaged, zonal-mean zonal wind (left column) and temperature (right column) profiles as functions of vertical pressure $P$ and latitude $\Phi$.  Top row: Held-Suarez dynamical benchmark.  Middle row: Earth-like model without convection.  Bottom row: Earth-like model with convection.  Contours are in units of m s$^{-1}$ (left column) and K (right column).}
\label{fig:earth_hs}
\vspace{-0.1in}
\end{figure}

\begin{figure}
\begin{center}
\includegraphics[width=0.48\columnwidth]{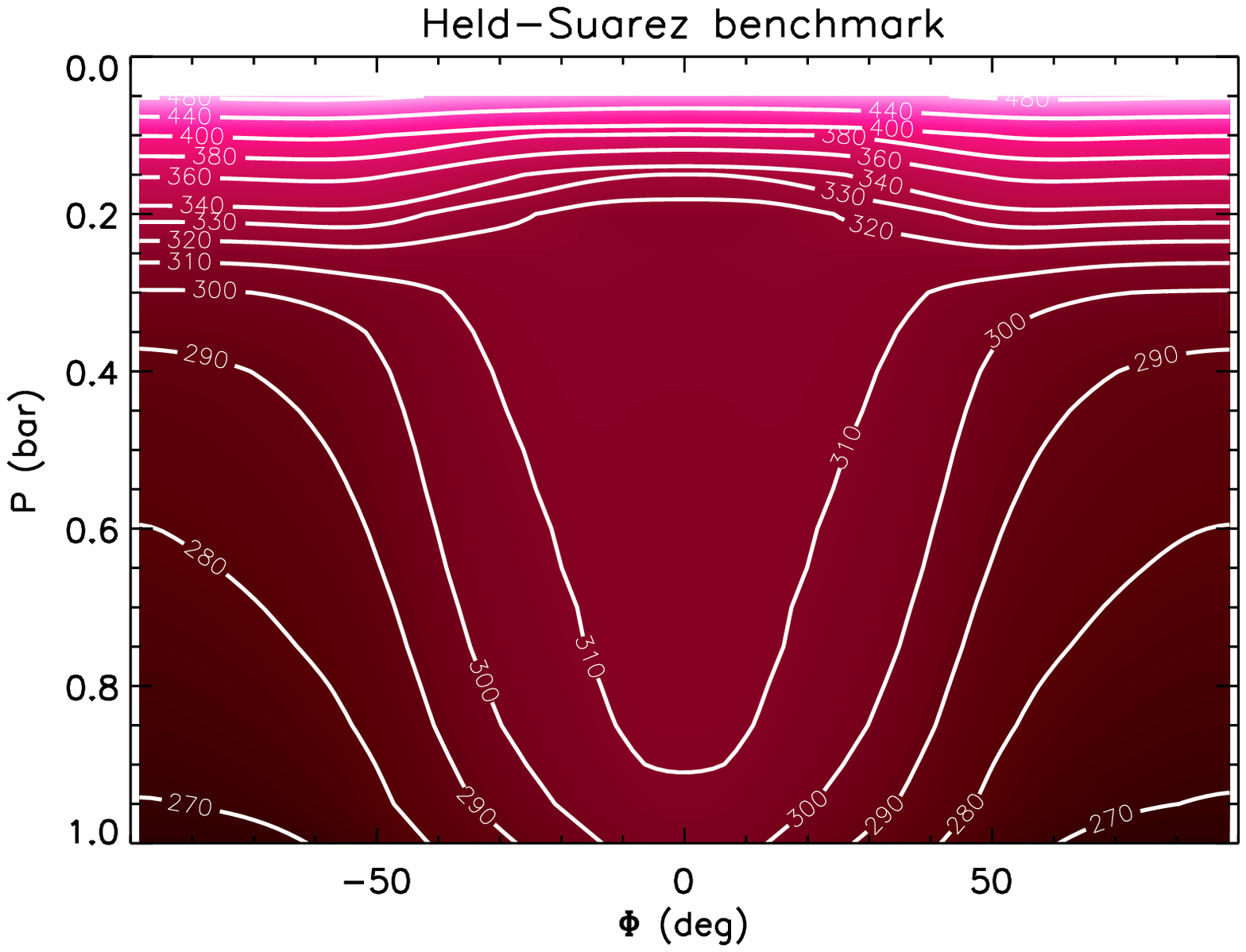}
\includegraphics[width=0.48\columnwidth]{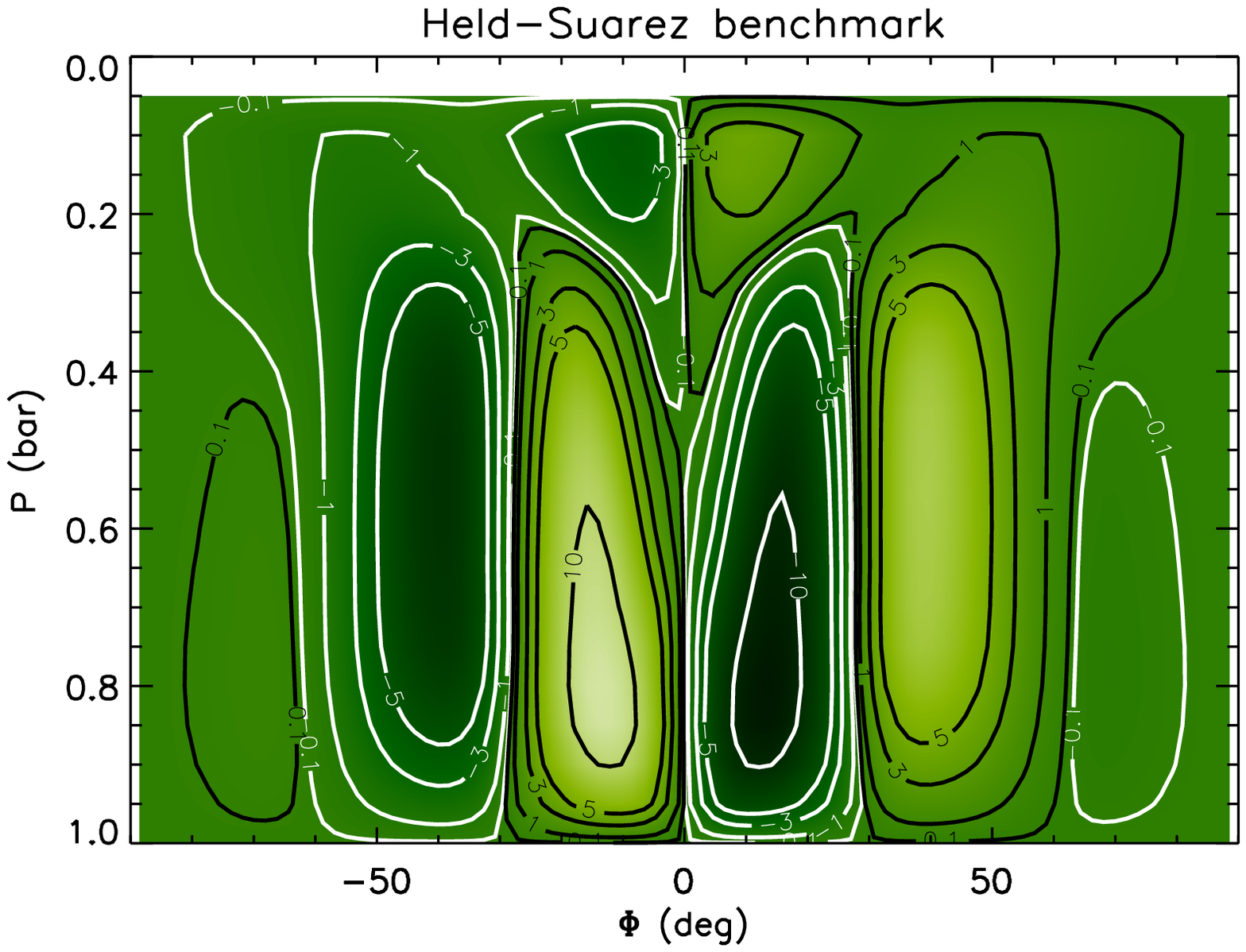}
\includegraphics[width=0.48\columnwidth]{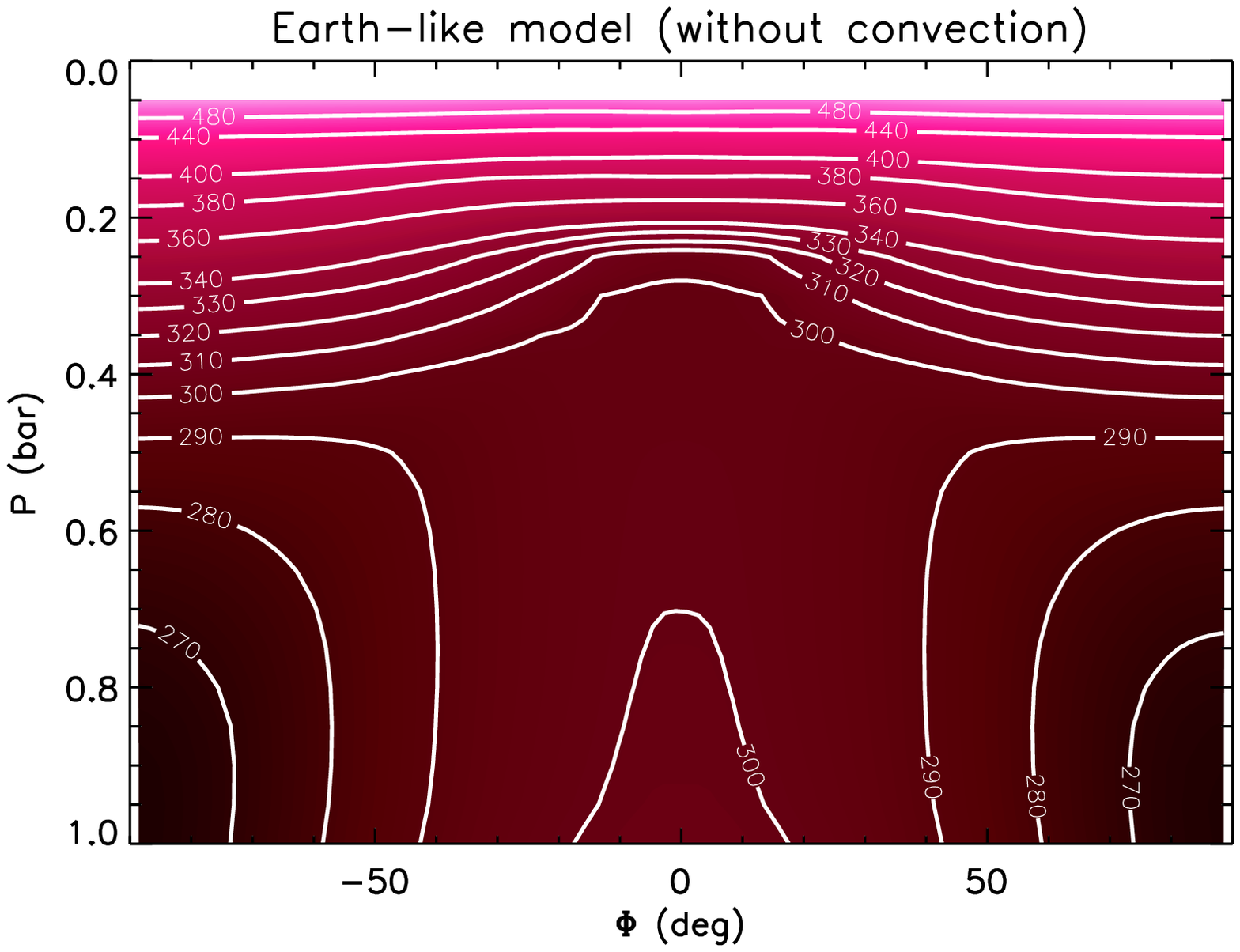}
\includegraphics[width=0.48\columnwidth]{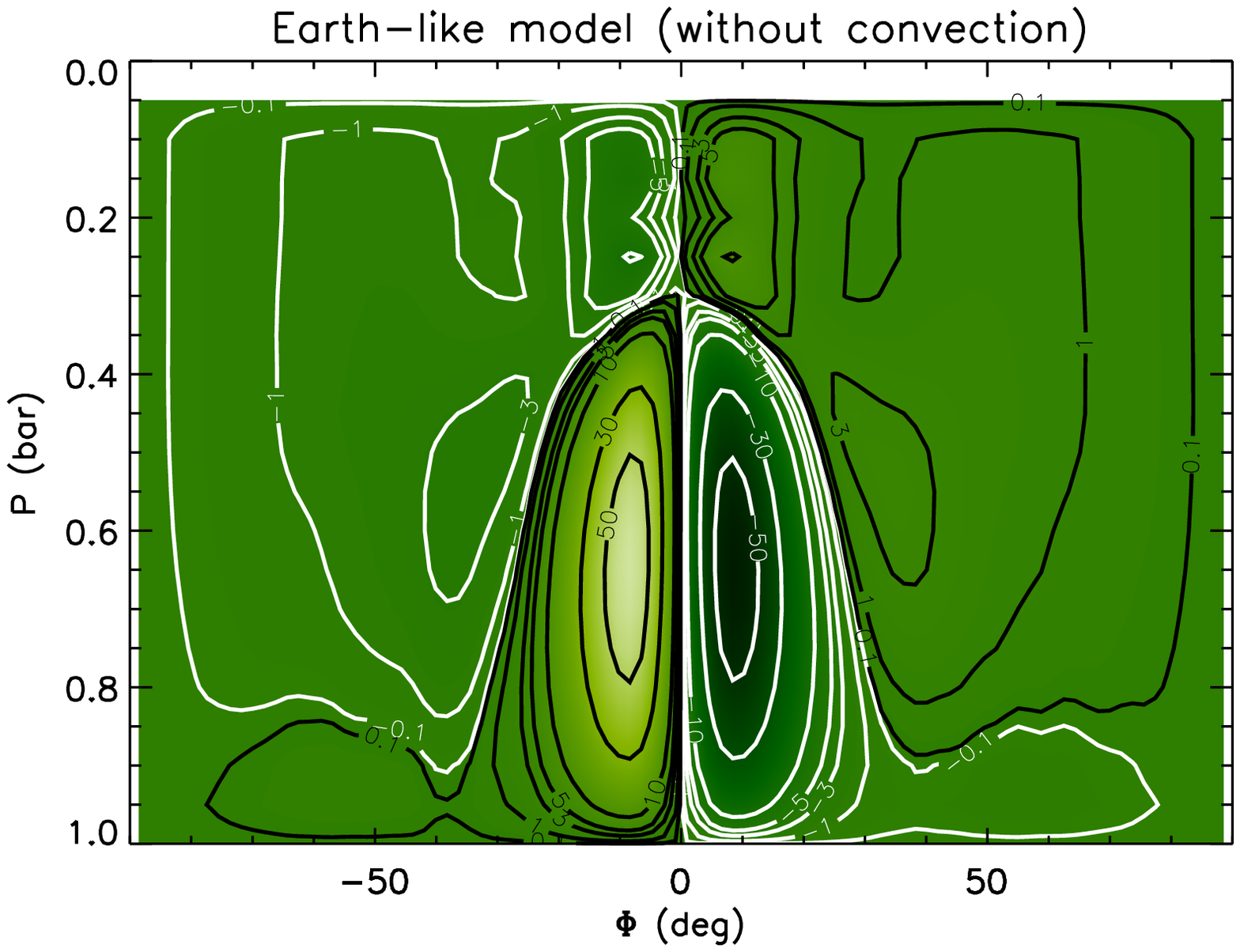}
\includegraphics[width=0.48\columnwidth]{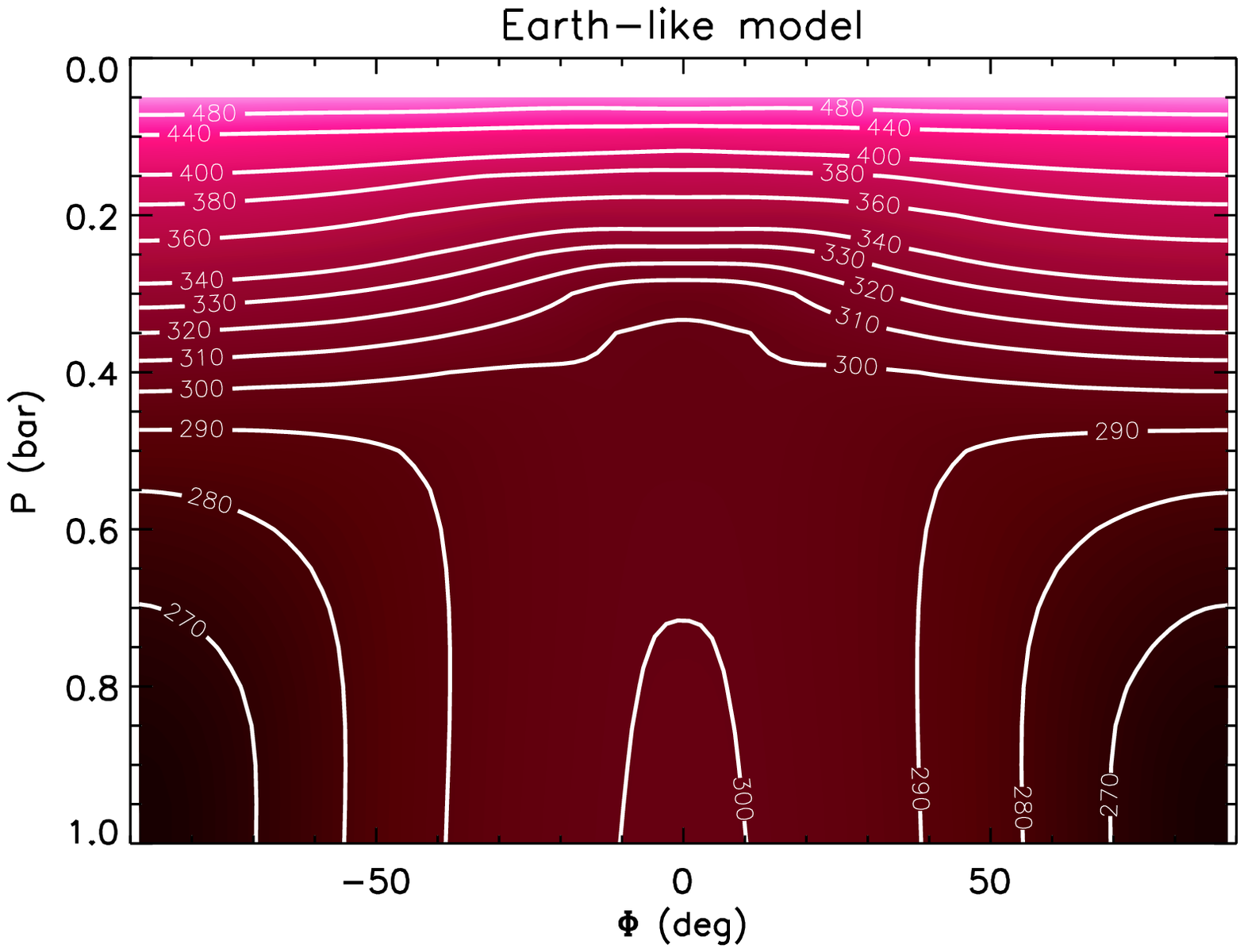}
\includegraphics[width=0.48\columnwidth]{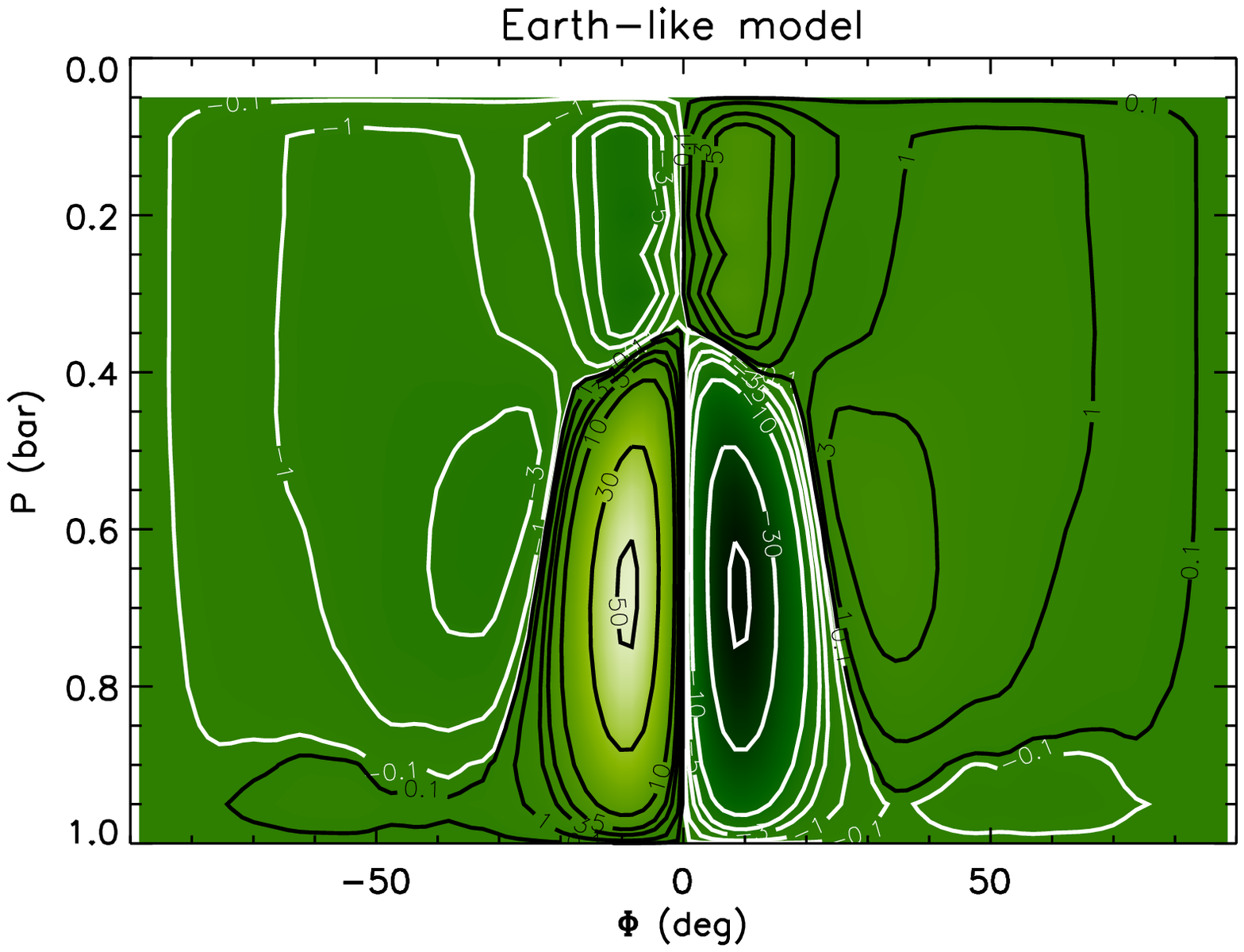}
\end{center}
\vspace{-0.2in}
\caption{Temporally-averaged, zonal-mean potential temperature (left column) and the Eulerian mean streamfunction (right column) profiles as functions of vertical pressure $P$ and latitude $\Phi$.  Top row: Held-Suarez dynamical benchmark.  Middle row: Earth-like model without convection.  Bottom row: Earth-like model with convection.  Contours are in units of K (left column) and $\times 10^9$ kg s$^{-1}$ (right column).}
\label{fig:earth_hs2}
\vspace{-0.15in}
\end{figure}

\begin{figure}
\begin{center}
\includegraphics[width=0.5\columnwidth]{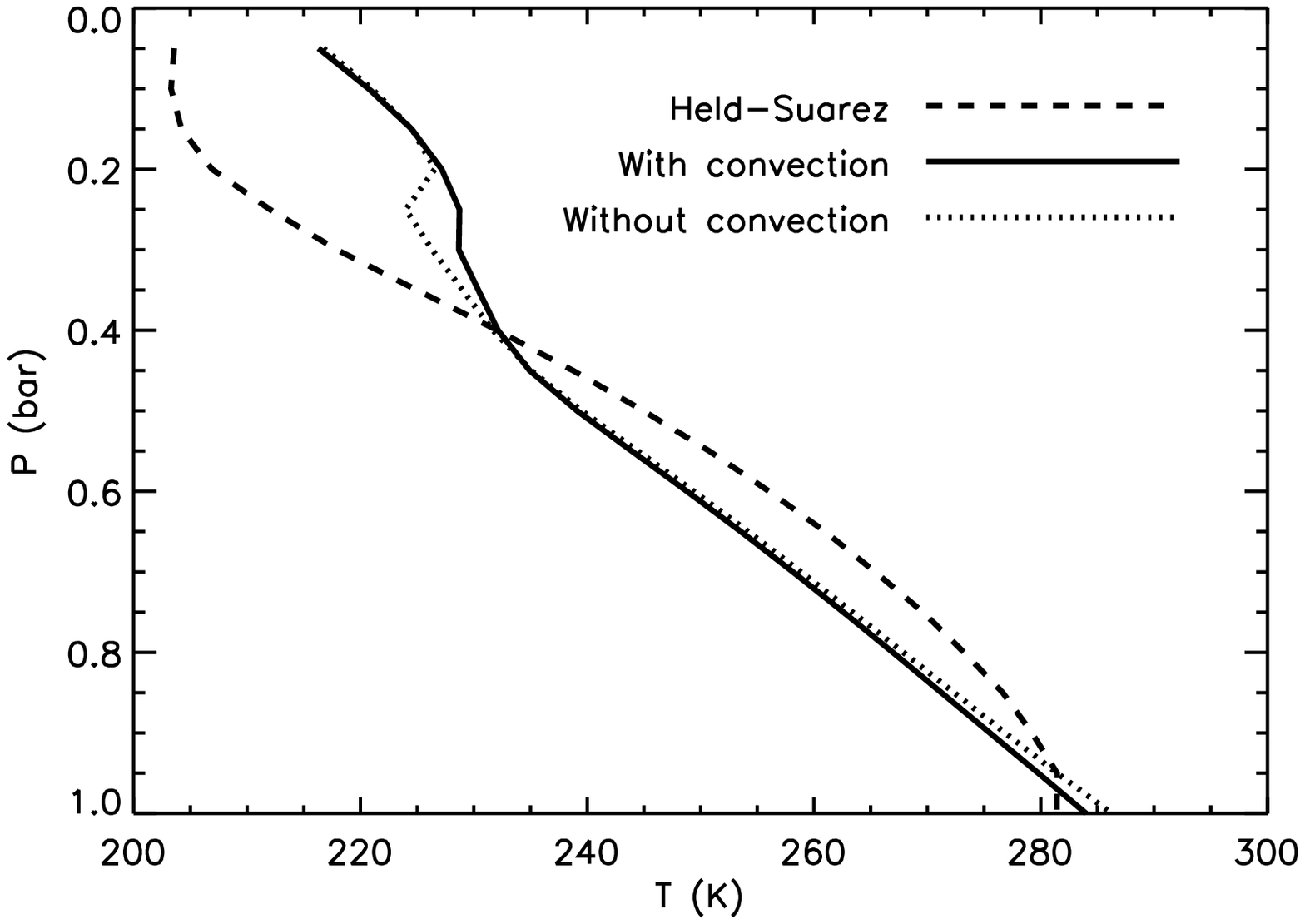}
\end{center}
\vspace{-0.2in}
\caption{Globally-averaged temperature-pressure profiles for the trio of Earth-like simulations presented in \S\ref{subsect:earth}.}
\label{fig:tp_earth}
\end{figure}

The boundary layer is the part of the atmosphere where the flow is strongly influenced by interaction with the surface.  It can be thought of as enforcing a ``no slip" boundary condition, such that large velocity gradients --- and hence shears --- exist across a small vertical height, which induces turbulent mixing (e.g., \S5 of \citealt{holton}).  The boundary layer may become convectively unstable, which induces mixing.  The end result is the same: eddies are produced which are smaller in size than those resulting from large-scale rotational flows, are typically unresolved in global simulations of atmospheric circulation and thus need to be accounted for using sub-grid models (e.g., \citealt{tm86}).  In other words, turbulence within the boundary layer may be driven mechanically or via buoyancy.  The eddies tend to have comparable horizontal and vertical length scales --- i.e., they are effectively three-dimensional --- and are responsible for heat and momentum exchanges between the atmosphere and the surface.  Consequently, the boundary layer affects the dynamics and thermodynamics of the terrestrial atmosphere and is responsible for a non-negligible fraction of kinetic energy loss by the flow \citep{g94}.  On Earth, the height of the boundary layer is $\sim 30$--3000 m and contains $\sim 10\%$ of the mass of the terrestrial atmosphere.

We find that the boundary layer scheme is needed if the bottom of the simulation domain is placed within the active part of the atmosphere, such as in an Earth-like simulation.  However, this scheme is not required to successfully execute hot Jupiter simulations.  Moreover, it is not entirely clear if the analogy with a terrestrial boundary layer can be carried over to mimic drag in the deep, inert layers of a hot Jupiter.  As such, although we include a description of our boundary layer scheme for completeness (since it is used for our Earth-like simulations), we relegate the technical details to Appendix \ref{append:pbl}.  As described therein, our atmospheric boundary layer scheme requires the specification of four additional parameters: the roughness length $z_{\rm rough}$, the critical bulk Richardson number ${\cal R}_{i,{\rm crit}}$, the von K\'{a}rm\'{a}n constant $\kappa_{\rm vK}$ and the surface layer fraction $f_b$.

\subsection{Additional physical quantities}
\label{subsect:additional}

\begin{figure}
\begin{center}
\includegraphics[width=0.48\columnwidth]{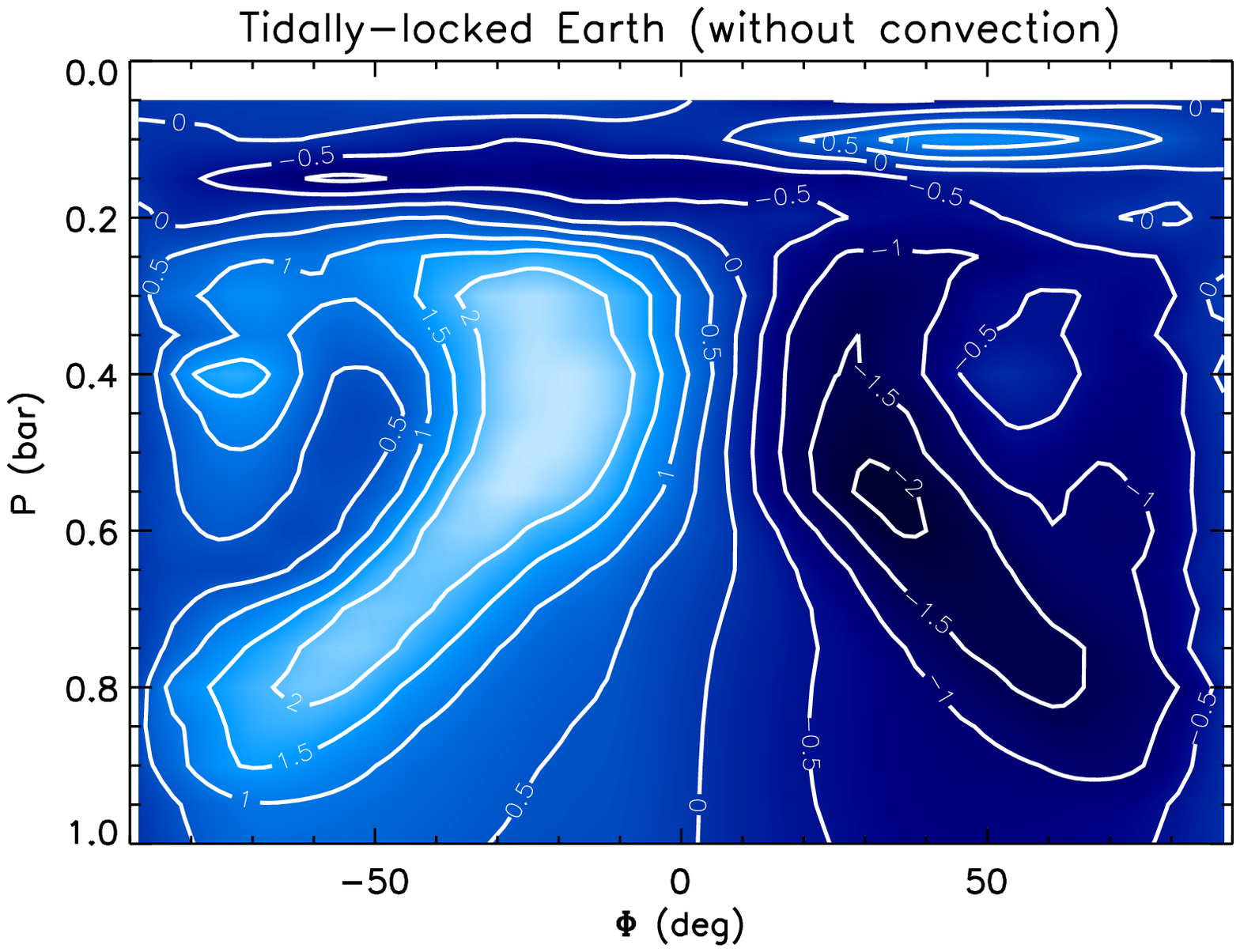}
\includegraphics[width=0.48\columnwidth]{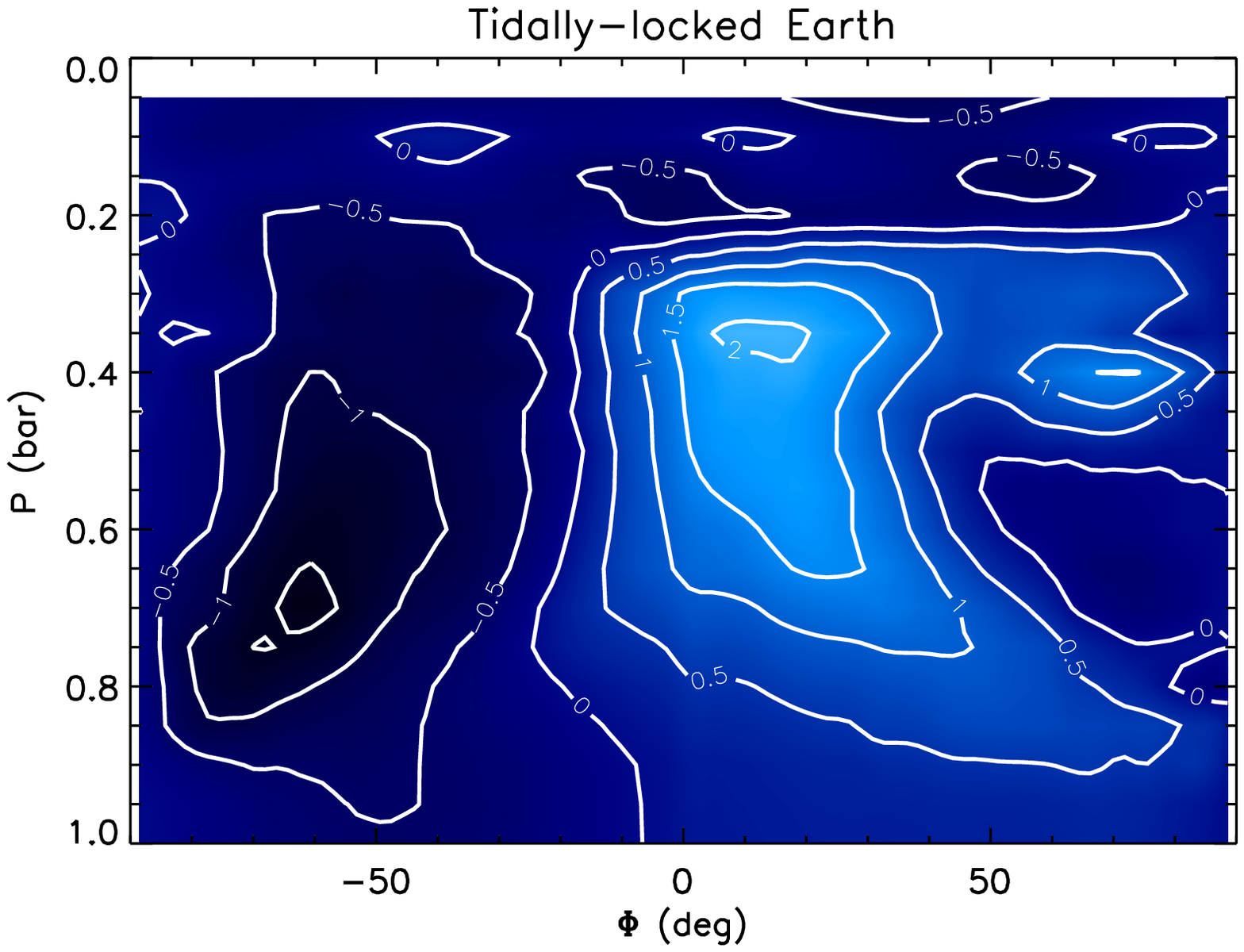}
\includegraphics[width=0.48\columnwidth]{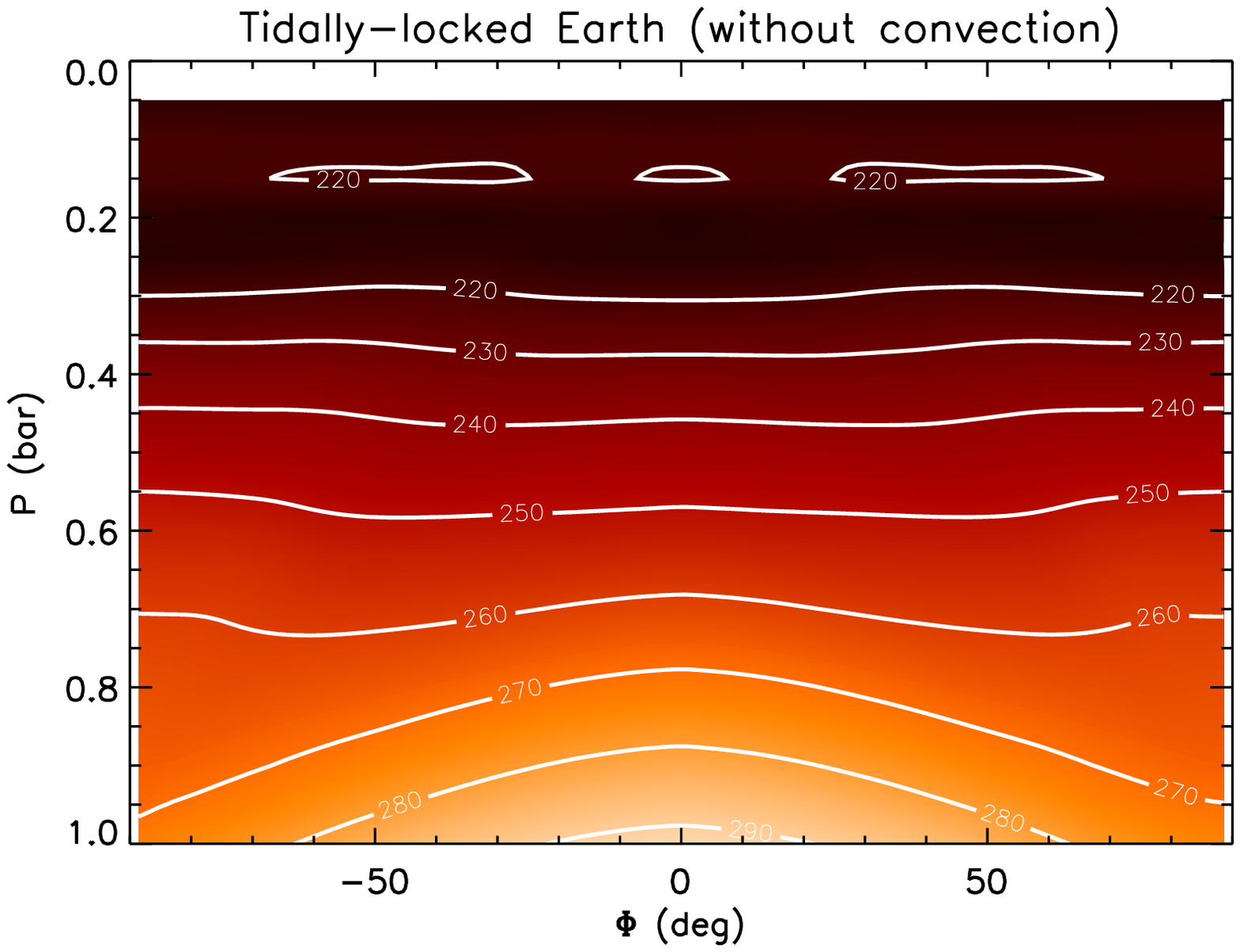}
\includegraphics[width=0.48\columnwidth]{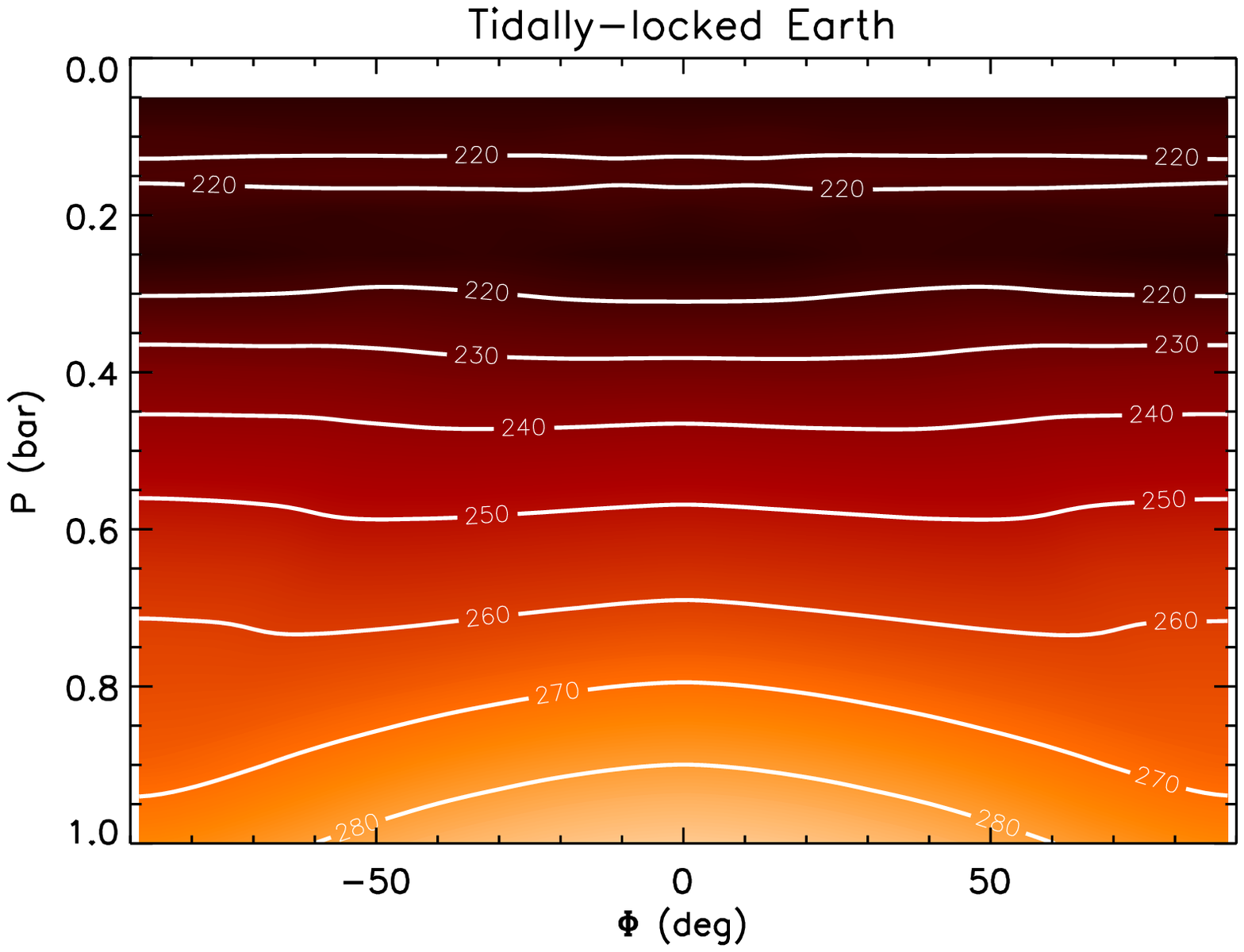}
\includegraphics[width=0.48\columnwidth]{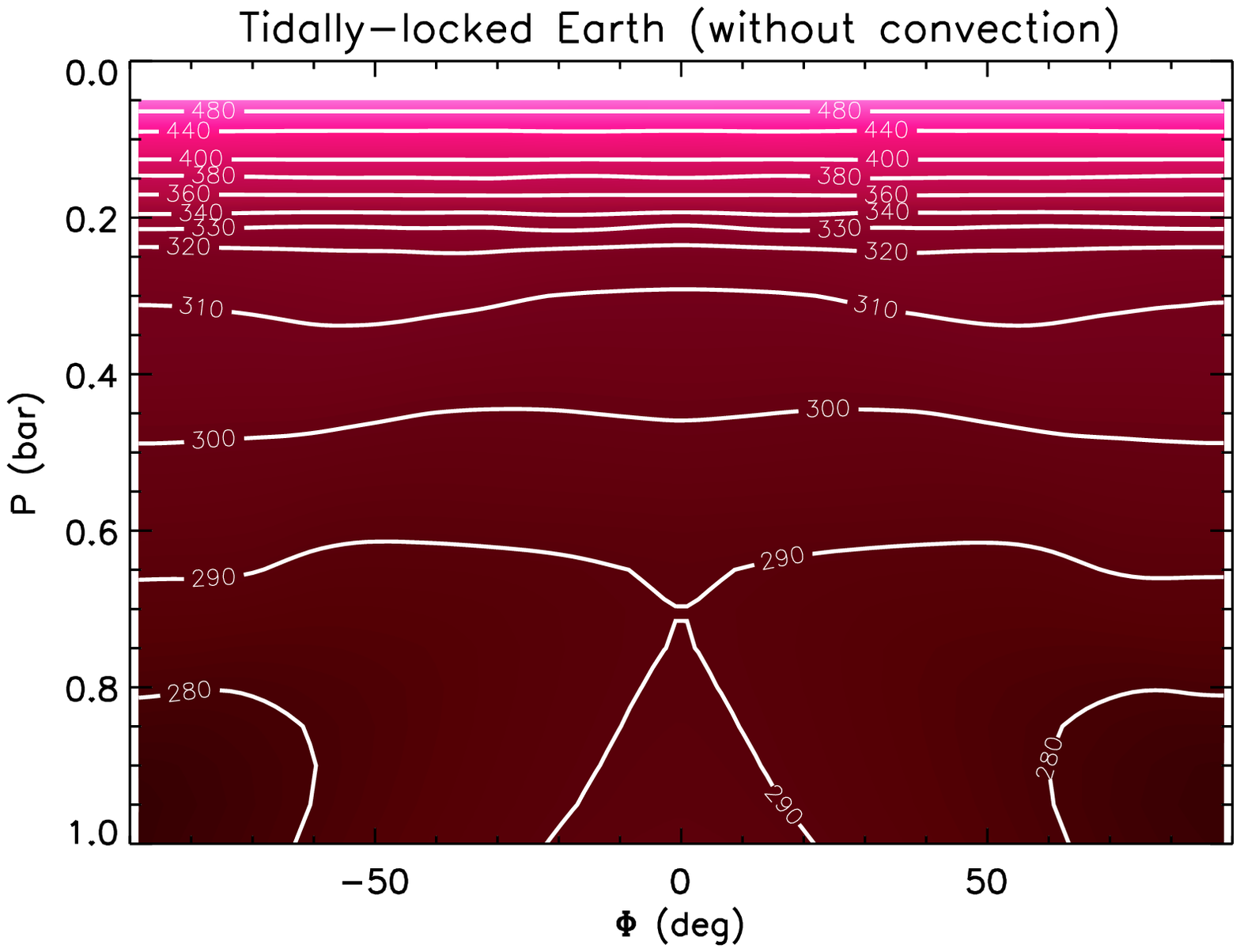}
\includegraphics[width=0.48\columnwidth]{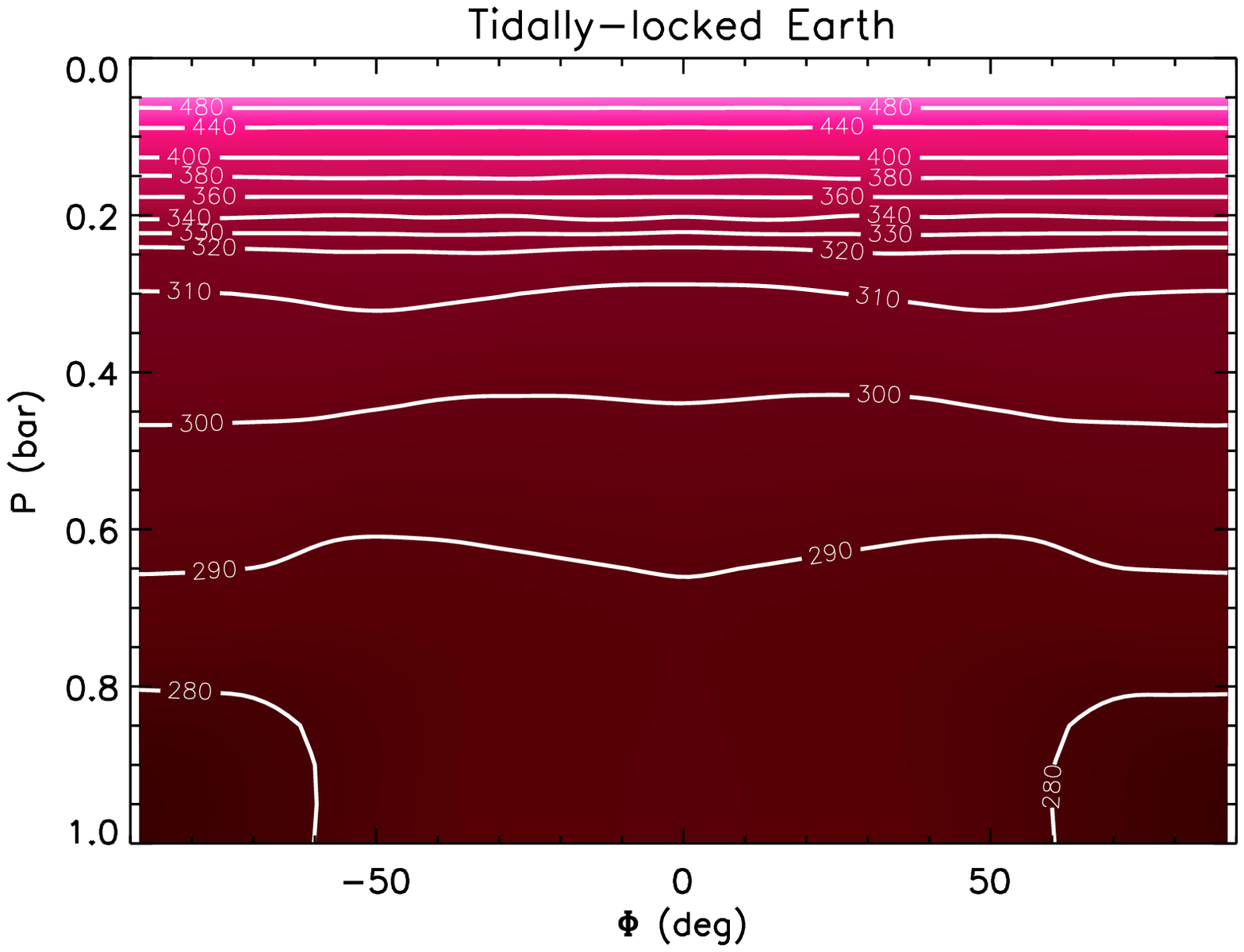}
\end{center}
\vspace{-0.2in}
\caption{Held-Suarez mean flow quantities for the tidally-locked Earth simulation.  Top row: zonal wind (m s$^{-1}$).  Middle row: temperature (K).  Bottom row: potential temperature (K).  Left column: without convection.  Right column: with convection.}
\label{fig:tidal_earth}
\vspace{-0.1in}
\end{figure}

\begin{figure}
\begin{center}
\includegraphics[width=0.48\columnwidth]{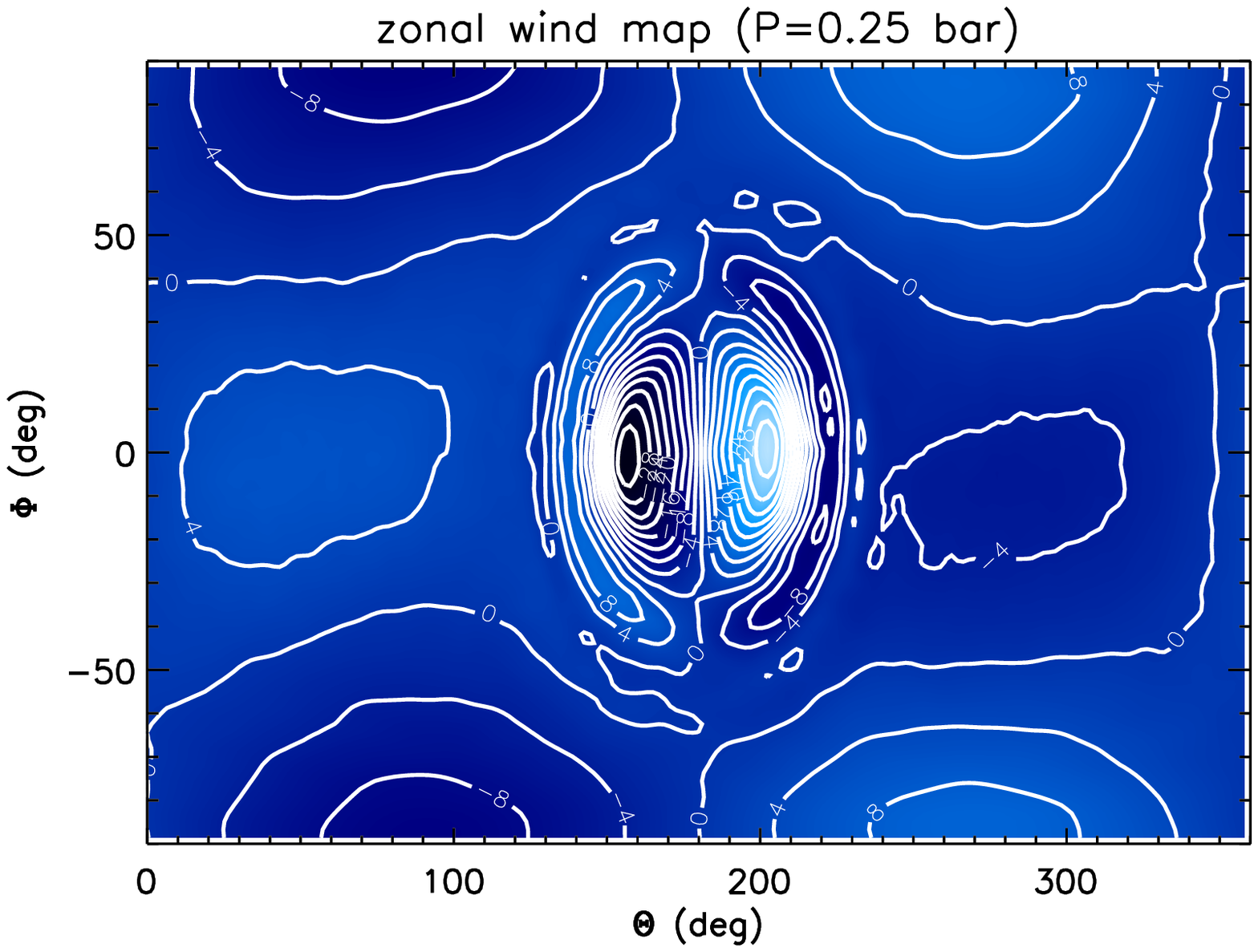}
\includegraphics[width=0.48\columnwidth]{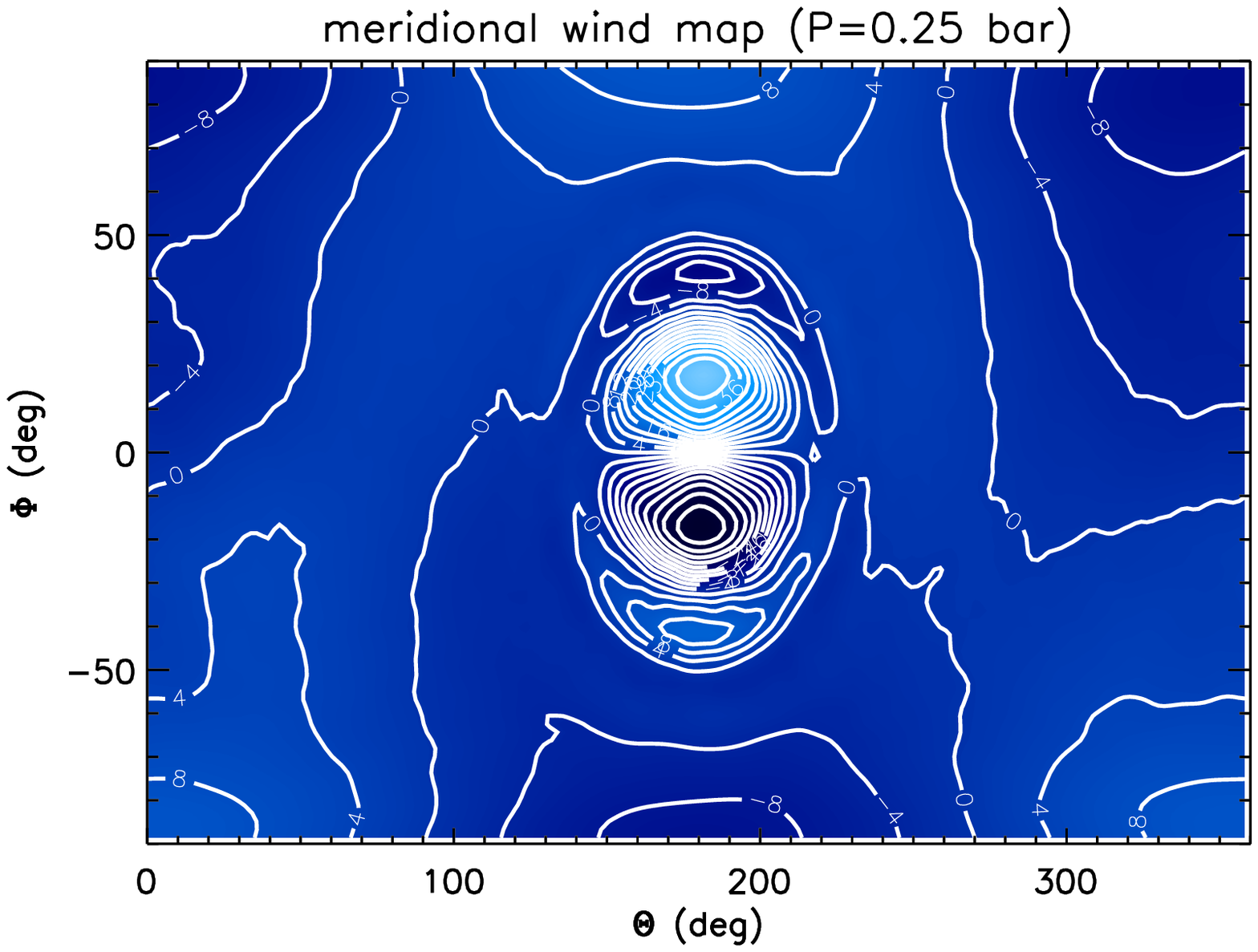}
\includegraphics[width=0.48\columnwidth]{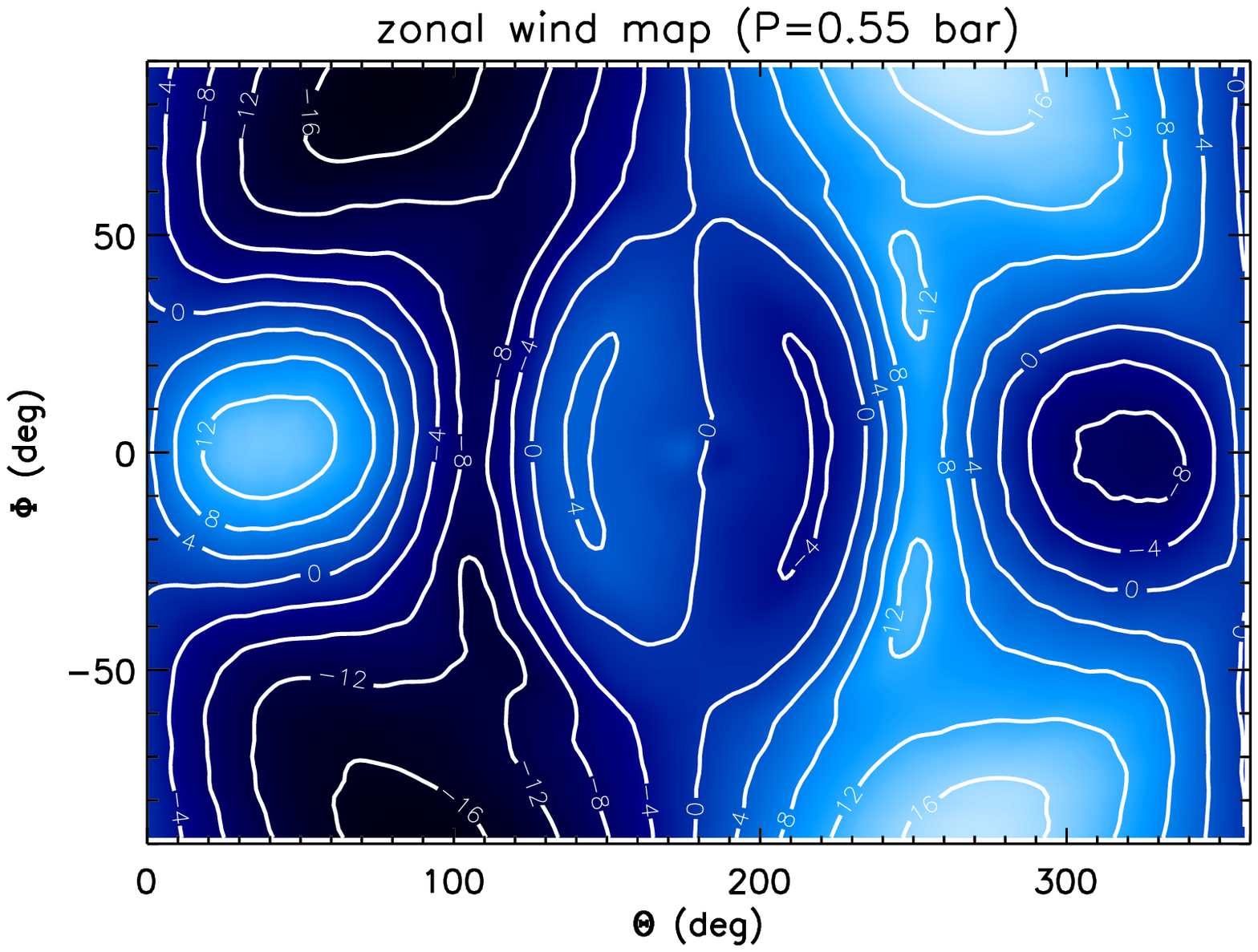}
\includegraphics[width=0.48\columnwidth]{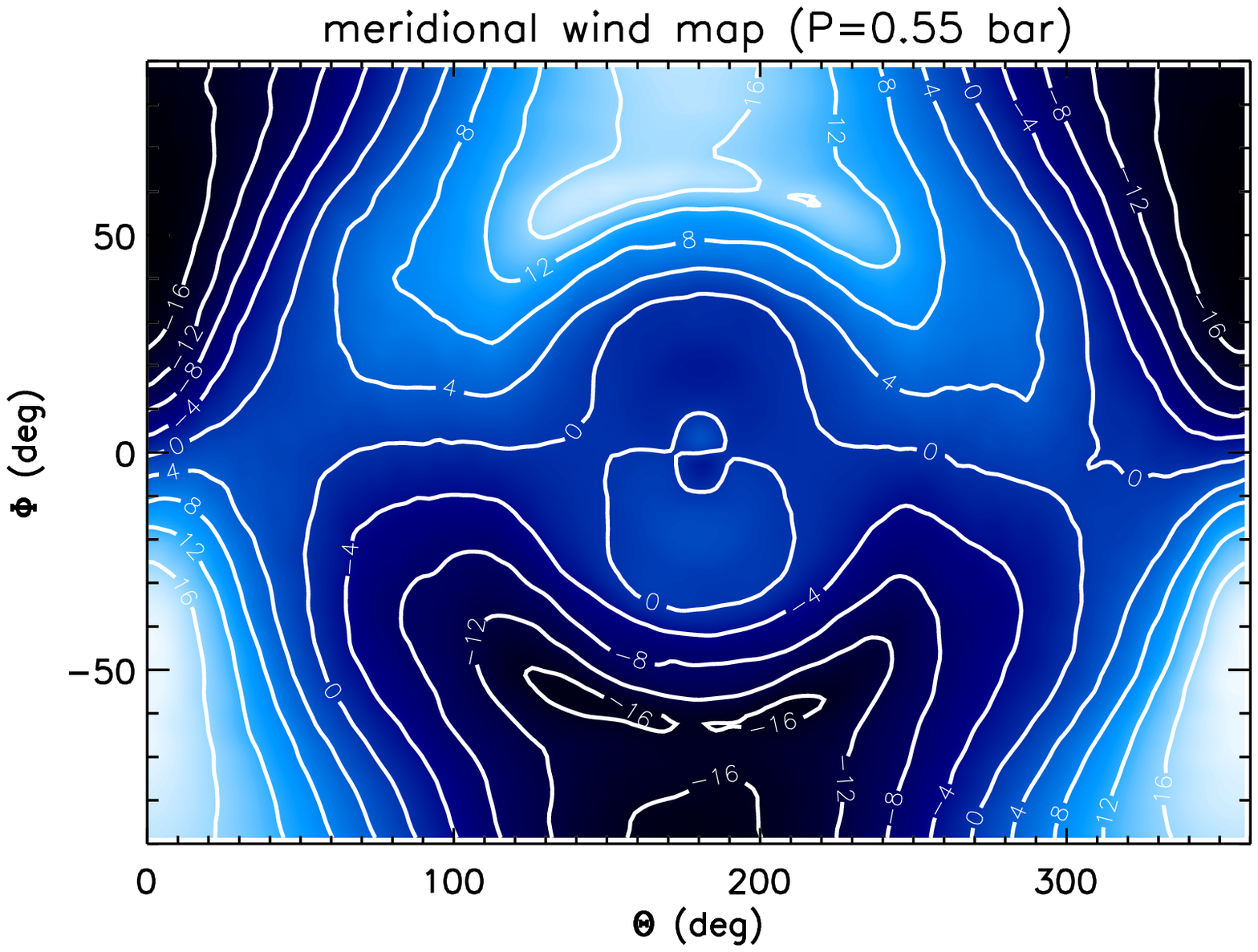}
\includegraphics[width=0.48\columnwidth]{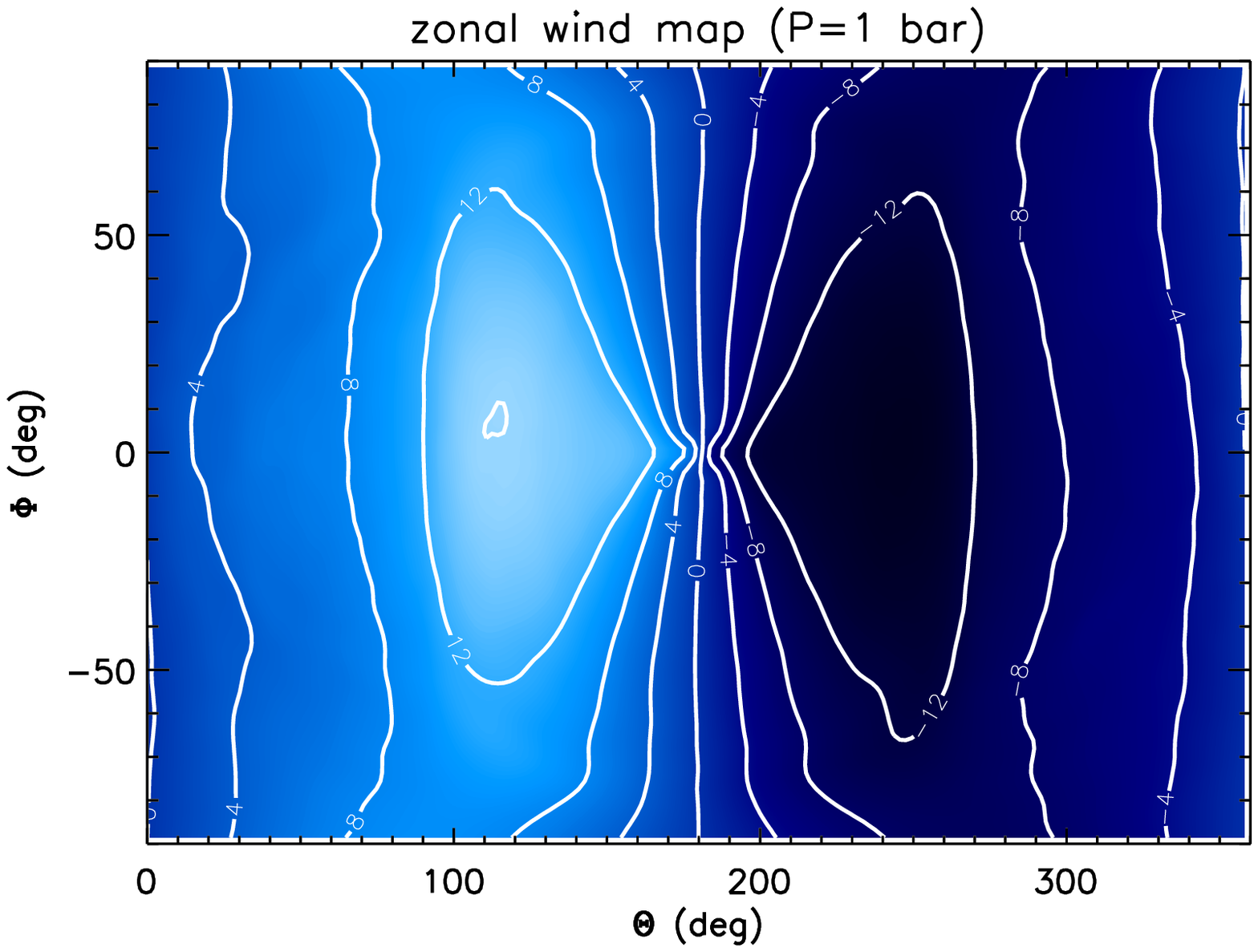}
\includegraphics[width=0.48\columnwidth]{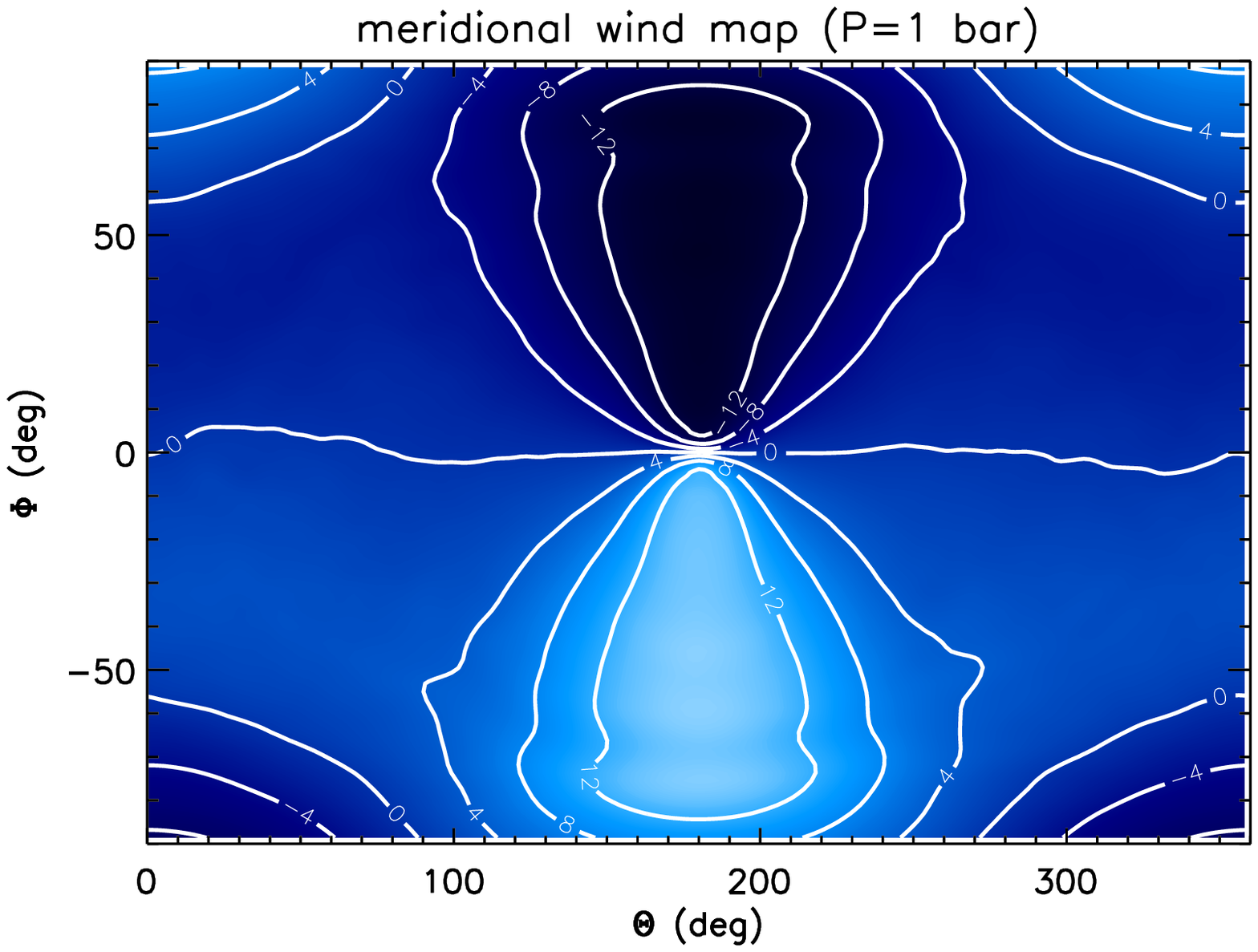}
\end{center}
\vspace{-0.2in}
\caption{Wind maps from the tidally-locked Earth simulation with convective adjustment.  Left column: zonal wind.  Right column: meridional wind.  The first, second and third rows show the maps at $P=0.25$, 0.55 and 1 bar, respectively.  Contours are in units of m s$^{-1}$.}
\label{fig:tidal_earth_maps}
\vspace{-0.1in}
\end{figure}

In the interest of clarity, we will define physical quantities, used in the present study, which may be unfamiliar within the astronomical/astrophysical literature.

The potential temperature $\theta_{\rm T}$ (equation [\ref{eq:potential_temp}]) is related to the entropy ${\cal S}$ by (e.g., \S1.6.1 of \citealt{vallis}, equation [3.7] of \citealt{po84})
\begin{equation}
{\cal S} = c_P \ln\theta_{\rm T}.
\label{eq:entropy}
\end{equation}
The preceding expression is only valid if $c_P$ is constant (otherwise it involves an integration over $\theta_{\rm T}$).  Therefore, examining profiles of the potential temperature is equivalent to determining where the surfaces of constant entropy (``isentropes") reside, which provides insight into the convective stability of the atmosphere.

Insight into the large-scale circulation patterns within the atmosphere may be obtained by examining the Euleran mean streamfunction, defined as \citep{po84,pck08}
\begin{equation}
\Psi \approx \frac{R_p P_0 ~\cos\Phi}{g_p} \int_{\sigma_0}^1 \bar{v}_\Phi\left(\Phi,\tilde{\sigma}_0\right) ~d\tilde{\sigma}_0,
\label{eq:streamfunction}
\end{equation}
where $\bar{v}_\Phi = \bar{v}_\Phi(\Phi,\sigma_0)$ is the temporally- and zonally-averaged meridional velocity.  On Earth, we have $\Psi \sim 10^9$ kg s$^{-1}$.  We note that the ``Sverdrup" (Sv), where 1 Sv $\equiv 10^9$ kg s$^{-1}$, is an alternative unit used in the oceanography community.  In the case of hot Jupiters (\S\ref{sect:hj}), we have $\Psi \sim 10^{14}$ kg s$^{-1}$.  Our definition of the streamfunction follows that of \cite{pck08}, such that $\Psi < 0$ for a circulation cell situated just above the equator in the northern hemisphere, there is southward flow at low altitudes (high $P$ values) and northward flow at high altitudes (low $P$ values).  It is worth noting that the presence of vertical motions due to atmospheric circulation does not violate the assumption of hydrostatic balance provided the vertical accelerations are small compared to the acceleration due to gravity (see \S2.5 of \citealt{p10}).

In Figures \ref{fig:tidal_earth2}, \ref{fig:hd209}, \ref{fig:baseline} and \ref{fig:baseline2}, we opt to split up the Eulerian mean streamfunction into the day and night sides of the exoplanet.  Strictly speaking, this implies that our use of the word ``streamfunction" to describe the relevant panels in these figures becomes invalid, since streamfunctions are normally associated with circulations which are non-divergent in the plane being considered --- an integration over all longitudes is strictly required.  However, our intention is to capture the differences between the atmospheric circulation on the day and night sides of a tidally locked exoplanet.  The reader should therefore be mindful of the way we use the term ``streamfunction" in these figures.  

\section{Earth-like Models}
\label{sect:earthlike}

\subsection{Earth}
\label{subsect:earth}

The first step is to establish a Earth-like model, both as a consistency check and as an operational baseline from which to generalize to tidally-locked exoplanets \citep{hmp11,hv11}.  The present model should be regarded as a dry, simplified version of the \cite{fhz06} model.  We omit latent heating effects for simplicity and thus the (dry) Earth-like models presented here have less fidelity in simulating the terrestrial climate than the models of \cite{fhz06} in aspects which are strongly influenced by water condensation, such as lapse rates and meridional overturning circulation strengths.  The \cite{fhz06} model with moisture has a significantly better fit to observations than the Held-Suarez model in aspects of the terrestrial circulation such as the strength of the Ferrel cell, meridional energy transports and radiative cooling rates.  Following \cite{fhz06,fhz07} and \cite{f07}, we define the stellar irradiation function to be
\begin{equation}
{\cal G} = \frac{1}{4} \left[1 + \frac{\Delta_{\rm T}}{4}\left( 1 - 3\sin^2\Phi \right) \right].
\end{equation}
The meridional temperature gradient can be adjusted by varying the value of the dimensionless constant $\Delta_{\rm T}$, which we take to be 1.4.  The Solar constant is taken to be ${\cal F}_0 = 938.4$ W m$^{-2}$, which corresponds to a global-mean albedo of about 30\%.  The seasonal cycle of insolation\footnote{In this study, we will use the terms ``irradiation" and ``insolation" interchangably.} is not considered, meaning the exoplanet is implicitly assumed to reside on an orbit with zero eccentricity and obliquity.  As an aside, we note that solar irradiation has been inferred to vary over much longer time scales \citep{bl91,c00}.  Other parameter values are listed in Table \ref{tab:params} and are largely adopted from \cite{fhz06} and \cite{f07}.

A key difference between the Held-Suarez model and our dry Earth-like model is that the radiative equilibrium of the latter is strongly unstable to convection.  Preliminary simulations without the boundary layer scheme (\S\ref{subsect:boundary}) implemented fail, suggesting that models in which the surface is located within the active --- as opposed to inert --- part of the atmosphere require a scheme to establish heat and momentum equilibrium between the atmosphere and the surface.  This expectation is borne out in our hot Jupiter simulations (\S\ref{sect:hj}) --- where the surface is located at $P \sim 100$ bar, well into the inert part of the atmosphere --- which do not require the boundary layer scheme to run to completion.  We therefore include the boundary layer schemes only in our Earth-like models.

In Figure \ref{fig:earth_hs}, we show the temporally- and zonally-averaged profiles of zonal wind and temperature as functions of the vertical pressure $P$ and the latitude $\Phi$, which we term the ``Held-Suarez mean flow quantities" \citep{hs94}.  For comparison, we include the Held-Suarez dynamical benchmark, as computed by \cite{hmp11}.  In the trio of simulations, the jet streams in the upper troposphere ($P \approx 0.3$--0.4 bar), at mid-latitudes, are clearly evident.  There are quantitative differences between each of the temperature profiles, but the general trends are that the temperature decreases away from the equator and also with increasing altitude.

More insight on the convective stability of the model atmospheres may be gleaned by examining the temporally- and zonally-averaged potential temperature profiles, as shown in Figure \ref{fig:earth_hs2}.  The potential temperature may be regarded as the ``height-adjusted" temperature (\S III of \citealt{po84}).  If it monotonically increases with height and does not depend on latitude, its structure is termed ``barotropic".  In this case, no convection occurs.  If there is a gradient in the potential temperature across latitude, then its structure is said to be ``baroclinic".  In such situations, horizontal or ``slanted" convective motions occur.  Tapping into the available potential energy of the atmosphere \citep{lorenz}, the baroclinic eddies resulting from these motions are responsible for transporting sensible heat from the equator to the poles \citep{ps95}.  The stratosphere and troposphere are thus regions of the atmosphere where the potential temperature structure is (largely) barotropic and baroclinic, respectively, at least within the temporally- and zonally-averaged context of our models.  In reality, the terrestrial stratosphere is not strictly barotropic --- along isobars, there are distinctly non-zero variations of temperature with latitude which are dynamically important.  The height or vertical pressure at which the troposphere transitions into the stratosphere is the tropopause (e.g., see Figure 4 of \citealt{f07}), located at $P \approx 0.2$ bar near the equator (the tropics) and $P \approx 0.3$--0.4 bar otherwise (the temperate and polar regions).

The middle row of Figure \ref{fig:earth_hs2} shows the simulation with radiative transfer but without convective adjustment.  Unlike in the Held-Suarez benchmark, the existence of a surface with a finite heat capacity (see \S\ref{subsect:rt}) and its heating by solar irradiation (see \S\ref{subsect:irradiation}) results in convectively unstable vertical columns, thus necessitating the treatment of convection.  The bottom row of Figure \ref{fig:earth_hs2} demonstrates how the dry convective adjustment scheme (see \S\ref{subsect:convection}) brings the model atmosphere to convective stability in the vertical direction, while preserving the baroclinic instabilities (i.e., horizontal convection) produced in the simulation.  Essentially, the role of convective adjustment is to straighten the isentropes such that they are vertically stable.

Atmospheric circulation of the terrestrial atmosphere can be generally regarded as having a three-cell structure (e.g., \citealt{po84}; \S2.1.5 of \citealt{wp05}).  The Hadley and polar cells, located near the equator and poles, respectively, are \emph{direct} circulation cells because they are driven by heating and cooling patterns of the Earth's surface.  Both cells are characterized by rising air in their warmer branches.  By contrast, the mid-latitudinal Ferrel cells are indirect because they are driven by the presence of baroclinic eddies; cooler air is forced to rise.  The presence of these cells may be revealed by examining the Eulerian mean streamfunction $\Psi$ (e.g., \citealt{f07,ms10}), as defined in equation (\ref{eq:streamfunction}).  The Held-Suarez benchmark, which mimics heating and cooling patterns by forcing the temperature field to relax to an ad hoc ``equilibrium temperature" profile on a specified ``cooling" time scale, manages to produce the Hadley, Ferrel and polar circulation cells, although the circulation associated with the Ferrel cells is too strong by a factor of two (top right panel of Figure \ref{fig:earth_hs2}).  As observed on Earth, the polar cells are relatively weak with $\Psi$ having values about an order of magnitude lower than for the other two cells.  The simulations with dual-band radiation (middle right and bottom right panels of Figure \ref{fig:earth_hs2}) manage to only produce the Hadley and Ferrel cells; there is a pair of cells near the poles, but they reside only at low altitudes ($P \approx 0.9$--1 bar).  The absence of the polar cells in our simulations and also those of \cite{fhz06} is unsurprising since the details concerning the polar temperatures (e.g., sea ice, the presence of Antarctica) are not modelled.  The Hadley cell is significantly too strong in the dual-band simulation due to the near-neutral convective stability in the tropics, while the Ferrel cell is confined to only high altitudes due to the presence of convective momentum fluxes in the lower troposphere (see \citealt{fhz06} for more details) --- these issues are ameliorated in simulations with moisture.

Figure \ref{fig:tp_earth} shows the globally-averaged temperature-pressure profiles for the trio of simulations.  By construction, the Held-Suarez benchmark does not produce the temperature inversion empirically observed in the stratosphere, since the stratospheric temperature is set to be constant in this model.  There are slight differences between the $T$-$P$ profiles for the simulations with and without convective adjustment, where the latter model produces a small temperature inversion in the stratosphere.

Our dry model neglects the fact that the terrestrial tropospheric temperature profile is strongly influenced by latent heat release from the condensation of water vapour.  In the tropics, the temperatures are approximately moist adiabatic \citep{xe89}, while in the extratropics they are determined by a coupling between latent heat release and baroclinic eddies (e.g., \citealt{fhz06}) with additional contributions from the land-sea contrast.  On average, these factors assure that the tropospheric lapse rate on Earth is less than the dry adiabatic value.  

In summary, we have generalized our Earth-like model away from the purely dynamical Held-Suarez benchmark and demonstrated the utility of examining profiles of the potential temperature and Eulerian mean streamfunction.  We have also demonstrated the necessity of including the treatment of convection alongside radiative transfer, at least for our Earth-like models.  We next extend the same simulation and analysis techniques to a hypothetical tidally-locked Earth.

\subsection{Tidally-locked Earth}
\label{subsect:tidal_earth}

\begin{figure}
\begin{center}
\includegraphics[width=0.48\columnwidth]{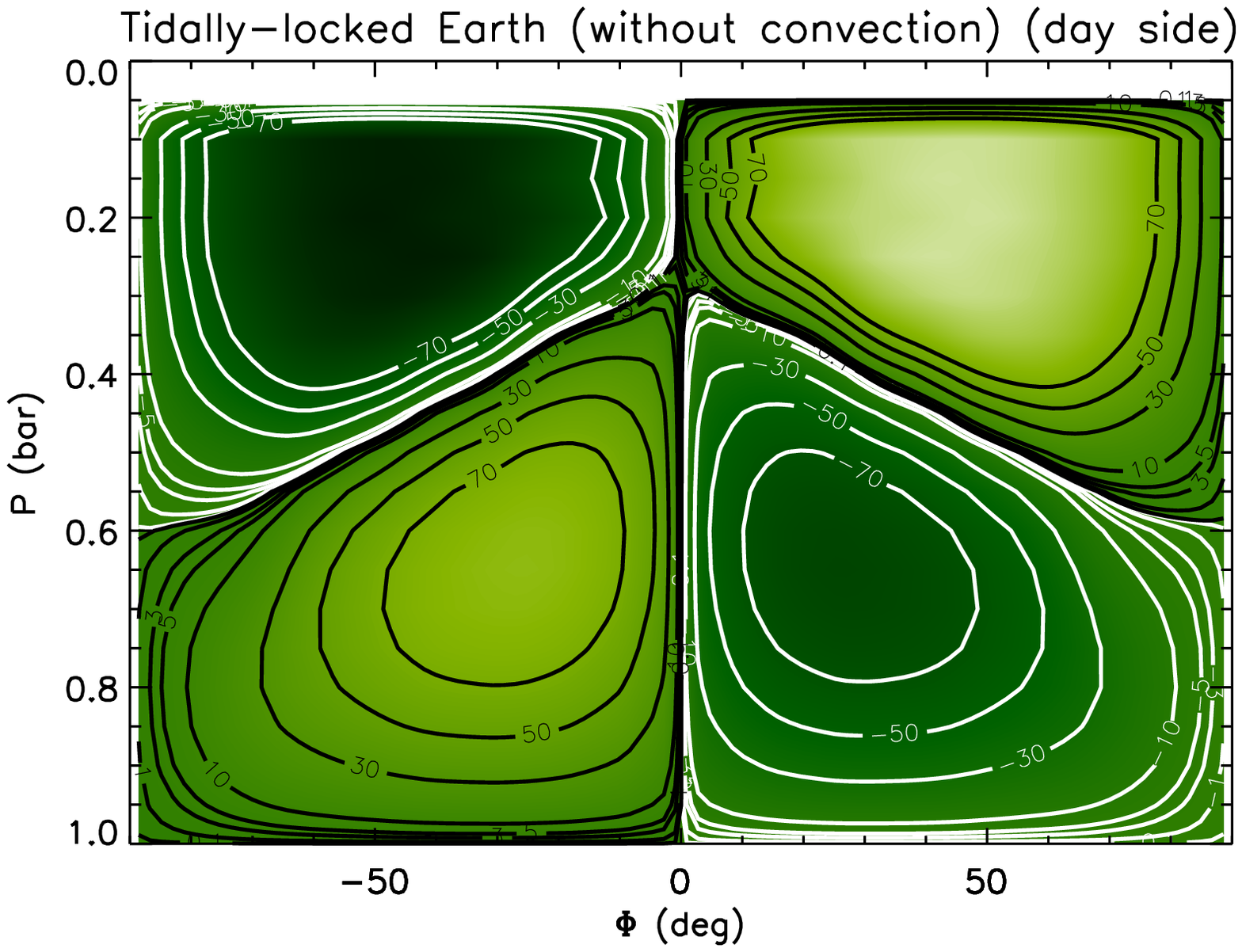}
\includegraphics[width=0.48\columnwidth]{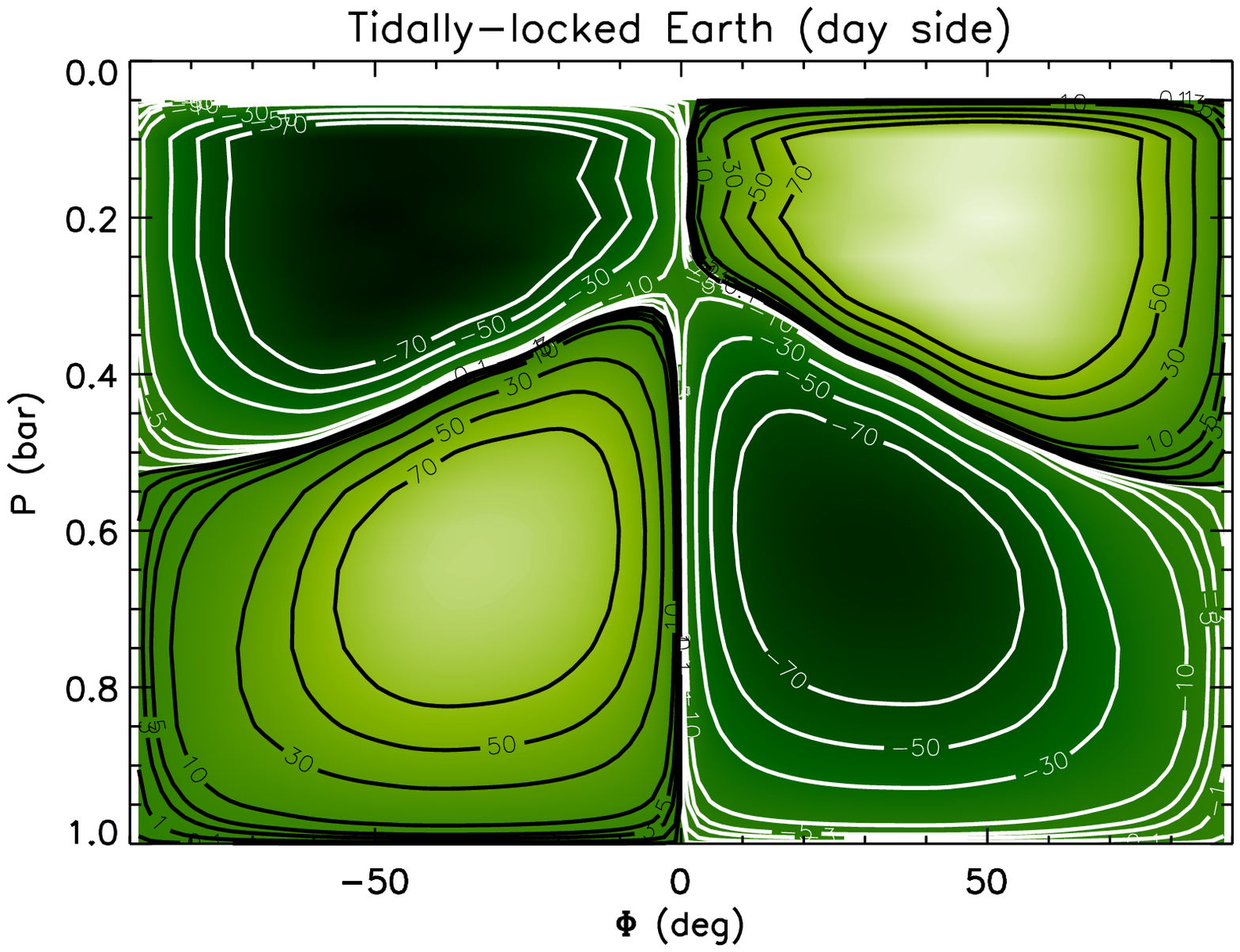}
\includegraphics[width=0.48\columnwidth]{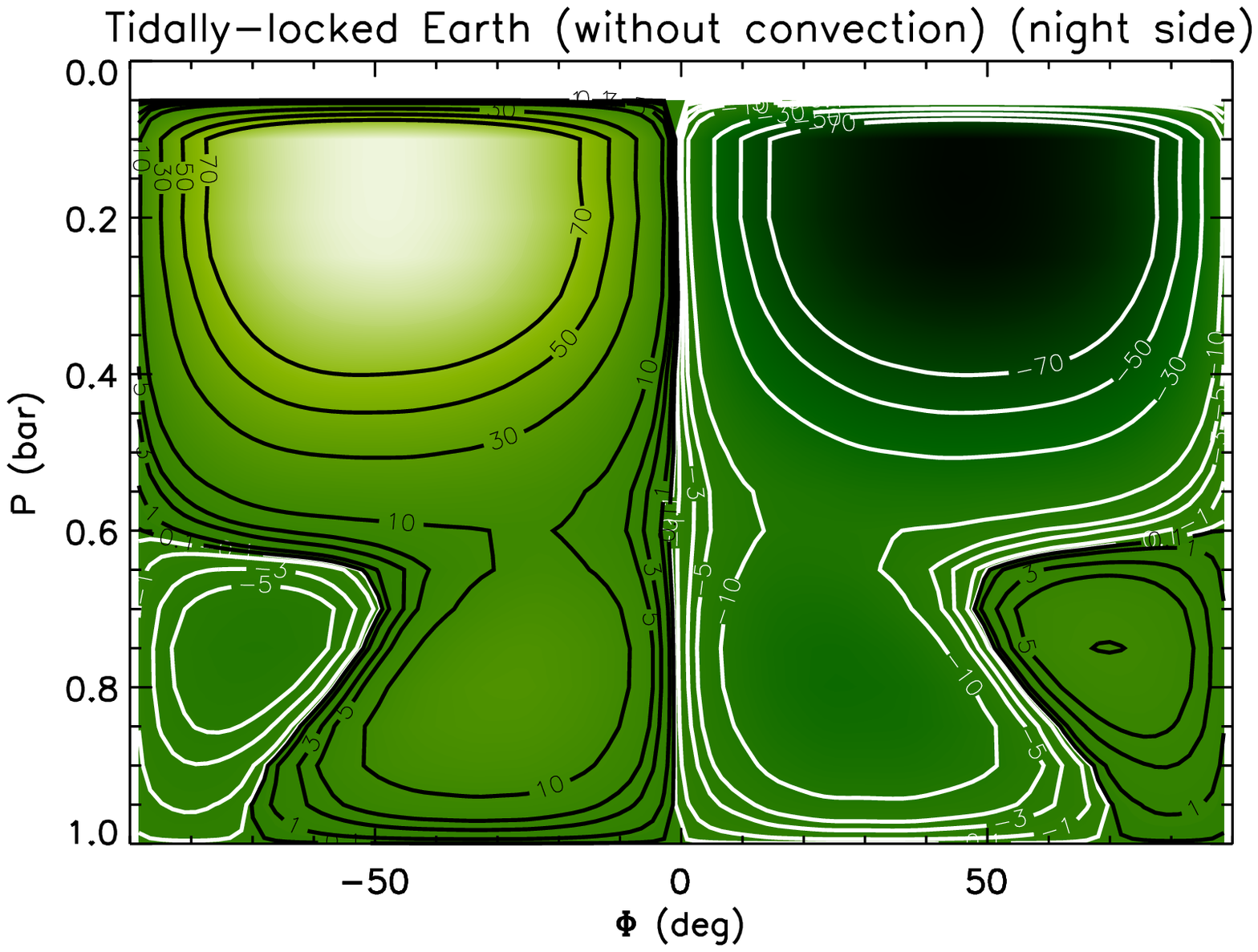}
\includegraphics[width=0.48\columnwidth]{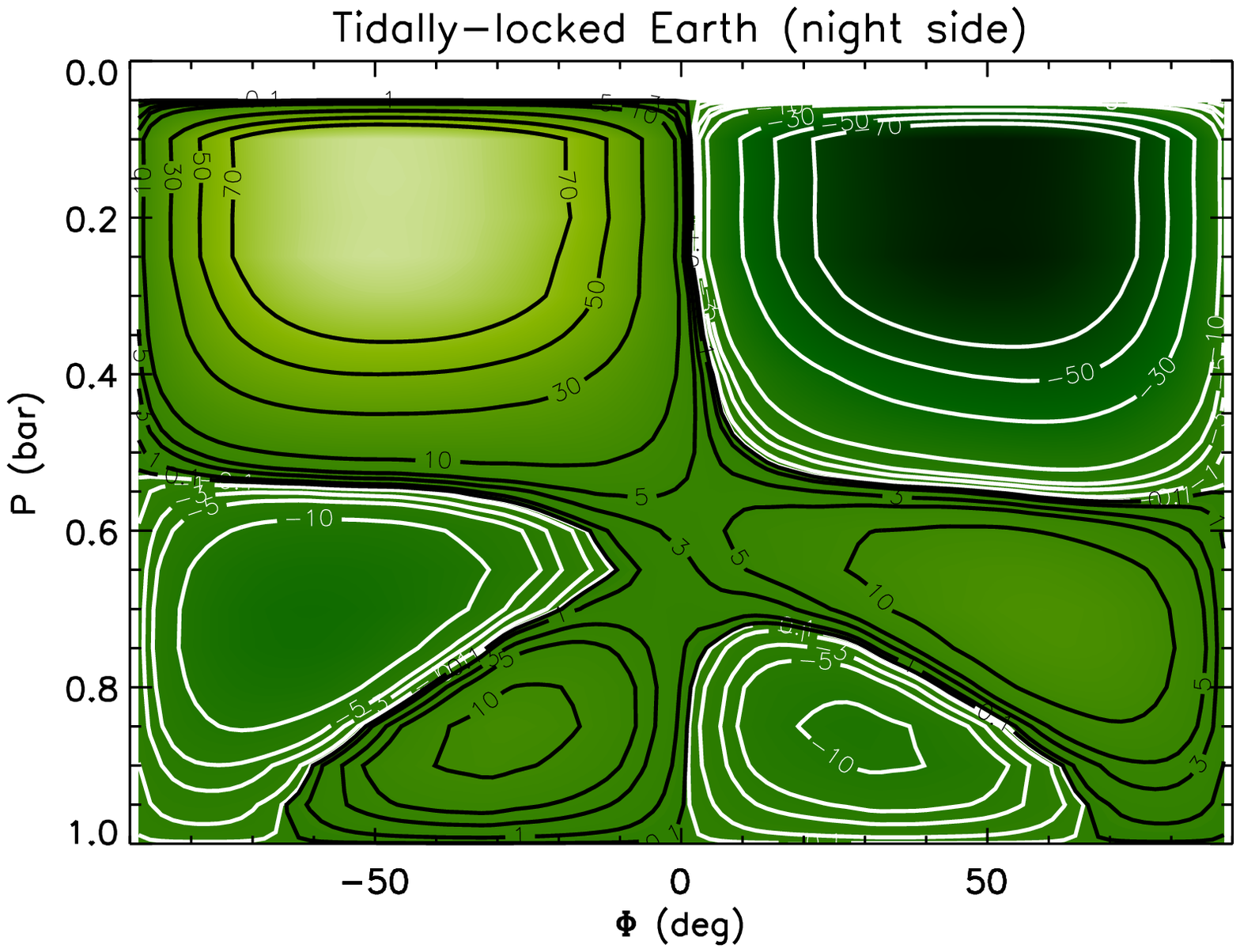}
\end{center}
\vspace{-0.2in}
\caption{Held-Suarez mean flow quantities for the Eulerian mean streamfunction (in units of $\times 10^9$ kg s$^{-1}$) in the case of the hypothetical tidally-locked Earth.  Top row: day side.  Bottom row: night side.  Left column: without convection.  Right column: with convection.  The irregular contour intervals are chosen to reveal the presence of weaker circulation cells.}
\label{fig:tidal_earth2}
\end{figure}

\begin{figure}
\begin{center}
\includegraphics[width=0.5\columnwidth]{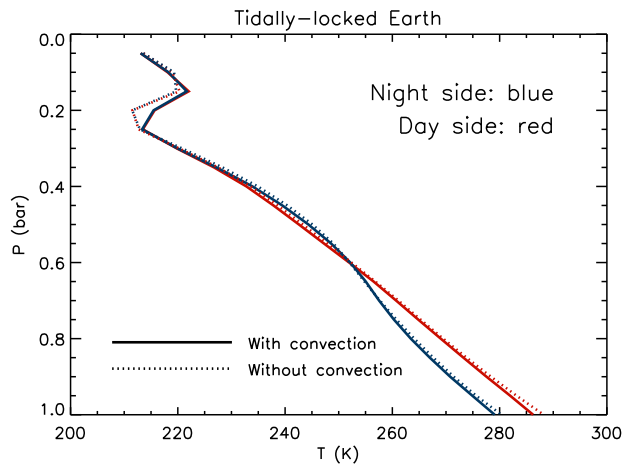}
\end{center}
\vspace{-0.2in}
\caption{Hemispherically-averaged temperature-pressure profiles for the day and night sides of a tidally-locked Earth (\S\ref{subsect:tidal_earth}).  Shown are the simulations with and without convective adjustment.}
\label{fig:tp_earth_tidal}
\end{figure}

The case of a hypothetical tidally-locked Earth is a useful and computationally efficient case study for operationally transitioning between the simulation of Earth and hot Jupiters \citep{hmp11}.  The stellar irradiation function is \citep{ms10},
\begin{equation}
{\cal G} = \mbox{max}\left\{ 0, \cos\Phi \cos\left( \Theta - \Theta_0 \right) \right\},
\label{eq:tidal_irradiation}
\end{equation}
where the substellar point is located at $\Phi = 0^\circ$ and $\Theta = \Theta_0$.  Our (arbitrary) choice is $\Theta_0 = 180^\circ$.  The parameter values are the same as those adopted in the Earth-like simulation, but the rotational frequency is slowed down such that the exoplanet takes one year to rotate on its axis \citep{ms10,hmp11},
\begin{equation}
\Omega_p \rightarrow \Omega_p/365.
\end{equation}

Our functional form for the normalization of the longwave optical depth, as described by equation (\ref{eq:tau_L0}), does not have a longitudinal dependence and is the same as that adopted by \cite{fhz06} and \cite{os08}.  \cite{ms10} have noted that a more realistic treatment is to allow for $\tau_{\rm L_0}$ to longitudinally vary due to longwave water vapour feedback.  However, since our intention is to implement a simplistic, dry model for a tidally locked exo-Earth merely as a sanity check, we ignore this refinement while acknowledging its need if the intention is to model tidally locked, Earth-like aquaplanets.

The Held-Suarez mean flow quantities for zonal wind, temperature and potential temperature are shown in Figure \ref{fig:tidal_earth}.  Our results for the zonal-mean zonal wind profile are similar to the top-left panel of Figure 6 of \cite{ms10}, who noted that the profile contains weak residuals of opposing contributions from various longitudes.  It is therefore unsurprisingly that the simulations with and without convective adjustment have zonal-mean zonal wind profiles which are somewhat different.  This phenomenon can be seen more clearly in the zonal and meridional wind maps as functions of longitude and latitude (Figure \ref{fig:tidal_earth_maps}), which indicate the presence of large circulation cells.  The potential temperature profile shows that the atmosphere is, on average, approximately barotropic down to $P \approx 0.5$ bar, implying that the tropopause is located farther down.  We show the zonal-mean potential temperatures without separating them into day and night side profiles as they are fairly similar; we will examine this issue in greater detail for the hot Jupiter simulations (\S\ref{sect:hj}).  Rather, our main intention is to demonstrate the effect of convective adjustment.

The Eulerian mean streamfunction in Figure \ref{fig:tidal_earth2} reveals the presence of multiple circulation cells on the day and night sides.  On the day side, a pair of cells exist in both the northern and southern hemispheres.  The cells are able to extend from the equator to the poles, because the reduced rate of rotation of the exoplanet weakens the Coriolis deflection.  On Earth, the relatively rapid rotation limits the Hadley cells to about $\Phi \le \pm 30^\circ$.  On the night side, a pair of large, direct cells reside at higher altitudes ($P \lesssim 0.6$ bar); a collection of four smaller circulation cells is evident at lower altitudes.  These features are robust to the application of convective adjustment.

The hemispherically-averaged temperature-pressure profiles for the day and night sides of the tidally-locked Earth are shown in Figure \ref{fig:tp_earth_tidal}.  On average, the temperature differences between the two sides are barely 10 K and are only evident at $P \gtrsim 0.6$ bar.  As in the Earth-like case (\S\ref{subsect:earth}), convective adjustment has little effect on the $T$-$P$ profiles.

\subsection{Other studies}

The computational setup described here has also been used for other studies of the terrestrial atmosphere, including the analysis of equatorial transients \citep{f07b}, the width of the Hadley circulation in different climates \citep{flc07}, the vertical temperature profile in mid-latitudes \citep{f08}, how the cross-equatorial Hadley circulation is affected by remote heating \citep{kang08} and the location of the jet streams \citep{lcf10}.  It was also applied to the study of the atmosphere on Titan \citep{mitchell09}.

\section{Hot Jupiter}
\label{sect:hj}

\begin{figure}
\begin{center}
\includegraphics[width=0.48\columnwidth]{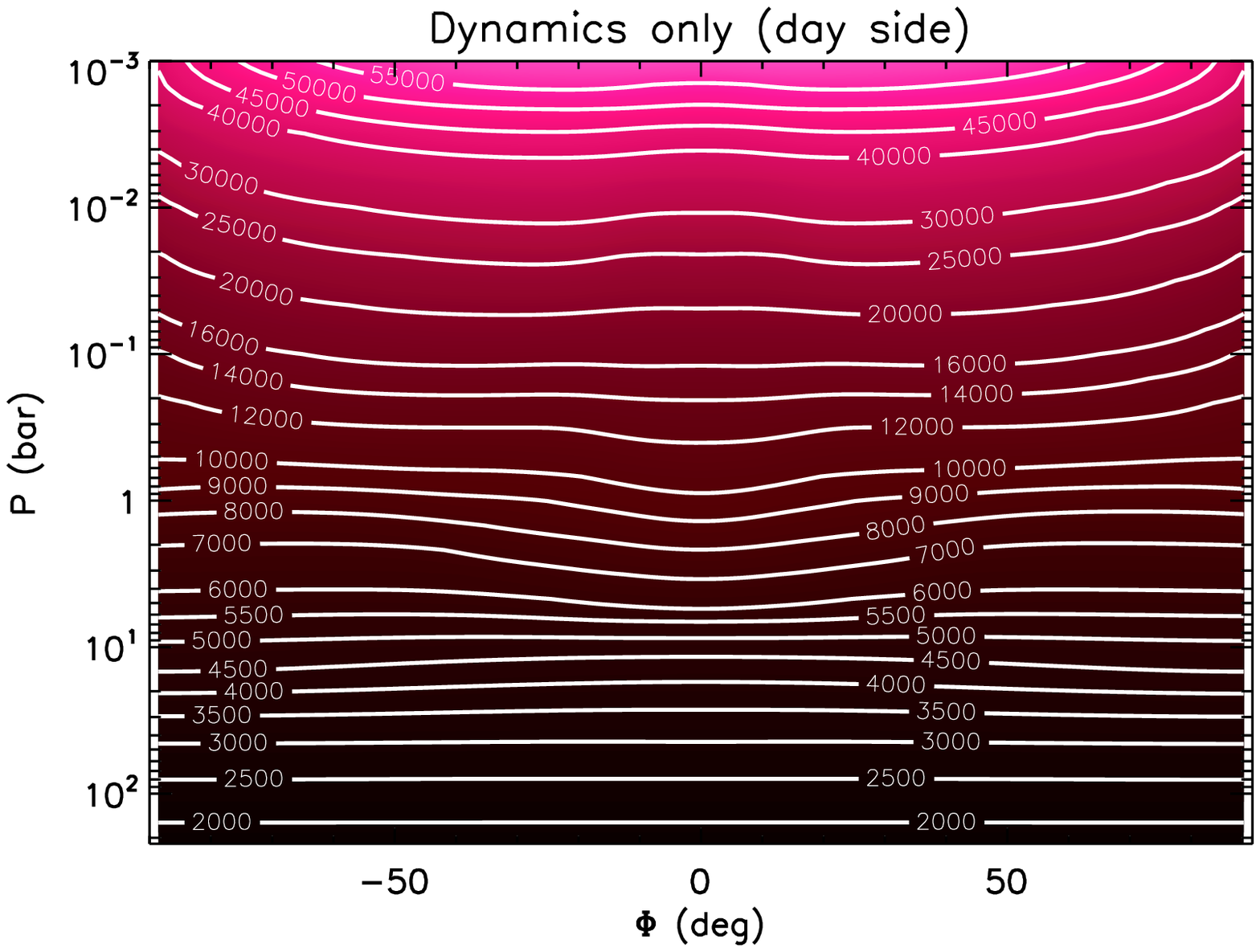}
\includegraphics[width=0.48\columnwidth]{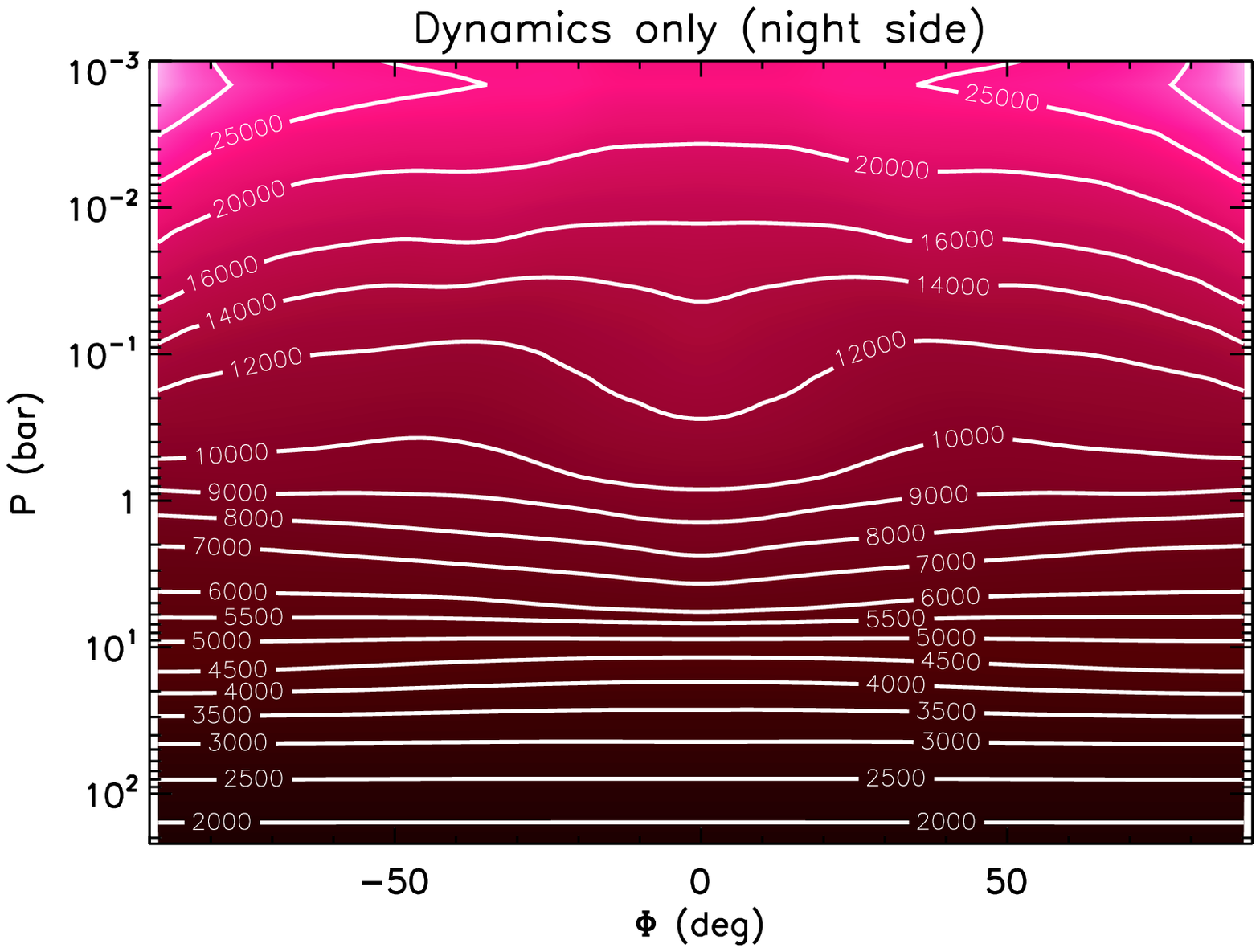}
\includegraphics[width=0.48\columnwidth]{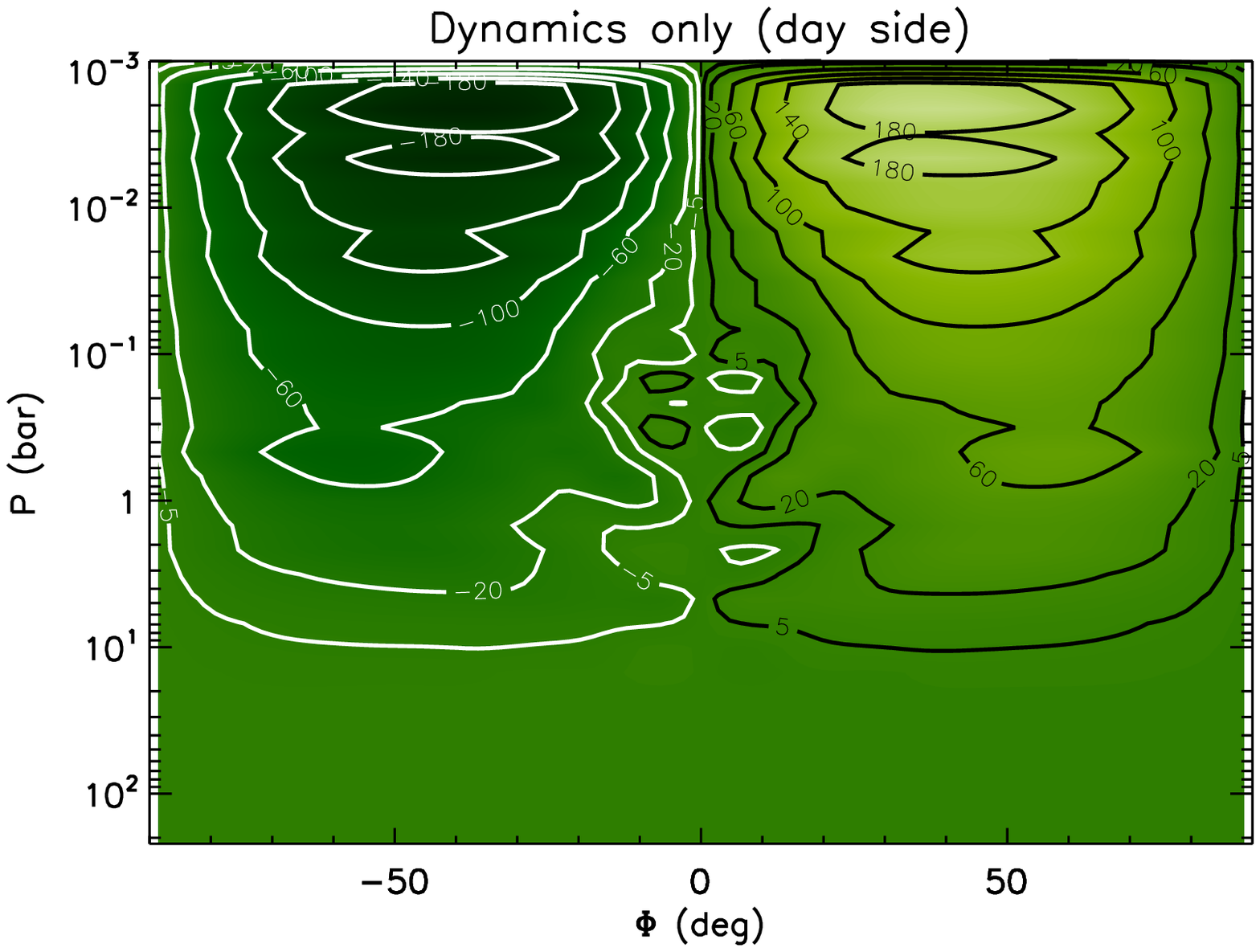}
\includegraphics[width=0.48\columnwidth]{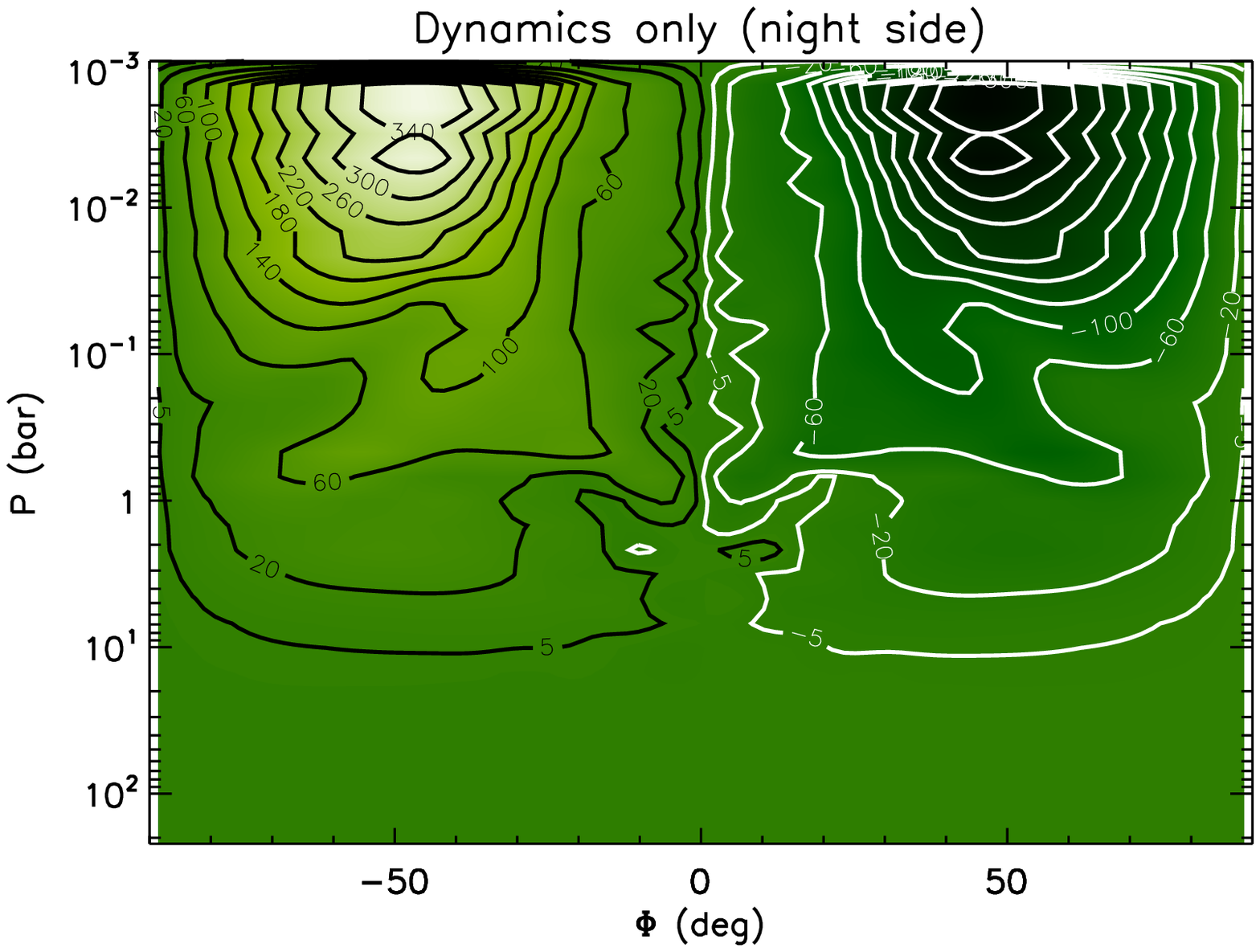}
\end{center}
\vspace{-0.2in}
\caption{Held-Suarez mean flow quantities for the T63L33 deep model of HD 209458b (see text).  Top row: potential temperature (K).  Bottom row: Eulerian mean streamfunction ($\times10^{13}$ kg s$^{-1}$).  Left column: day side.  Right column: night side.  For the potential temperature profile, contour intervals are made uneven in the interest of clarity.  For the Eulerian mean streamfunction profile, one contour interval is irregular so as to reveal the presence of weaker streamlines.}
\label{fig:hd209}
\end{figure}

To facilitate comparison with our previous work, most of the parameter values we adopt for our hot Jupiter simulations are culled from the deep model of HD 209458b presented in \cite{cs05,cs06} and \cite{hmp11}.  As in \cite{hmp11}, we use a T63L33 resolution with a range of pressures of $1 \mbox{ mbar} \lesssim P \le 220$ bar.  A noteworthy difference from the Earth-like models is the assumption that stellar irradiation impinges upon a well-mixed shortwave absorber of unspecified chemistry ($n_{\rm S}=1$).  The adiabatic coefficient used is $\kappa = 0.321$ ($n_{\rm dof} \approx 4.2$).

\subsection{Previous work: purely dynamical model}

As a prelude to our results, we re-analyze the simulation output from the purely dynamical, deep model of HD 209458b as computed by \cite{hmp11}.  Figure \ref{fig:hd209} shows the Held-Suarez mean flow quantities for the potential temperature (using equation [\ref{eq:potential_temp}]) and Eulerian mean streamfunction (using equation [\ref{eq:streamfunction}]) separated into the day and night sides of the hot Jupiter.  Near the top of the atmosphere ($P \approx 1$--10 mbar) on the day side, the atmosphere is slightly baroclinic but with flow converging towards the equator (unlike in the case of the terrestrial troposphere).  On the night side, the atmosphere is noticeably more baroclinic.  On both the day and night sides, the atmosphere becomes barotropic for $P \ge 10$ bar by construction, since the Newtonian relaxation time is defined to be infinite at these pressures.  The relative weakness of Coriolis deflection again allows for a pair of circulation cell extending from the equator to the poles, as indicated by the Eulerian mean streamfunction.  Unlike in the Earth-like (\S\ref{subsect:earth}) and tidally-locked Earth (\S\ref{subsect:tidal_earth}) simulations, the atmosphere \emph{sinks} at the equator and rises at the poles on the day side; the reverse happens on the night side.  As demonstrated by \cite{sp10,sp11} using a hierarchy of analytical models and simulations, there is downward transport of eddy momentum at the equators of hot Jovian atmospheres, which causes counter-rotating, equatorial flow.  However, the horizontal convergence of eddy momentum yields super-rotating, equatorial flow and tends to dominate the net flow (as shown in Figure 4 of \citealt{sp11}).  Our simulations are consistent with this picture.  It is somewhat counter-intuitive that the circulation is stronger on the night side (i.e., higher values of $\Psi$).  There is virtually no meridional circulation in the inert ($P \ge 10$ bar) part of the atmosphere.  

\subsection{Setup}
\label{subsect:setup}

A number of conceptual challenges exist when generalizing our setup to simulate hot Jovian atmospheres.  Firstly, as discussed in \cite{showman09}, it is presently unclear how to physically describe the frictional dissipation resulting from the differences in flow structures between the atmosphere and its deeper counterparts within a hot Jupiter.  Therefore, while the boundary layer scheme described in \S\ref{subsect:boundary} offers a tempting option to simulate frictional dissipation, we choose to switch it off for our hot Jupiter simulations.  Essentially, our models assumes ``free slip" lower boundary conditions.  A different approach is adopted by \cite{ls10}, who simulated the zonal wind structures of Jupiter, Saturn, Uranus and Neptune by varying the (Rayleigh) drag at the lower boundary in order to better fit observations, with the justification that such a formulation mimics the Ohmic dissipation caused by the induced magnetic fields opposing the zonal flows.  Since there are virtually no empirical constraints on the zonal wind profiles of hot Jupiters, unlike for the giant planets in our Solar System, we do not adopt the approach of \cite{ls10}.

Secondly, the bottom of the simulation domain (\S\ref{subsect:rt}) does not lend itself to easy physical interpretation, unlike in the Earth-like cases where it mimics the mixed layer ocean.  A plausible approach is to demand that the radiative time scale \citep{sg02},
\begin{equation}
t_{\rm rad} \sim \frac{c_P P}{4 g_p \sigma_{\rm SB} T^3} \approx 350 \mbox{ days} ~\left( \frac{c_P}{14308.4 \mbox{ J K}^{-1} \mbox{ kg}^{-1}} \frac{P}{220 \mbox{ bar}} \right) \left( \frac{g_p}{9.42 \mbox{ m s}^{-2}} \right)^{-1} \left( \frac{T}{1700 \mbox{ K}}\right)^{-3},
\label{eq:trad}
\end{equation}
is continuous at the bottom.  At and beneath the bottom of the simulation domain, the hot Jupiter has a finite thermal inertia if the heat capacity is non-zero ($C_{\rm int} \ne 0$ J K$^{-1}$ m$^{-2}$) --- both longwave and intrinsic heat are not instantaneously (re-)emitted.  The thermal inertia time scale is
\begin{equation}
t_{\cal I} \sim \left( \frac{C_{\rm int}}{{\cal I}} \right)^2,
\label{eq:thermal_inertia}
\end{equation}
where the thermal inertia is defined as (e.g., \citealt{pk81})
\begin{equation}
{\cal I} \equiv \left( k_{\rm con} ~\rho_{\rm int} ~c_{P,{\rm int}} \right)^{1/2},
\end{equation}
with $k_{\rm con}$, $\rho_{\rm int}$ and $c_{P,{\rm int}}$ being the thermal conductivity, mass density and specific heat capacity at constant pressure associated with the bottom, respectively.  The thermal conductivity ranges from $\sim 0.1$ W K$^{-1}$ m$^{-1}$ for gaseous hydrogen (and helium) to $\sim 10^3$ W K$^{-1}$ m$^{-1}$ for metallic hydrogen \citep{hubbard68}; the latter is not expected to exist below pressures $\sim 10^6$ bar \citep{stevenson82,hbl02}.

To ensure continuity between the simulated atmosphere and the bottom, we demand that $c_{P,{\rm int}}=c_P$.  Furthermore, $\rho_{\rm int}$ may be eliminated in favour of $P$ and $T$ via the ideal gas law (e.g., \S2.3.1 of \citealt{p10}),
\begin{equation}
\rho_{\rm int} = \frac{P}{{\cal R} T} \approx 3 \times 10^3 \mbox{ g cm}^{-3} ~\left( \frac{P}{220 \mbox{ bar}} \right) \left( \frac{{\cal R}}{4593 \mbox{ J K}^{-1} \mbox{ kg}^{-1}} \frac{T}{1700 \mbox{ K}} \right)^{-1},
\end{equation}
where the specific gas constant is ${\cal R} = {\cal R}^\ast/\mu$, ${\cal R}^\ast = 8314.5$ J K$^{-1}$ kg$^{-1}$ is the universal gas constant and $\mu$ is the mean molecular weight of the atmospheric molecules.  The specific gas constant may be related to more familiar quantities via
\begin{equation}
{\cal R} = \frac{k_{\rm B}}{\bar{m}},
\end{equation}
where $k_{\rm B}$ is the Boltzmann constant, $\bar{m} = \mu m_{\rm H}$ is the mean mass of the atmospheric molecules and $m_{\rm H}$ denotes the mass of a hydrogen atom.  For example, if ${\cal R} = 4593$ J K$^{-1}$ kg$^{-1}$ (as we are assuming for HD 209458b), then $\mu \approx 1.8$, close to the value for an atmosphere dominated by molecular hydrogen.  By equating equations (\ref{eq:trad}) and (\ref{eq:thermal_inertia}), we obtain a plausible estimate for the thermal inertia,
\begin{equation}
\begin{split}
C_{\rm int} &\sim \frac{c_P P}{T^2} \left( \frac{k_{\rm con}}{4 g_p \sigma_{\rm SB} {\cal R}} \right)^{1/2} \\
&\sim 10^5 \mbox{ J K}^{-1} \mbox{ m}^{-2} ~\left(\frac{c_P}{14308.4 \mbox{ J K}^{-1} \mbox{ kg}^{-1}} \frac{P}{220 \mbox{ bar}} \right) \left( \frac{T}{1700 \mbox{ K}} \right)^{-2} \left( \frac{k_{\rm con}}{0.1 \mbox{ W K}^{-1} \mbox{ m}^{-1}}  \right)^{1/2} \left( \frac{g_p}{9.42 \mbox{ m s}^{-2}}  \frac{{\cal R}}{4593 \mbox{ J K}^{-1} \mbox{ kg}^{-1}} \right)^{-1/2}.
\end{split}
\end{equation}
Besides the order-of-magnitude nature of the estimate, the main uncertainty lies in the value to assume for $k_{\rm con}$.  We will see later that our results are insensitive to these uncertainties.

\subsection{Analytical formalism for temperature-pressure profiles: generalization of Guillot (2010)}
\label{subsect:tinit}

The most efficient way to initiate a given simulation is with a temperature-pressure profile in radiative equilibrium, as may be computed using the models of \cite{hubeny03}, \cite{hansen08} or \cite{guillot10}.  We simplify our initial condition even further: we assume a constant temperature $T_{\rm init}$ which does not depend on latitude, longitude or pressure.  Since the radiative time scale increases with pressure \citep{sg02,iro05}, the temperatures at low pressures rapidly equilibrate to values consistent with dynamical and radiative equilibrium, whereas the temperatures at depth ($T_\infty$) remain close to radiative equilibrium.  It is therefore plausible to select $T_{\rm init} = T_\infty$.  In this sub-section, we generalize the analytical results of \cite{guillot10} to obtain the temperature-pressure profile with both $n_{\rm L}=1$ and 2, and subsequently use it to estimate the values for $T_\infty$.

By retaining the assumption of a constant shortwave opacity $\kappa_{\rm S}$, but allowing for an arbitrary functional form for the longwave opacity $\kappa_{\rm L}$, equation (29) of \cite{guillot10} may be generalized as
\begin{equation}
T^4 = \frac{3}{4} T_{\rm int}^4 \left( \frac{2}{3} + \int^x_0 \kappa_{\rm L} ~dx^\prime \right) + 3 f T^4_{\rm eq} \left[ \frac{2}{3} + \frac{\kappa_{\rm S}}{\sqrt{3} \kappa_{\rm L}} \exp{\left( -\sqrt{3} \kappa_{\rm S} x \right)} + \int^x_0 \kappa_{\rm L} \exp{\left( -\sqrt{3} \kappa_{\rm S} x^\prime \right)} ~dx^\prime \right],
\label{eq:tguillot}
\end{equation}
where $T_{\rm int}$ is the blackbody temperature of the internal heat flux, $\kappa_{\rm L} = \kappa_{\rm L}(x^\prime)$, $dx = \rho d\tilde{z}$, $\rho$ is the mass density, $\tilde{z} = \tilde{R}_p - z$ and $z$ is the vertical spatial variable (with the centre of the exoplanet as a zero reference point).  The distance from $z=0$ to the top of the atmosphere is $\tilde{R}_p$; we expect $\tilde{R}_p \approx R_p$.  The quantity $x$ is thus the column mass, per unit area, measured from the top of the atmosphere.

The quantity $f \approx 0.25$ is a dilution factor meant to mimic stellar irradiation being the strongest at the substellar point, since equation (\ref{eq:tguillot}) assumes isotropic irradiation.  The equilibrium temperature of the hot Jupiter, assuming zero albedo and no heat redistribution, is
\begin{equation}
T_{\rm eq} = \left( \frac{R_\star}{2a} \right)^{1/2} T_\star = \left( \frac{{\cal F}_0}{4\sigma_{\rm SB}} \right)^{1/4}.
\end{equation}
For the values of $R_\star$, $a$ and $T_\star$ associated with HD 209458b (see equation [\ref{eq:stellar_constant}]), we have $T_{\rm eq} \approx 1432$ K.  The corresponding irradiation temperature is $T_{\rm irr} = \sqrt{2} T_{\rm eq} \approx 2025$ K.

If $\kappa_{\rm L}$ is constant, then equation (\ref{eq:tguillot}) reduces to equation (29) of \cite{guillot10}.  With $T_{\rm int}=0$ K, it follows that the temperature at depth is
\begin{equation}
T_\infty \approx T_{\rm eq} \left[ f \left( 2 + \frac{\sqrt{3}}{\gamma} \right) \right]^{1/4} \approx 2129 \mbox{ K} ~f^{1/4},
\label{eq:tdeep1}
\end{equation}
where 
\begin{equation}
\gamma \equiv \frac{\kappa_{\rm S}}{\kappa_{\rm L}} = \frac{\tau_{\rm S_0}}{\tau_{\rm L_0}}.
\end{equation}
Note that the equality in the preceding definition only holds for the assumption of constant opacities.  Equation (49) of \cite{guillot10}, which averages over latitude and longitude, provides a more accurate estimate for the (mean) temperature at depth,\footnote{If $f=0.25$, equation (\ref{eq:tdeep1}) yields $T_\infty \approx 1506$ K.}
\begin{equation}
\bar{T}_\infty \approx T_{\rm eq} \left( \frac{1 + \gamma^{-1}}{2} \right)^{1/4} \approx 1539 \mbox{ K}.
\label{eq:tdeep2}
\end{equation}
Equating (\ref{eq:tdeep1}) and (\ref{eq:tdeep2}) allows for an estimate of the dilution factor required: $f \approx 0.273$.  Following \cite{guillot10}, we have adopted $\gamma=0.6$.

\begin{figure}
\begin{center}
\includegraphics[width=0.48\columnwidth]{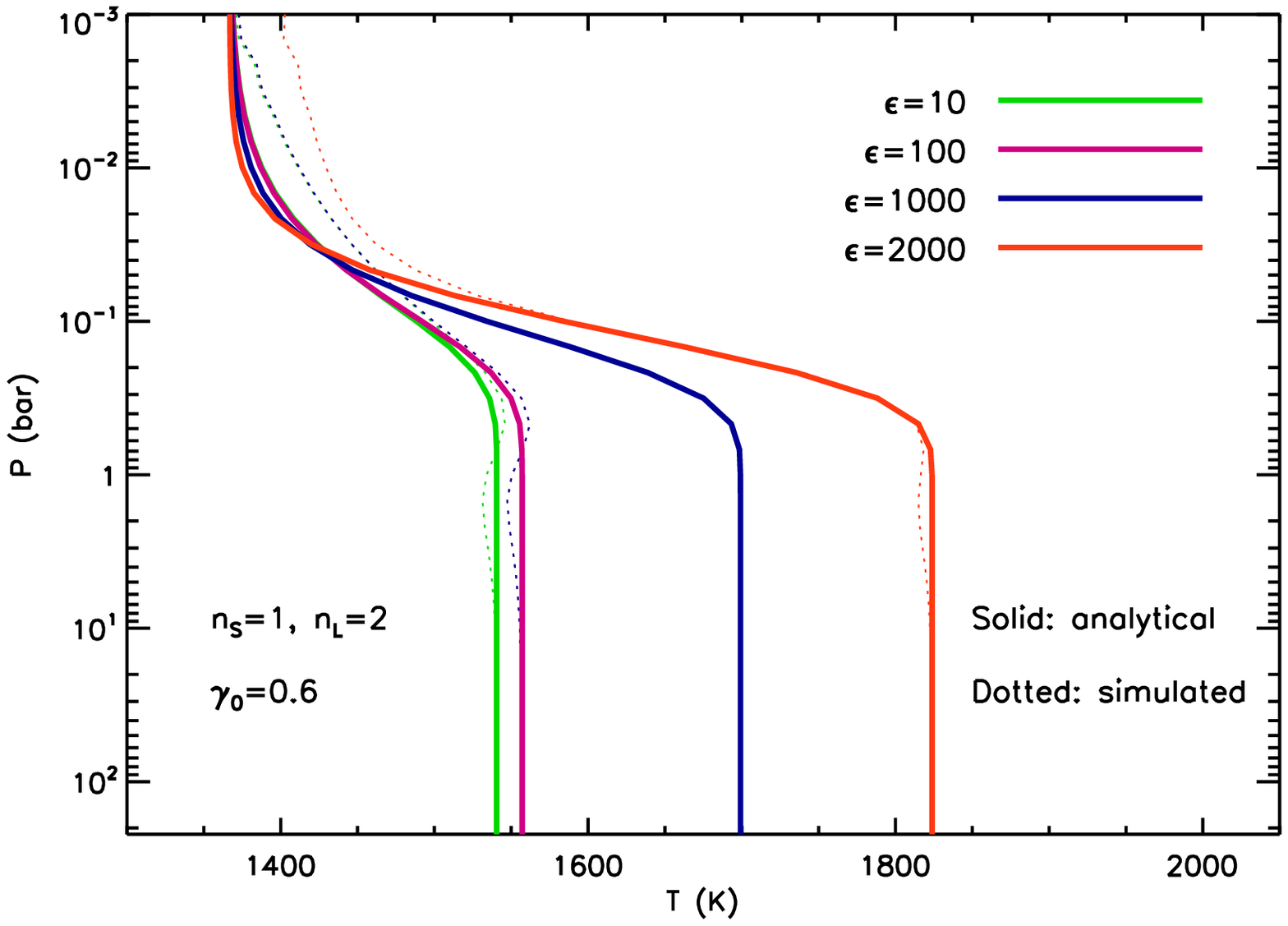}
\includegraphics[width=0.48\columnwidth]{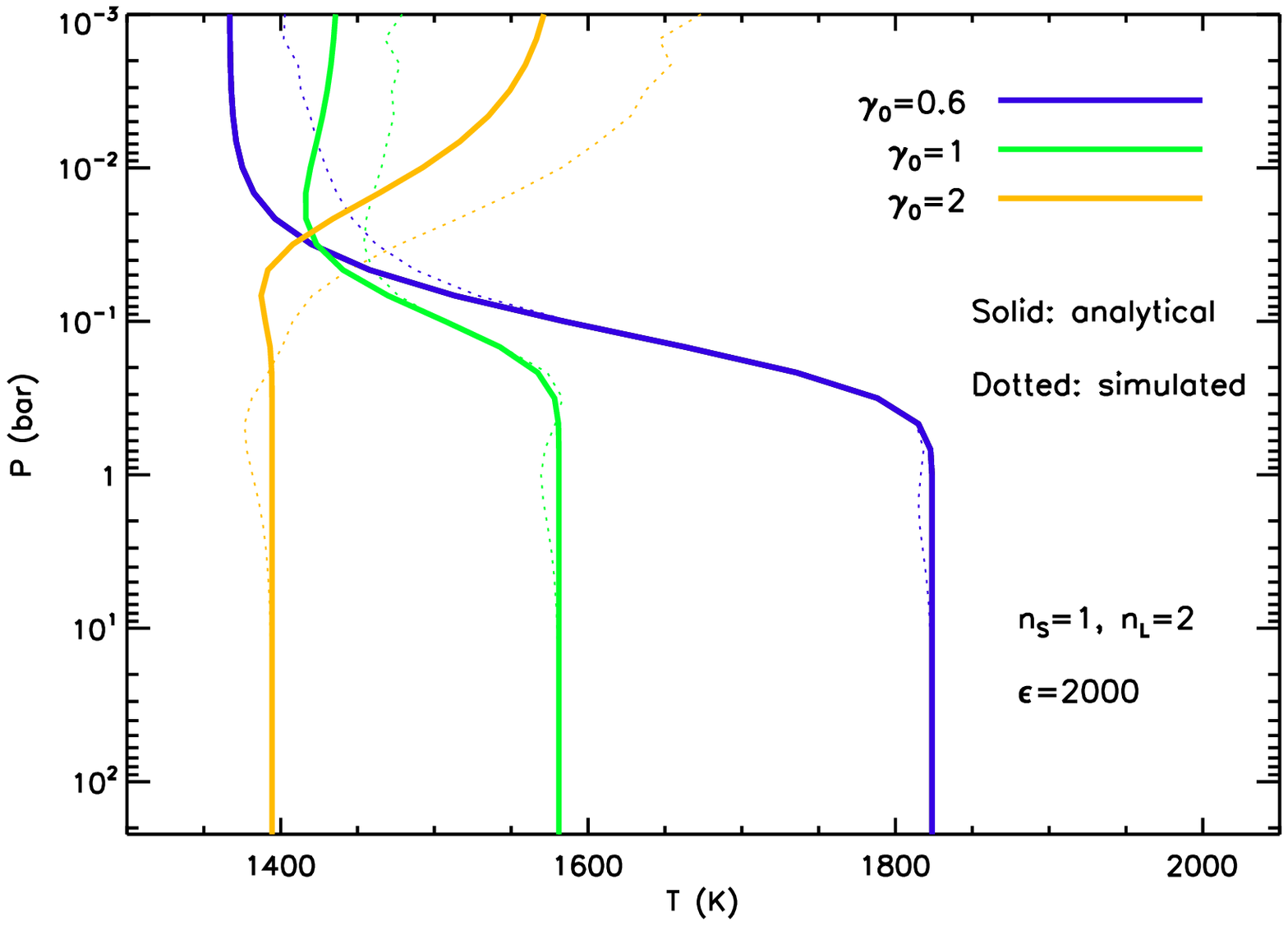}
\end{center}
\vspace{-0.2in}
\caption{Analytical temperature-pressure profiles calculated using our formalism in \S\ref{subsect:tinit} (solid curves), compared against simulated ones with only one Earth day of integration (dotted curves).  Note that the simulated profiles are hemispherically-averaged over the day side only.  Left panel: the increases in the temperatures at depth, due to the effect of collision-induced absorption, are described by a single parameter $\epsilon$.  Right panel: varying the ratio of shortwave to longwave opacity normalizations $\gamma_0$ for a fixed value of $\epsilon=2000$.}
\label{fig:guillot3}
\end{figure}

As noted by \cite{guillot10}, the assumption of $n_{\rm L}=1$ becomes inadequate deep inside the atmosphere, because collision-induced absorption becomes a non-negligible effect.  The longwave opacity then scales linearly as pressure, which implies that a correction term with $n_{\rm L}=2$ needs to be included.  We demand that
\begin{equation}
\begin{split}
\tau_{\rm L_w} + \tau_{\rm L_s} = \tau_{\rm L_0} = \epsilon \tau_0 &~\mbox{ when } P = P_0 \mbox{ or } \sigma_0 = 1, \\
\tau_{\rm L_w} \approx \tau_0 &~\mbox{ when } P \ll P_0 \mbox { or } \sigma_0 \ll 1,
\end{split}
\label{eq:tau_conditions}
\end{equation}
whence
\begin{equation}
\begin{split}
&f_l = \frac{1}{\epsilon}, \\
&\tau_{\rm L_s} = \left( \epsilon-1 \right) \tau_0.
\end{split}
\end{equation}
The quantity $\tau_0$ is the normalization for the longwave optical depth in the absence of collision-induced absorption, while the dimensionless, multiplicative factor $\epsilon \ge 1$ accounts for the increase of the longwave optical depth at $P=P_0$ due to this effect.  Collision-induced absorption becomes important when 
\begin{equation}
P > \frac{P_0}{\epsilon-1}.
\end{equation}

We write the longwave opacity as
\begin{equation}
\kappa_{\rm L} =  \kappa_0 \left[ 1 + 2\left( \epsilon  - 1 \right) \sigma_0 \right] =  \kappa_0 \left[ 1 + 2\left( \epsilon  - 1 \right) \left( \frac{x}{x_0} \right) \right],
\label{eq:kappa_long}
\end{equation}
where $\kappa_0 \equiv g_p \tau_0/P_0$, $x_0 \equiv P_0/g_p$ and the second equality follows from the assumption of hydrostatic equilibrium,
\begin{equation}
\frac{dP}{d\tilde{z}} = \rho g_p \implies P = x g_p.
\end{equation}
Evaluating equation (\ref{eq:tguillot}) in conjunction with equation (\ref{eq:kappa_long}), we get
\begin{equation}
\begin{split}
T^4 &= \frac{3}{4} T^4_{\rm int} \left( \frac{2}{3} + \tau_{\rm L} \right) \\
&+ f T^4_{\rm eq} \left\{ 2 + \sqrt{3} \gamma \exp\left( -\sqrt{3} \gamma_0 \tau \right) + \frac{\sqrt{3}}{\gamma_0} \left[ 1 - \exp\left( -\sqrt{3} \gamma_0 \tau \right) \right] + \frac{2 \left( \epsilon-1 \right)}{\gamma_0 \kappa_{\rm S} x_0} \left[ 1 - \left( 1 + \sqrt{3} \gamma_0 \tau \right) \exp\left( -\sqrt{3} \gamma_0 \tau \right) \right] \right\},
\end{split}
\label{eq:tp_general}
\end{equation}
where we have defined $\tau \equiv \kappa_0 x$.  Note that
\begin{equation}
\gamma_0 \equiv \frac{\kappa_{\rm S}}{\kappa_0}
\end{equation}
is defined such that it is constant and has no dependence on either $P$ or $\epsilon$.  When $\epsilon=1$, we recover $\kappa_{\rm L} = \kappa_0$, $\tau_{\rm L} = \tau$, $\gamma = \gamma_0$ and equation (\ref{eq:tp_general}) reduces to equation (29) of \cite{guillot10}.

It follows that the temperature at depth ($T_{\rm int}=0$ K, $\tau \gg 1$, $\gamma_0 \sim 1$) becomes
\begin{equation}
T_\infty \approx T_{\rm eq} \left\{ f \left[ 2 + \frac{1}{\gamma_0} \left( \sqrt{3} + \frac{2\left(\epsilon-1\right)}{\kappa_{\rm S} x_0} \right) \right] \right\}^{1/4}.
\label{eq:tdeep_general}
\end{equation}
The additional term in equation (\ref{eq:tdeep_general}) represents the increase in temperature at depth due to the effect of collision-induced absorption.  It is non-negligible only when $\epsilon \gg 1$.  Specifically, when
\begin{equation}
2 \epsilon \sim \frac{ \kappa_{\rm S} P_0}{g_p} = \tau_{\rm S_0} = 1401 \left(\frac{\kappa_{\rm S}}{0.006 \mbox{ cm}^2 \mbox{ g}^{-1}} \frac{P_0}{220 \mbox{ bar}} \right) \left( \frac{g_p}{9.42 \mbox{ m s}^{-1}} \right)^{-1}.
\end{equation}
Even with a wide range of values in $\epsilon$, we get $T_\infty \approx 1500$--1800 K.  For example, setting $\epsilon = 10, 100, 1000$ and 2000 corresponds to $T_\infty \approx 1541, 1557, 1699$ and 1824 K, respectively.  For these respective values of $\epsilon$, collision-induced absorption dominates the long-wave optical depth at pressures greater than 24, 2.2, 0.22 and 0.11 bar.  Figure \ref{fig:guillot3} plots the temperature-pressure profiles corresponding to these values of $\epsilon$, as well as three examples of profiles with different values of $\gamma_0$.  We have chosen $\epsilon=2000$ for the three cases with different $\gamma_0$ values so as to emphasize the differences between them (which are enhanced with increasing $\epsilon$).

Note that in deriving the results in this sub-section, we have retained the values of the first and second (longwave) Eddington coefficients (1/3 and 1/2, respectively) as adopted by \cite{guillot10}, as well as the closure relation that the ratio of the second to the zeroth moments of the shortwave intensity is 1/3.  Detailed modelling of the opacity sources present in hot Jovian atmospheres (e.g., \citealt{gf03}) will inform us about the actual value of $\epsilon$, but for now we are content with being able to account for the effect of collision-induced absorption via a single parameter.  Readers interested in broader generalizations of \cite{guillot10} are referred to \cite{hhps11}.

\subsection{Baseline models}

\subsubsection{Constant opacities ($n_{\rm S}=n_{\rm L}=1$)}
\label{subsect:guillot}

\begin{figure}
\begin{center}
\includegraphics[width=0.5\columnwidth]{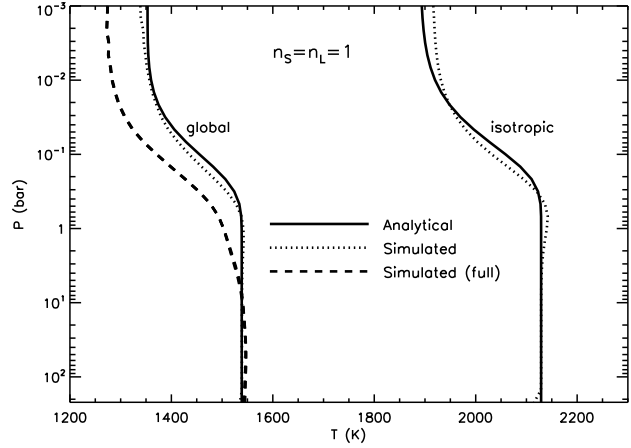}
\end{center}
\vspace{-0.2in}
\caption{Comparing analytical and simulated temperature-pressure profiles for the $n_{\rm S}=n_{\rm L}=1$ hot Jupiter models.  The profiles labelled ``global" and ``isotropic" correspond to simulations with globally-averaged temperatures and uniform irradiation, respectively (see text).  The dashed curve (``Simulated (full)") is taken from a globally-averaged, 1500-day simulation where the first 500 days of initialization have been disregarded, while the dotted curves (``Simulated") are taken from only 1 day of integration (with no output being disregarded).}
\label{fig:guillot}
\end{figure}

\begin{figure}
\begin{center}
\includegraphics[width=0.5\columnwidth]{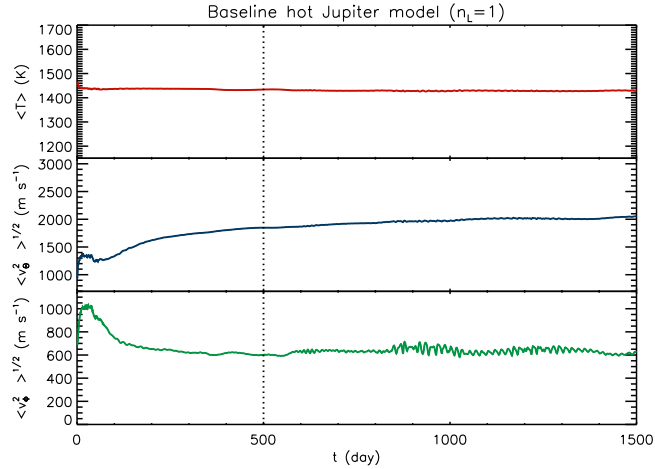}
\end{center}
\vspace{-0.2in}
\caption{Temperature (top panel), zonal wind velocity (middle panel) and meridional wind velocity (bottom panel), averaged over the entire $n_{\rm L}=1$, hot Jupiter simulation domain, as functions of the simulation time (in Earth days).  For the velocities, we have computed the root mean square (RMS) values as functions of time.  The dotted vertical line marks the end of the initialization (``spin up") phase of the simulation.}
\label{fig:spinup}
\end{figure}

As a first step, we test our computational setup against the $n_{\rm S}=n_{\rm L}=1$ analytical models of \cite{guillot10}, as stated in equations (29) and (49) of that study.  (See also \citealt{hansen08}).  To be consistent with the constant shortwave and longwave opacities of $\kappa_{\rm S}=0.006$ cm$^2$ g$^{-1}$ and $\kappa_{\rm L}=0.01$ cm$^2$ g$^{-1}$, respectively, as adopted by \cite{guillot10}, we use $\tau_{\rm S_0}=1401$ and $\tau_{\rm L_0}=2335$ such that $\gamma=\gamma_0=0.6$.  One can choose either $\epsilon=1$ or $f_l=0$.  In the absence of robust empirical constraints, we assume that the longwave optical depth does not depend on latitude,
\begin{equation}
\tau_{\rm L_0} = \tau_{\rm L_{eq}} = \tau_{\rm L_{pole}},
\end{equation}
an assumption we will retain for the rest of the study.

Figure \ref{fig:guillot} compares the analytical versus simulated temperature-pressure profiles.  The isotropic profile is obtained from a simulation where the stellar irradiation is uniform everywhere on the surface.  This is equivalent to assuming $f=1$ in equation (29) of \cite{guillot10} --- which corresponds to the situation where the exoplanet is uniformly irradiated with the substellar intensity everywhere --- and accounts for the higher temperature ($T_\infty \approx 2129$ K) at depth.  For the profiles labelled ``global", the stellar irradiation is described by equations (\ref{eq:flux_toa}) and (\ref{eq:tidal_irradiation}) and the resulting temperature-pressure profiles are averaged over the entire globe, thus resulting in a lower temperature ($T_\infty \approx 1539$ K) at depth.  The temperatures become isothermal at depths at which the stellar irradiation becomes completely absorbed ($\tau_{\rm S} \gtrsim 1$).  We have extracted temperature-pressure profiles only from the first Earth day of integration, such that the zonal winds in the simulations have not had time to increase to their quasi-equilibrium values (labelled ``Simulated").  Recall that our simulations are initiated with a constant temperature $T_{\rm init}$ and not from a $T$-$P$ profile in radiative equilibrium (see \S\ref{subsect:tinit}).  Thus, the reasonable agreement between the dotted and solid curves in Figure \ref{fig:guillot} demonstrates that the code is implementing the initial radiative condition correctly --- it is \textit{not} a rigourous test of the radiative transfer scheme.  The agreement is imperfect because the analytical expressions of \cite{guillot10} use approximate closure conditions (i.e., the Eddington approximations) and also because the vertical grid points in the simulated profiles are taken to be the larger of the pressure half levels in a Simmons-Burridge scheme (see \S3.1 of \citealt{hmp11}).  The dashed curve in Figure \ref{fig:guillot} (labelled ``Simulated (full)") is the globally-averaged temperature-pressure profile where the simulation is executed for 1500 days with the first 500 days being disregarded.  Its general agreement with the analytical profile constitutes a weak test of the radiative transfer scheme at lower pressures: at $P \lesssim 10$ bar, the radiative time scale is $t_{\rm rad} \lesssim 20$ days, which is more than an order of magnitude shorter than our nominal initialization period of 500 days, implying that radiative equilibrium has been attained at these pressures.  At higher altitudes ($P \lesssim 10$ bar), the fact that the simulated profile (dashed curve) in Figure \ref{fig:guillot} is generally $\sim 50$--100 K cooler than the analytical one (solid curve) suggests that thermal energy has been converted to mechanical energy in the form of winds, predominantly in the zonal and meridional directions.

Figure \ref{fig:spinup} shows the temperature, zonal wind velocity and meridional wind velocity, averaged over the entire simulation domain, as functions of the simulation time.  The zonal wind is absent during the first few Earth days of simulation, but gradually builds up as the simulation attains quasi-equilibrium (sometimes termed ``spin up" in the literature).  When the zonal wind is accelerated from rest, its equilibrium value is reached when the accelerations due to vertical and horizontal eddy momentum convergence attain steady values, which is expected to happen within 100 Earth days \citep{sp11}.\footnote{The spin up of the zonal wind is shown in Figure 11 of \cite{sp11}, but this specific three-dimensional model was constructed for the case of HD 189733b.  However, we do not expect dramatic changes in the spin up period for the case of HD 209458b.}  It is apparent that 500 days is a reasonable period for the initialization phase.  The mean temperature plateaus at about 1430 K, which is approximately the equilibrium temperature of the exoplanet.  It is possible that the very deepest atmospheric layers ($P \gtrsim 100$ bar) have not completely ``spun up" yet, but in simulations which we executed for longer than 1500 days we witness that the qualitative trends in our Held-Suarez mean flow quantities are invariant, although the associated quantitative details may change slightly.

An uncertainty of our technique is the value of $C_{\rm int}$ to adopt: for $k_{\rm con} = 0.1$--$10^3$ W K$^{-1}$ m$^{-1}$, we have $C_{\rm int} \sim 10^5$--$10^7$ J K$^{-1}$ m$^{-2}$.  We thus executed simulations with $C_{\rm int} = 10^5, 10^6$ and $10^7$ J K$^{-1}$ m$^{-2}$.  The resulting mean temperature-pressure profiles are virtually identical with differences of at most $\sim 1$ K.  We further verified that simulated profiles associated with $C_{\rm int} = 10^4$ J K$^{-1}$ m$^{-2}$ and $C_{\rm int} = 10^{10}$ J K$^{-1}$ m$^{-2}$ deviate substantially ($\gtrsim 50$ K) from the analytical ones at depth.

The zonal-mean zonal wind, temperature, potential temperature and Eulerian mean streamfunction are shown in Figure \ref{fig:baseline}.  It is re-assuring that the zonal wind profile resembles Figure 5 of \cite{showman09}, where a super-rotating, equatorial jet reaches down to about 10 bar, flanked by counter-rotating jets at mid-latitudes.  The super-rotating and counter-rotating jets have maximum speeds of about 5.9 km s$^{-1}$ and -0.6 km s$^{-1}$, respectively.  From examining the potential temperature profiles, one can infer that the atmosphere is baroclinic for $P < 10$ bar (and barotropic at greater pressures), in agreement with the purely dynamical results presented in Figure \ref{fig:hd209}.  A difference from Figure \ref{fig:hd209} is that there are now two pairs of circulation cells in each hemisphere (as is evident from examining the Eulerian mean streamfunction).  We note that convective adjustment is negligibly invoked, which is unsurprising given the deep radiative zone --- which encompasses both the baroclinic and barotropic atmospheric components --- of hot Jupiters.

\subsubsection{Collision-induced longwave absorption ($n_{\rm L}=2$, $\epsilon > 1$)}
\label{subsect:baseline}

\begin{figure}
\begin{center}
\includegraphics[width=0.48\columnwidth]{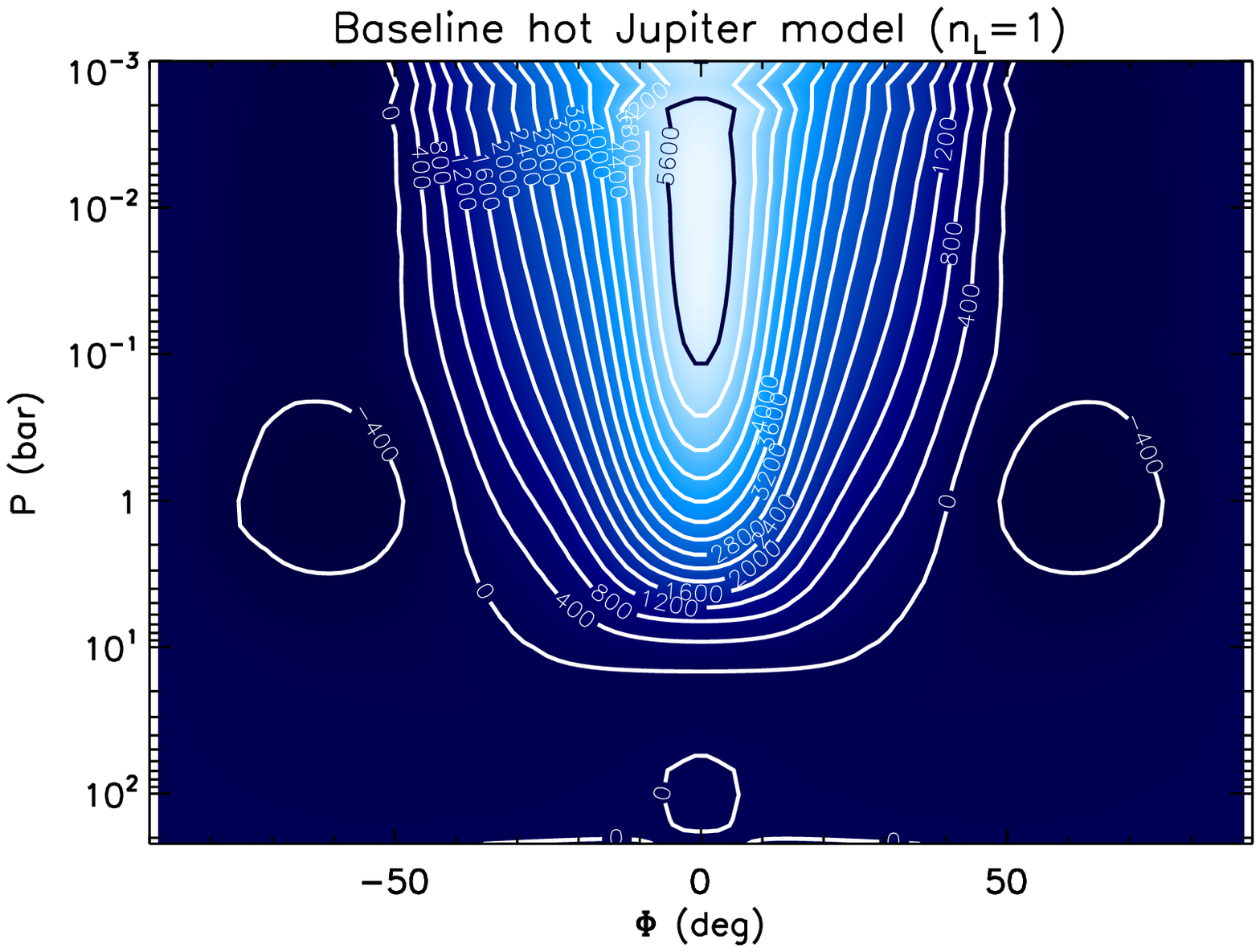}
\includegraphics[width=0.48\columnwidth]{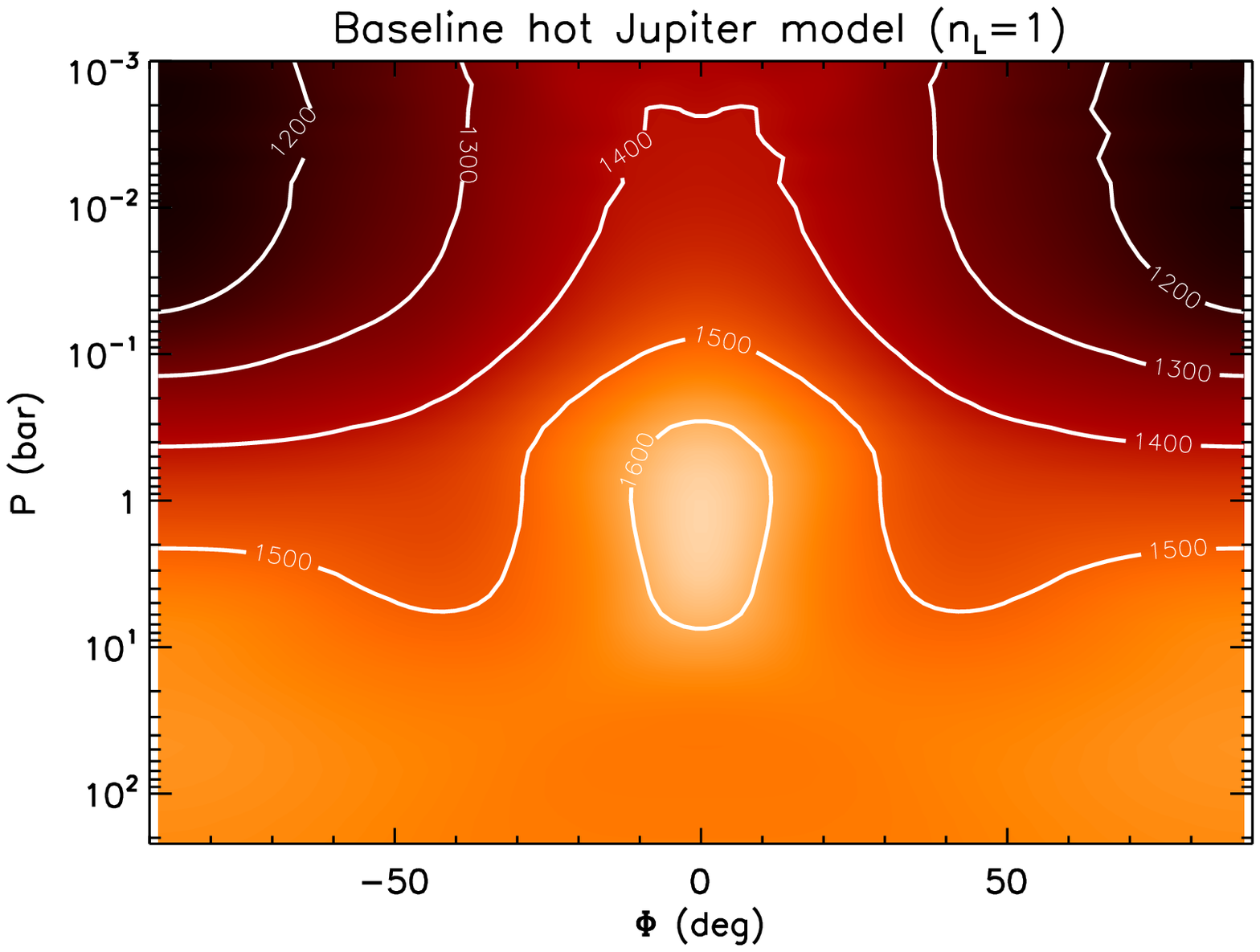}
\includegraphics[width=0.48\columnwidth]{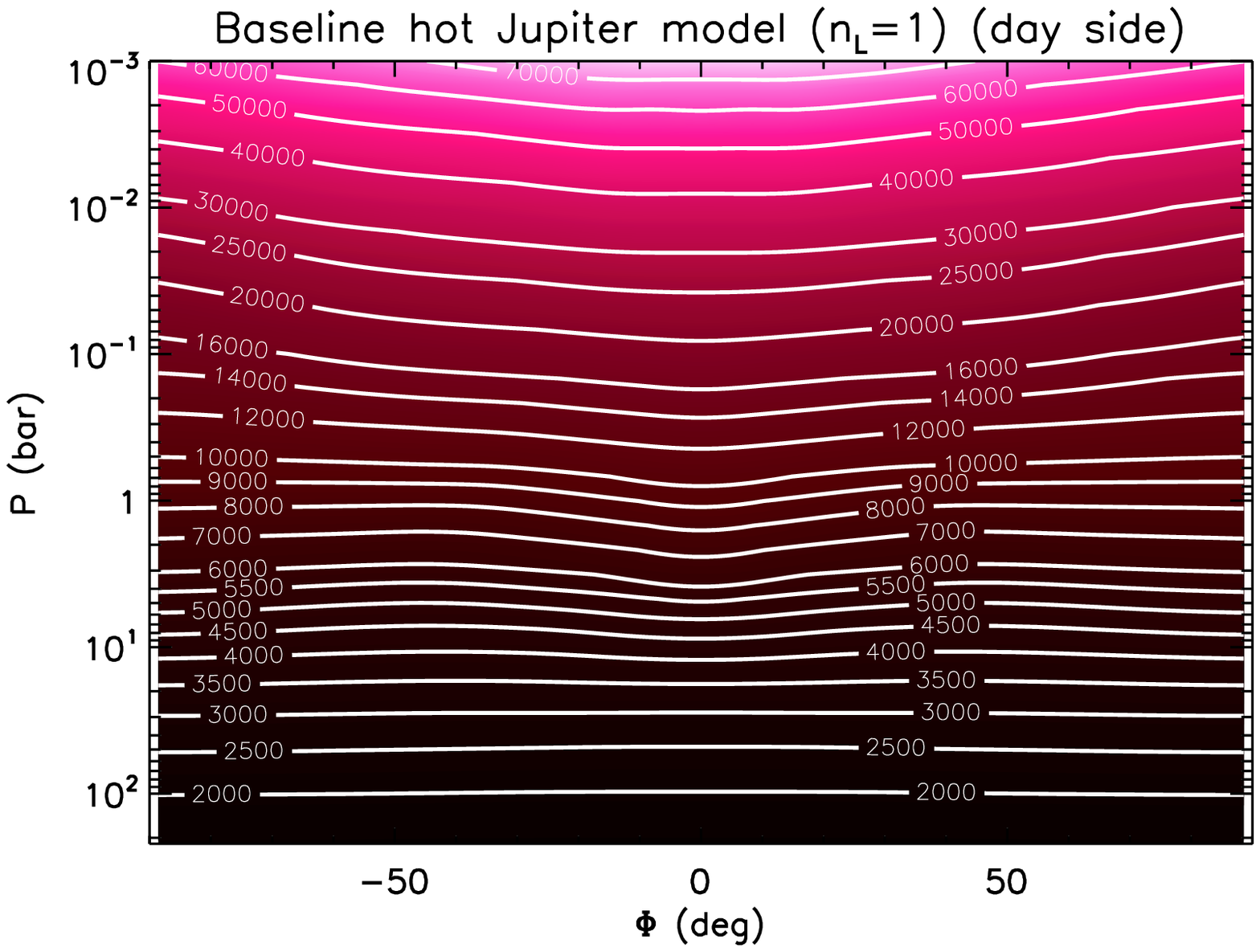}
\includegraphics[width=0.48\columnwidth]{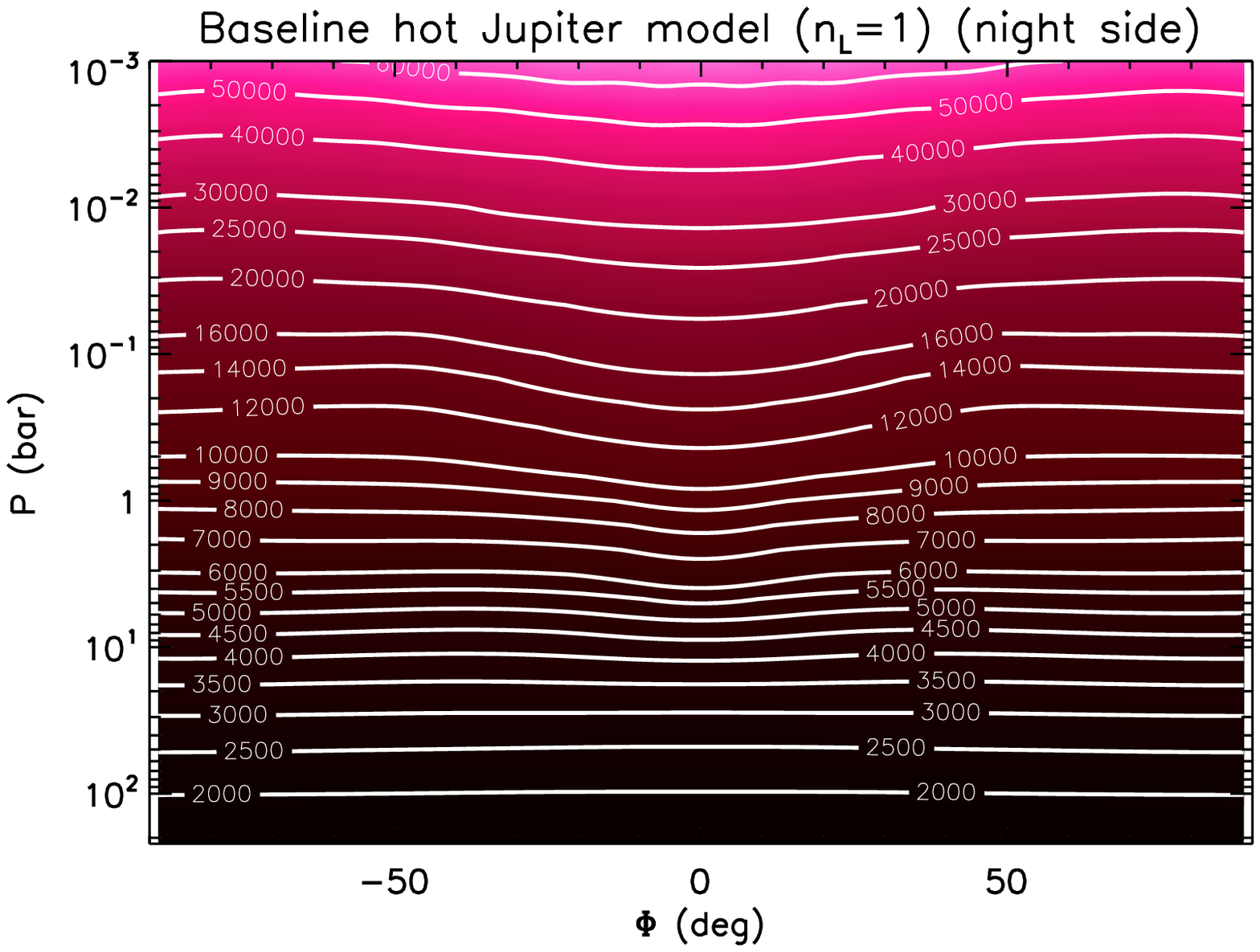}
\includegraphics[width=0.48\columnwidth]{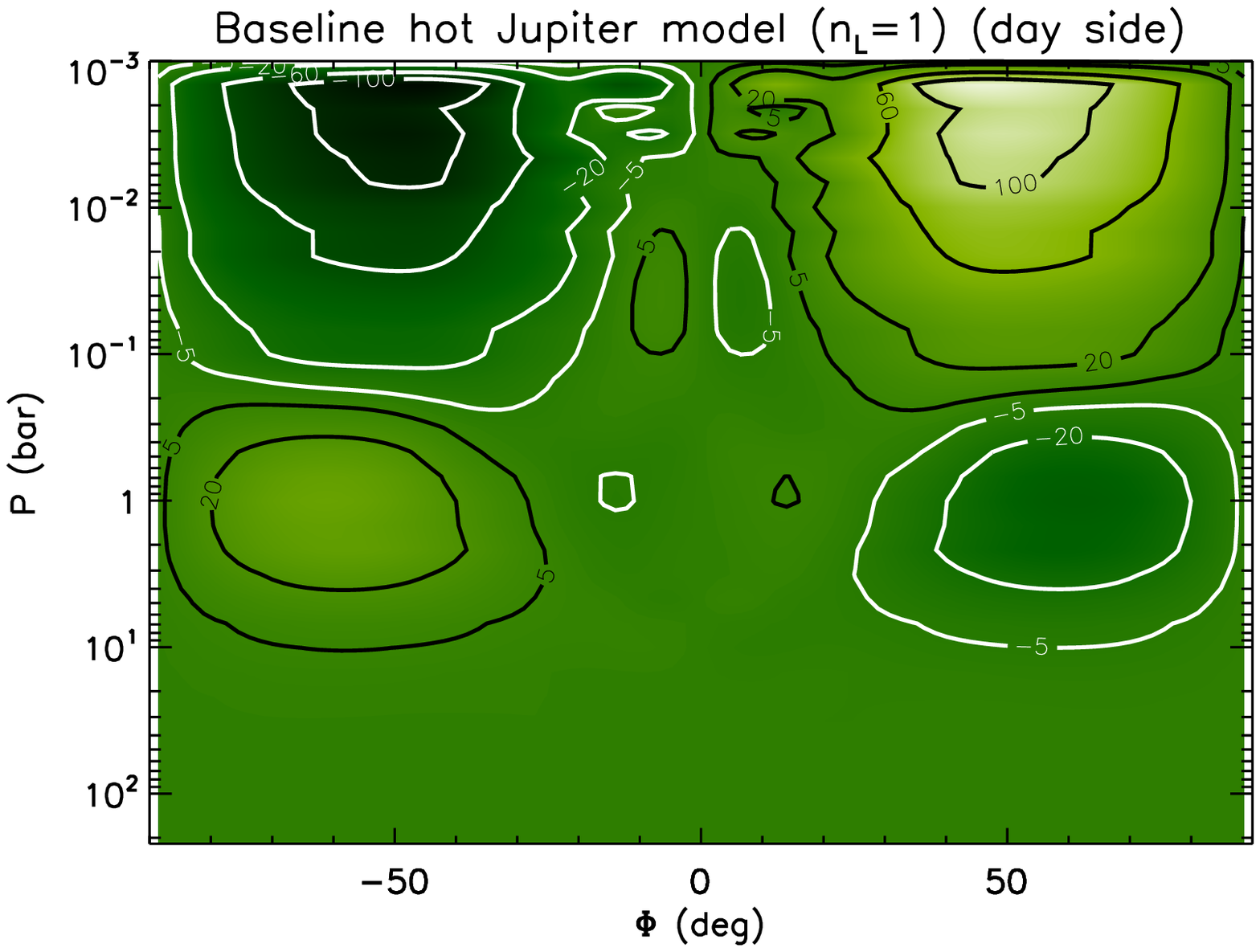}
\includegraphics[width=0.48\columnwidth]{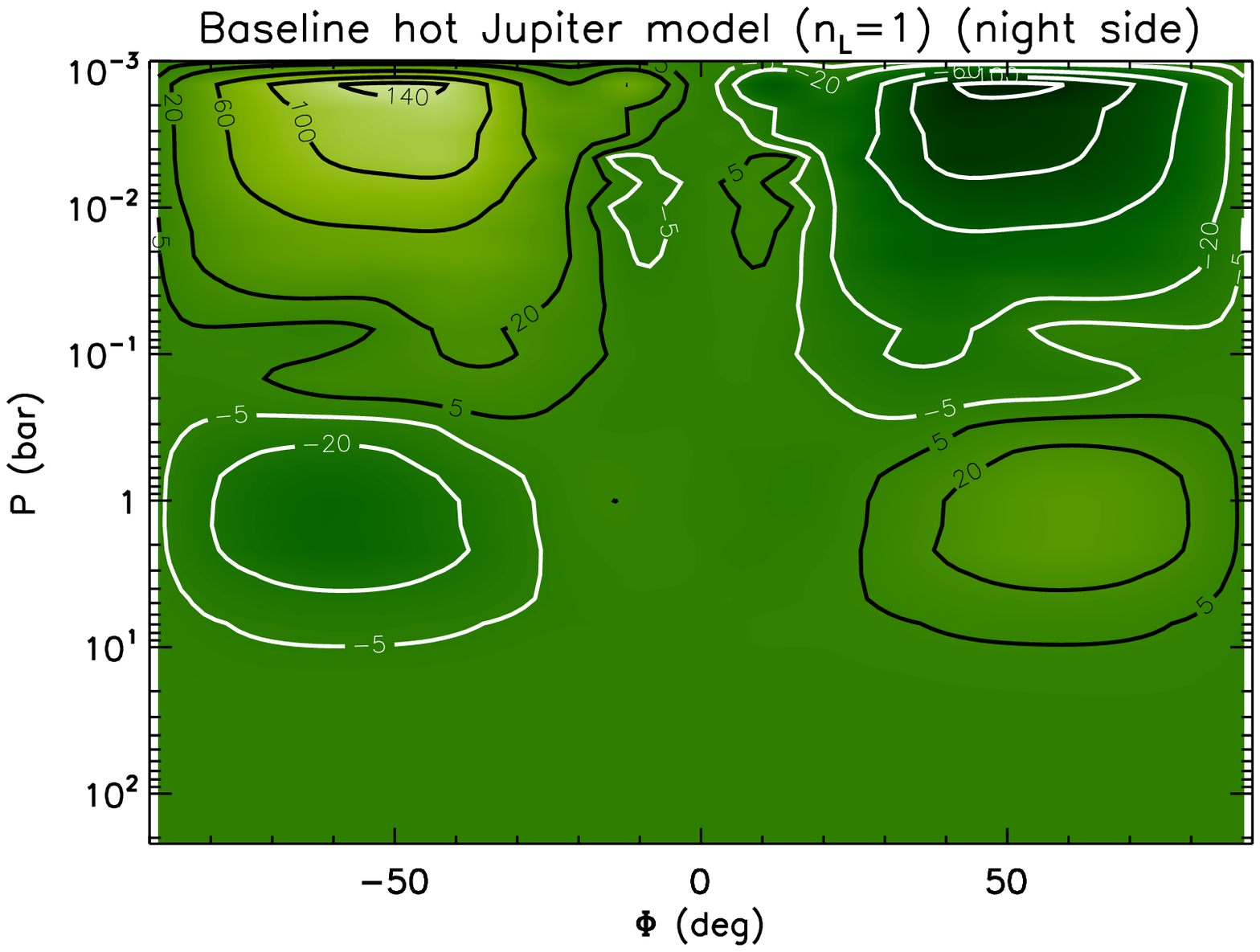}
\end{center}
\vspace{-0.2in}
\caption{Held-Suarez mean flow quantities for the baseline hot Jupiter simulation with $n_{\rm L}=1$ (see \S\ref{subsect:guillot}).  Top left panel: zonal wind (m s$^{-1}$).  Top right panel: temperature (K).  Middle row: potential temperature (K).  Bottom row: Eulerian mean streamfunction ($\times 10^{13}$ kg s$^{-1}$).}
\label{fig:baseline}
\end{figure}

\begin{figure}
\begin{center}
\includegraphics[width=0.5\columnwidth]{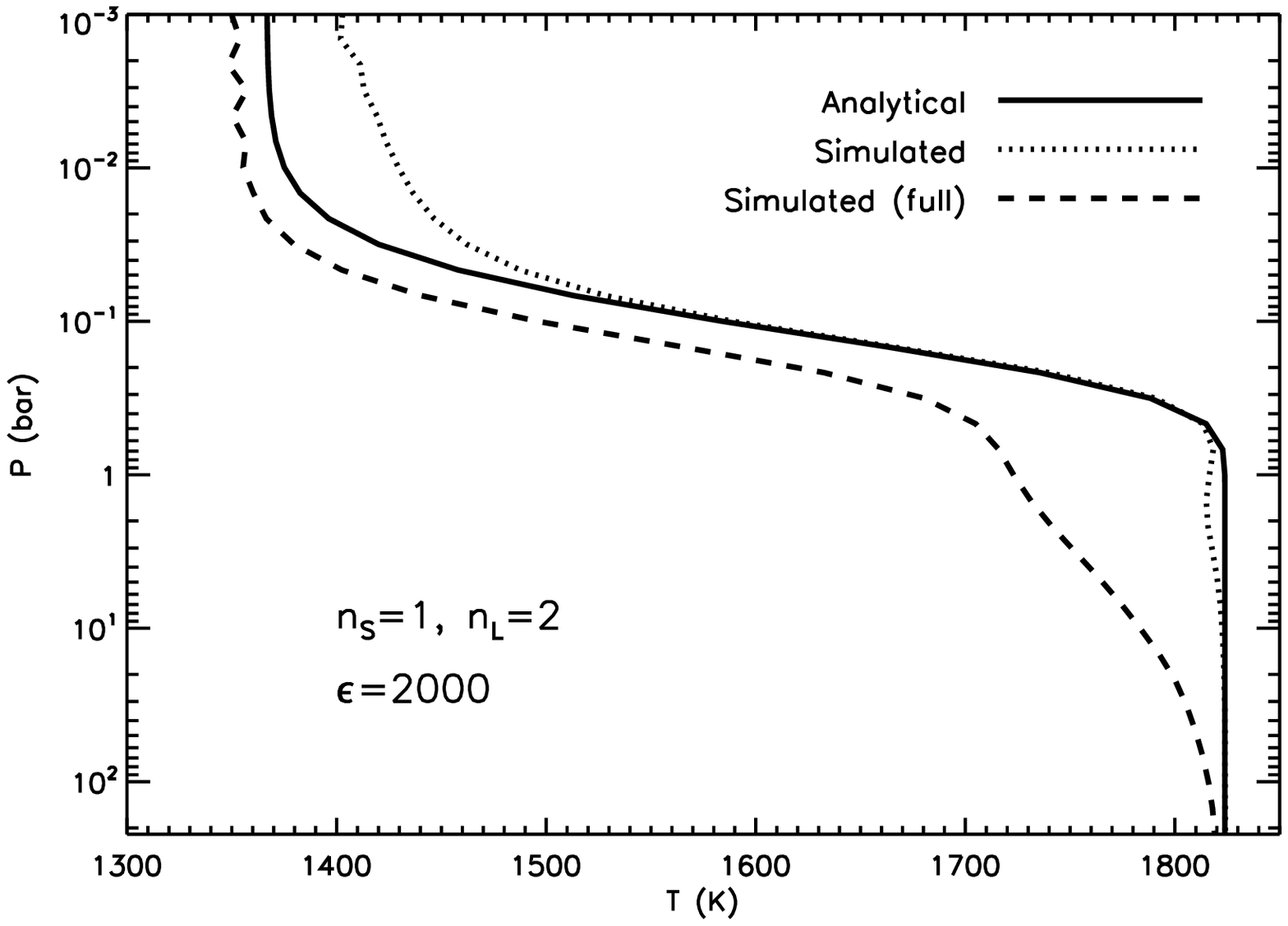}
\end{center}
\vspace{-0.2in}
\caption{Comparing analytical and simulated temperature-pressure profiles for the case of $n_{\rm L}=2$ and $\epsilon=2000$.  The solid curve is based on our generalization of the Guillot (2010) analytical profile (see \S\ref{subsect:tinit}).  The dotted curve (``Simulated") is taken from a hemispherically-averaged, 1-day simulation, while the dashed curve (``Simulated (full)") is taken from a hemispherically-averaged, 1500-day simulation where the first 500 days of initialization have been disregarded.  Note that both of the simulated profiles are extracted from the day side only.}
\label{fig:guillot2}
\end{figure}

Collision-induced absorption,\footnote{For example, the hydrogen molecule in isolation possesses no dipole moment, which implies it cannot absorb photons.  However, it may form transient ``super molecules" with other hydrogen molecules under high pressure and the resulting, weak dipole moment allows for absorption.} a phenomenon first discovered by \cite{herzberg52} to be at work within the atmospheres of Neptune and Uranus, becomes a non-negligible effect at large pressures.  We account for this effect using the formalism derived in \S\ref{subsect:tinit}.  As before, our fiducial model assumes $\tau_{\rm S_0}=1401$, but we now have $\tau_0=2335$ and $\tau_{\rm L_0}=\epsilon \tau_0$ where $\epsilon > 1$.  We first check that our simulated and hemispherically-averaged (day side only) temperature-pressure profiles are consistent with the analytical ones as derived in \S\ref{subsect:tinit} (Figure \ref{fig:guillot3}).  The reasonable agreement between them for $\epsilon=10$--2000 again demonstrates that our code is implementing the initial radiative condition correctly (but does not constitute a rigourous test of the radiative transfer scheme).

We next focus on a specific value of the correction factor to the longwave optical depth, at the bottom of the simulation domain, due to collision-induced absorption: $\epsilon=2000$.  Figure \ref{fig:guillot2} shows the temperature-pressure profiles from our analytical formalism (solid curve), a 1-day simulation meant to mimic the case of no atmospheric dynamics (dotted curve) and a 1500-day simulation with the first 500 days being disregarded (dashed curve).  The simulated profiles are hemispherically-averaged over the day side only.  The agreement between the solid and dotted curves is reasonable considering that equation (\ref{eq:tp_general}) was formulated for isotropic stellar irradiation (with a dilution factor $f$).  We have executed simulations with $C_{\rm int}=10^5, 10^6$ and $10^7$ J K$^{-1}$ m$^{-2}$ and witnessed differences of at most 1 K in the globally-averaged temperature-pressure profiles, again demonstrating that uncertainties in the value of $C_{\rm int}$ do not significantly affect our results.

\begin{figure}
\begin{center}
\includegraphics[width=0.48\columnwidth]{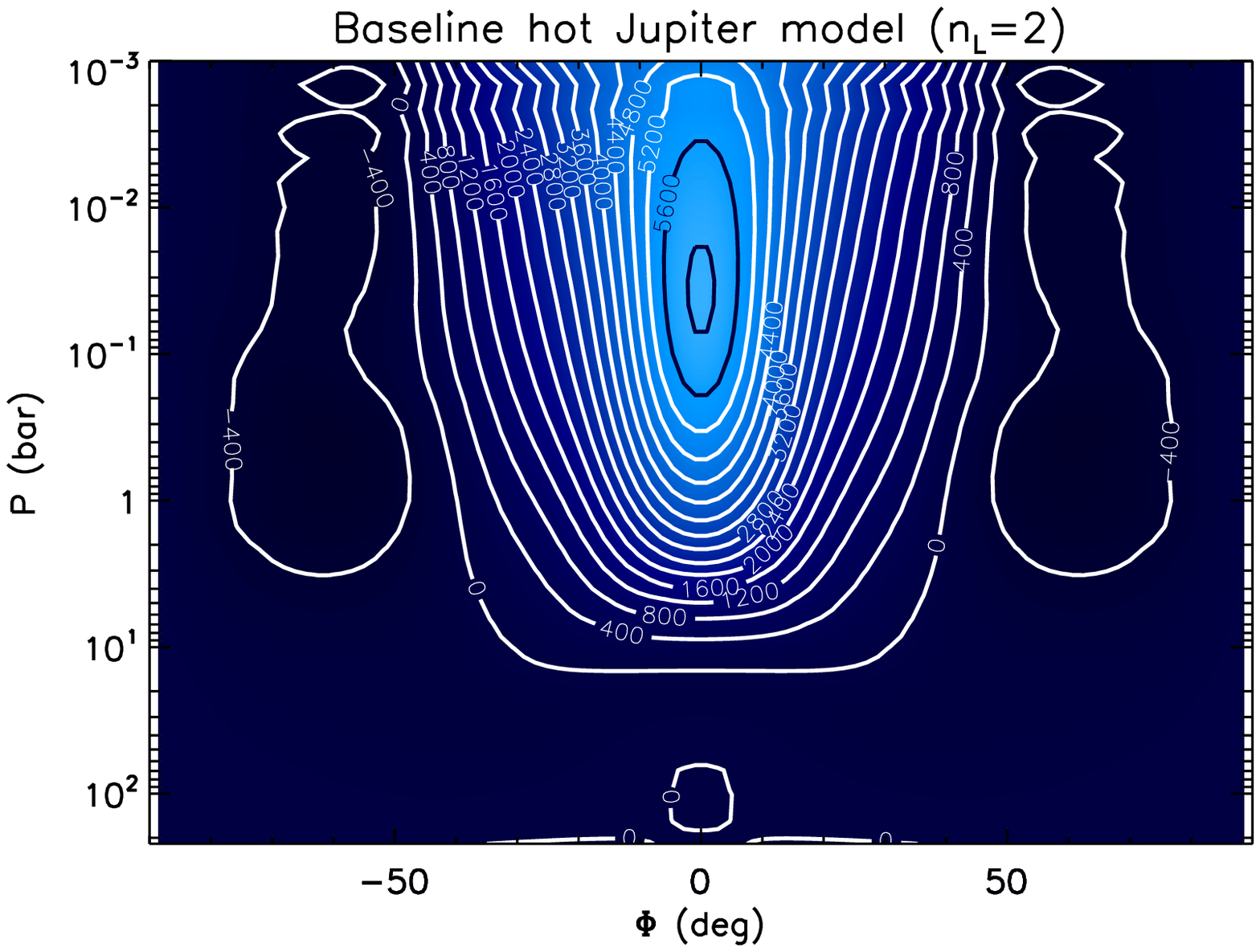}
\includegraphics[width=0.48\columnwidth]{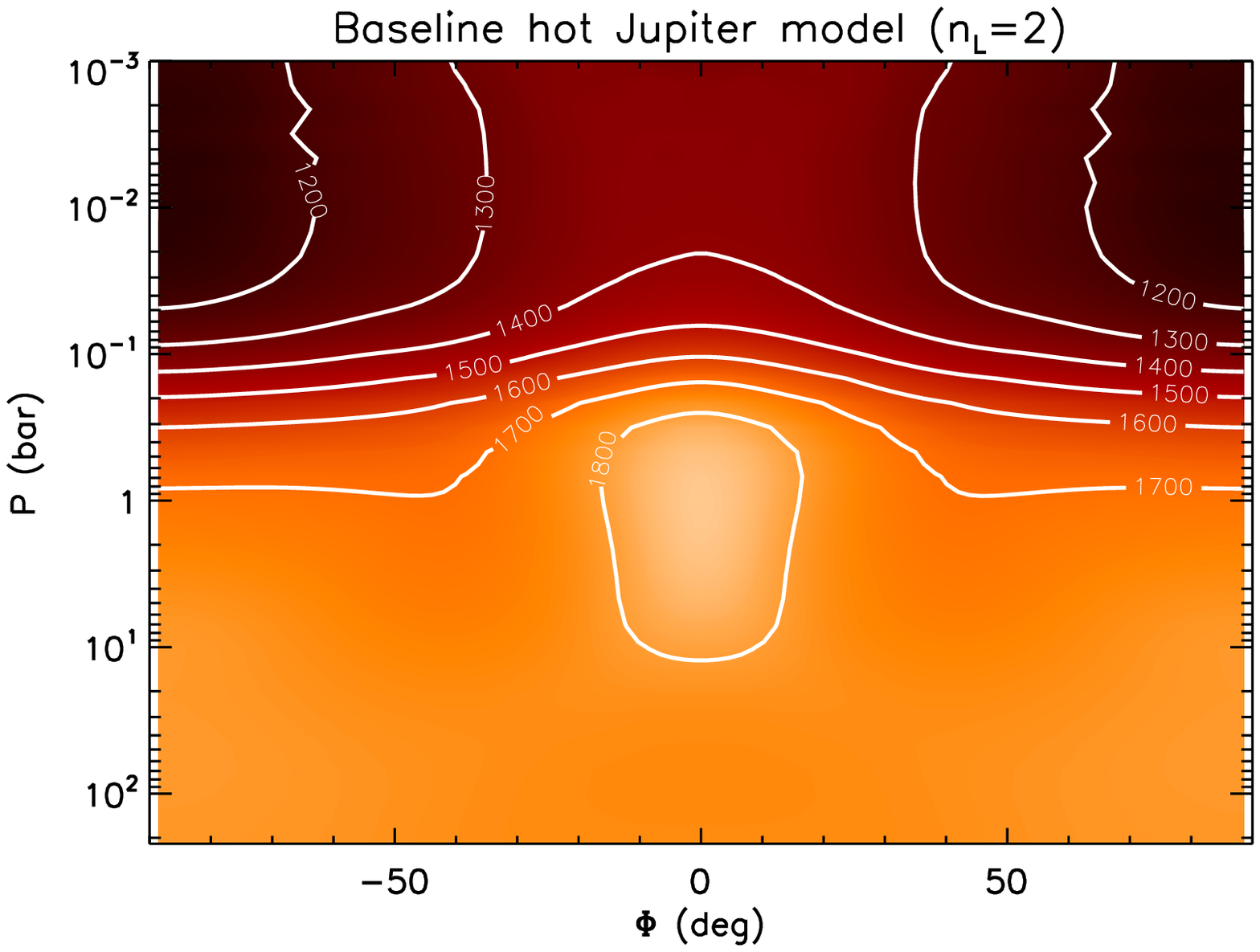}
\includegraphics[width=0.48\columnwidth]{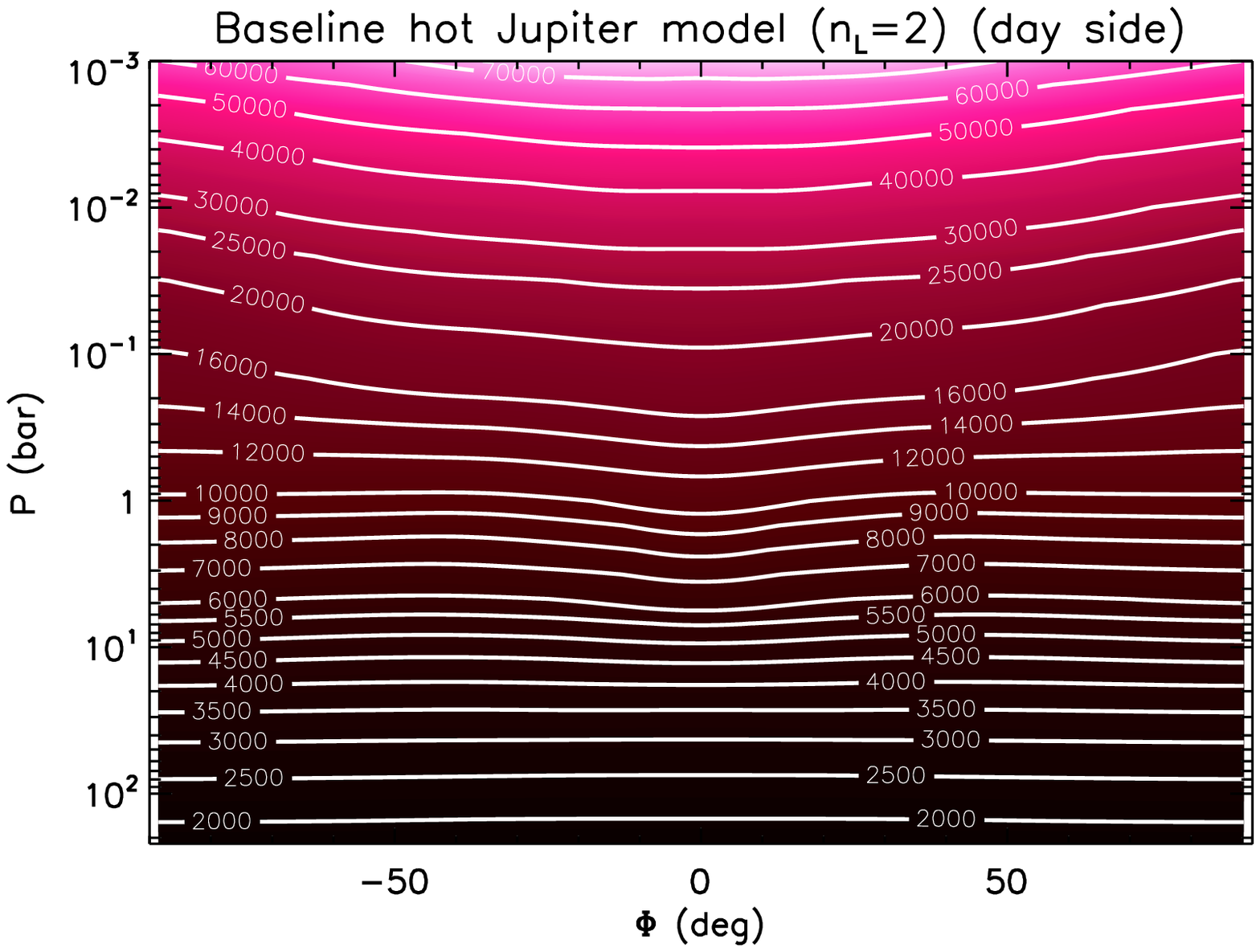}
\includegraphics[width=0.48\columnwidth]{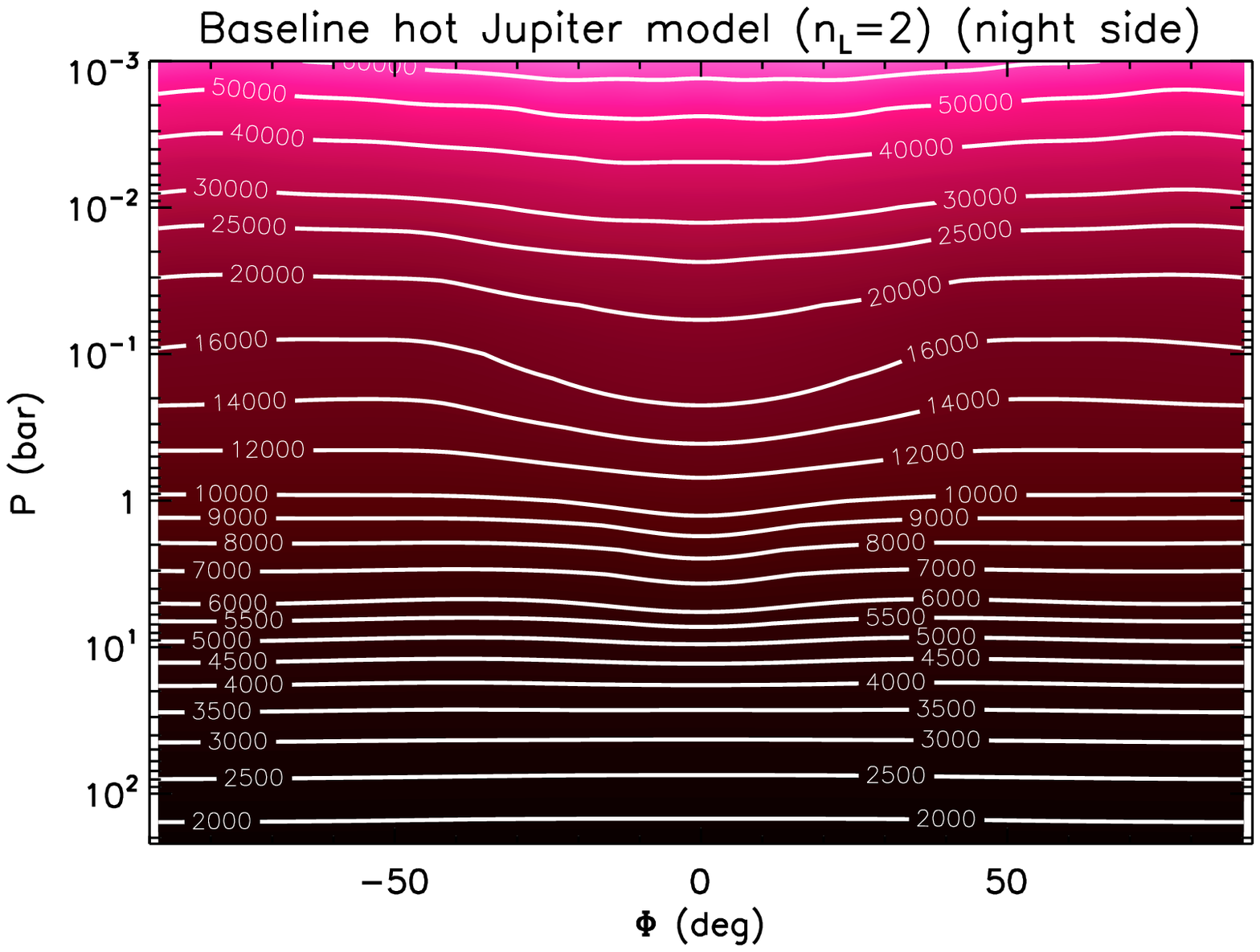}
\includegraphics[width=0.48\columnwidth]{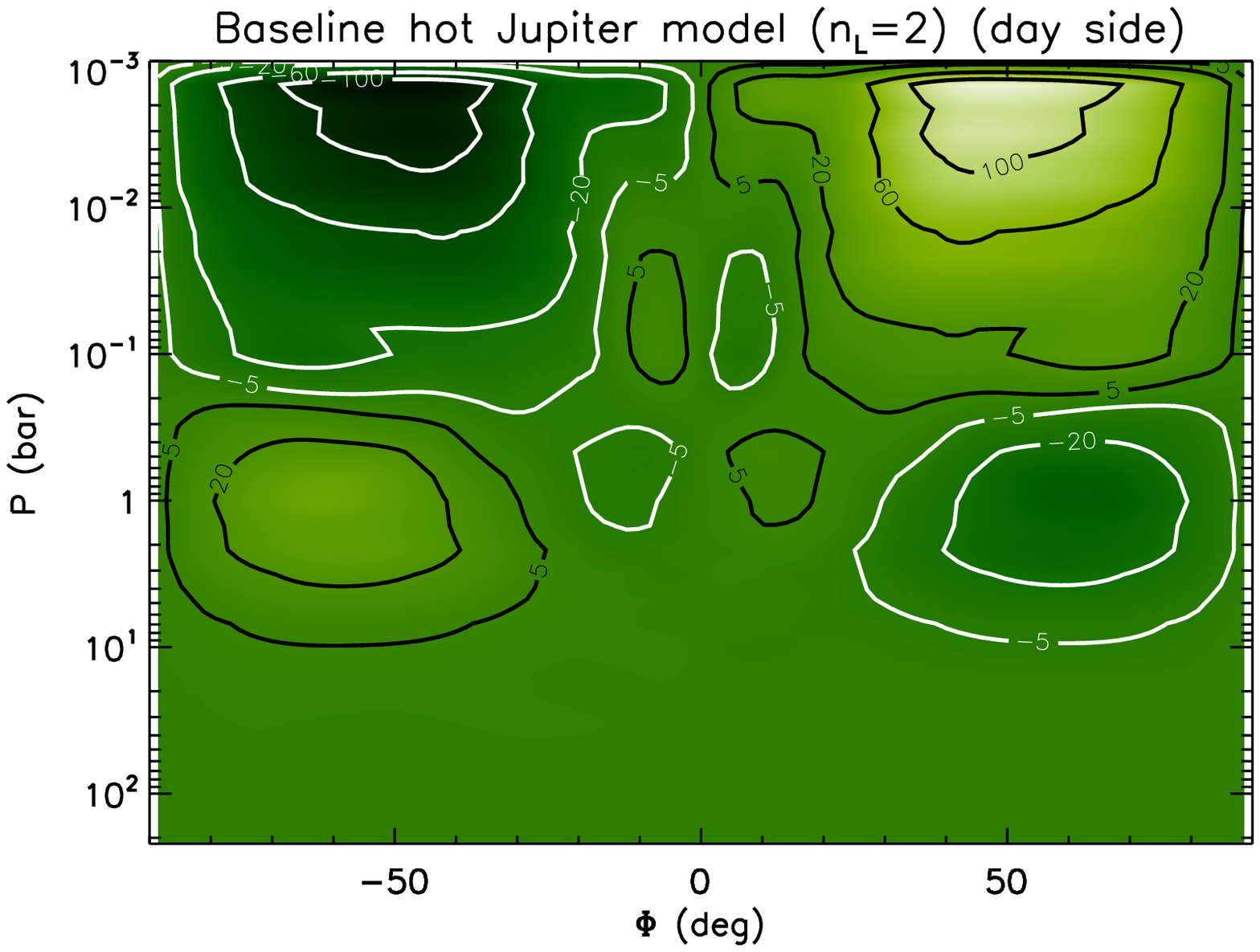}
\includegraphics[width=0.48\columnwidth]{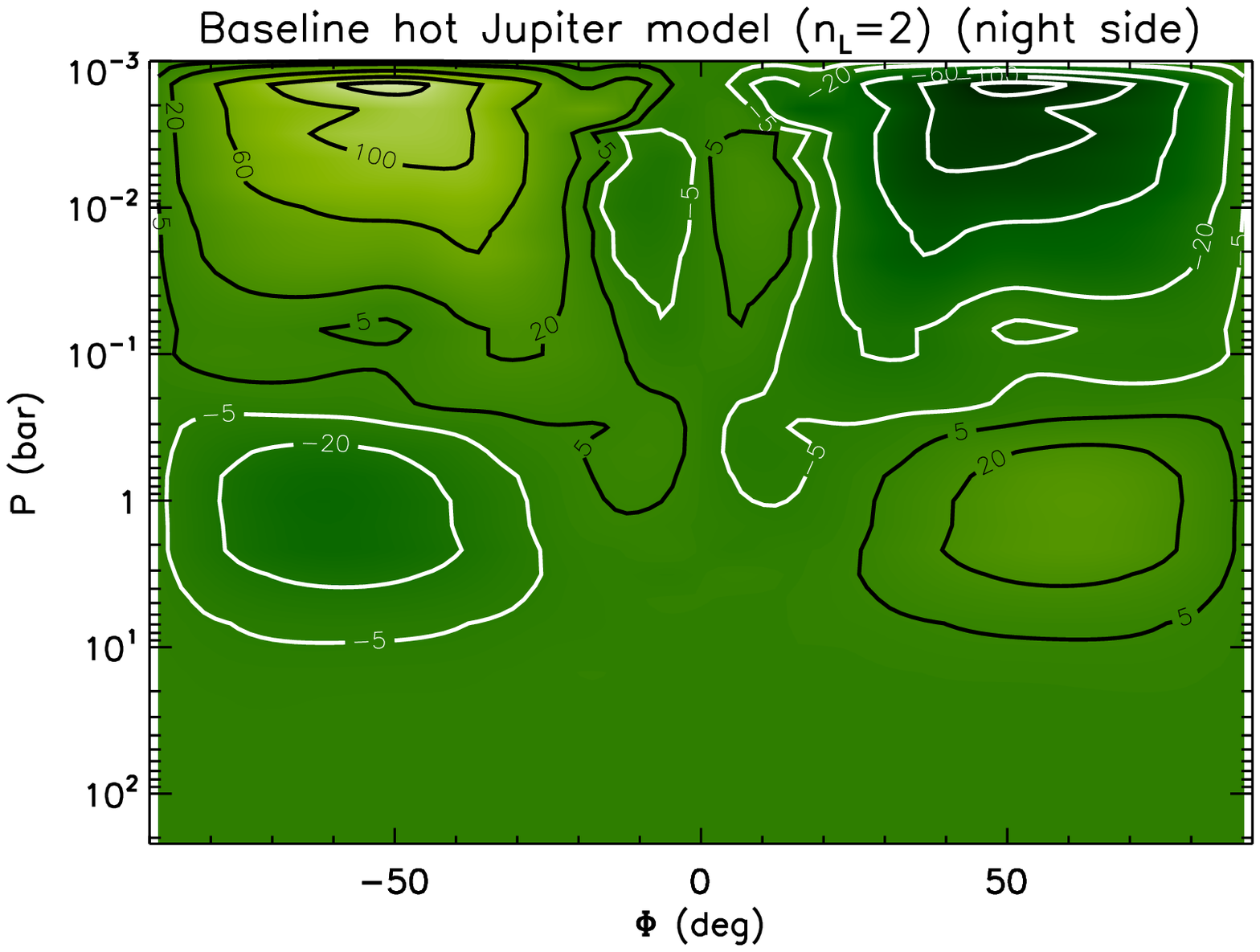}
\end{center}
\vspace{-0.2in}
\caption{Same as Figure \ref{fig:baseline}, but for a simulation with $n_{\rm L}=2$ and $\epsilon=2000$.}
\label{fig:baseline2}
\end{figure}

The Held-Suarez mean flow quantities from the full simulation are shown in Figure \ref{fig:baseline2}, where it is apparent that they are qualitatively similar to those from Figure \ref{fig:baseline}.  The zonal-mean zonal wind profile resembles that obtained from the purely dynamical simulation of \cite{hmp11} (see the top left panel of their Figure 12).  An equatorial, super-rotating jet with a maximum speed of about 6.1 km s$^{-1}$ is again flanked by slower, counter-rotating jets ($-0.8$ km s$^{-1}$) at mid-latitudes.  At $P \gtrsim 10$ bar, the zonal wind structure becomes uniform across latitude, which is consistent with the uniform temperature and barotropic structure present.  Two pairs of circulation cells extend from the equator to the poles on both the day and night sides, in disagreement with the purely dynamical results, presented in Figure \ref{fig:hd209}, where only a single pair of cells exists.  Unlike in the terrestrial atmosphere, our baseline model shows a baroclinic upper atmosphere sitting atop a barotropic lower atmosphere.  It is thus somewhat awkward to prescribe the terms ``stratosphere" and ``troposphere" to the upper and lower atmospheres, respectively.  The analogue of the tropopause sits at $P \approx 10$ bar.

\begin{figure}
\begin{center}
\includegraphics[width=0.48\columnwidth]{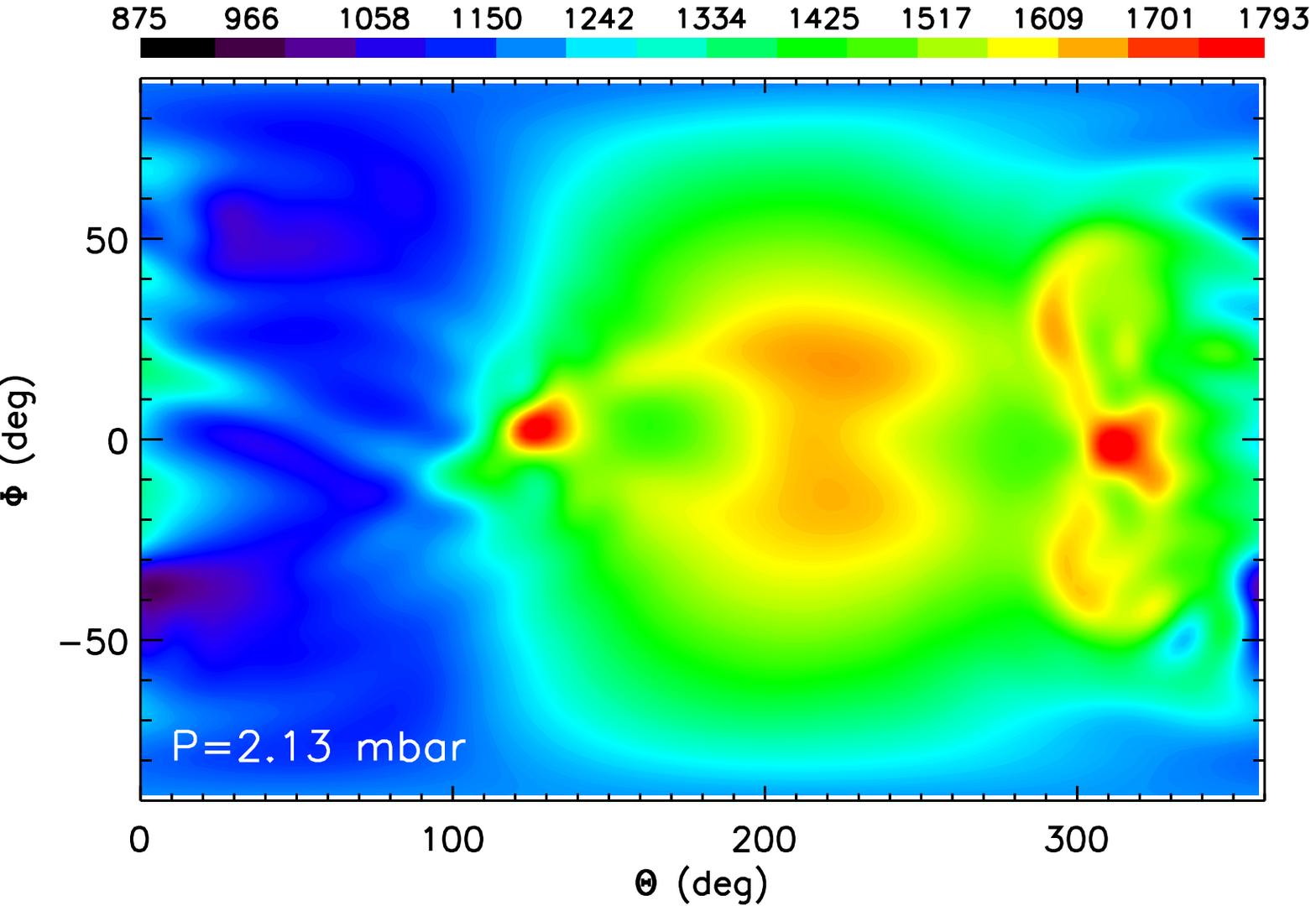}
\includegraphics[width=0.48\columnwidth]{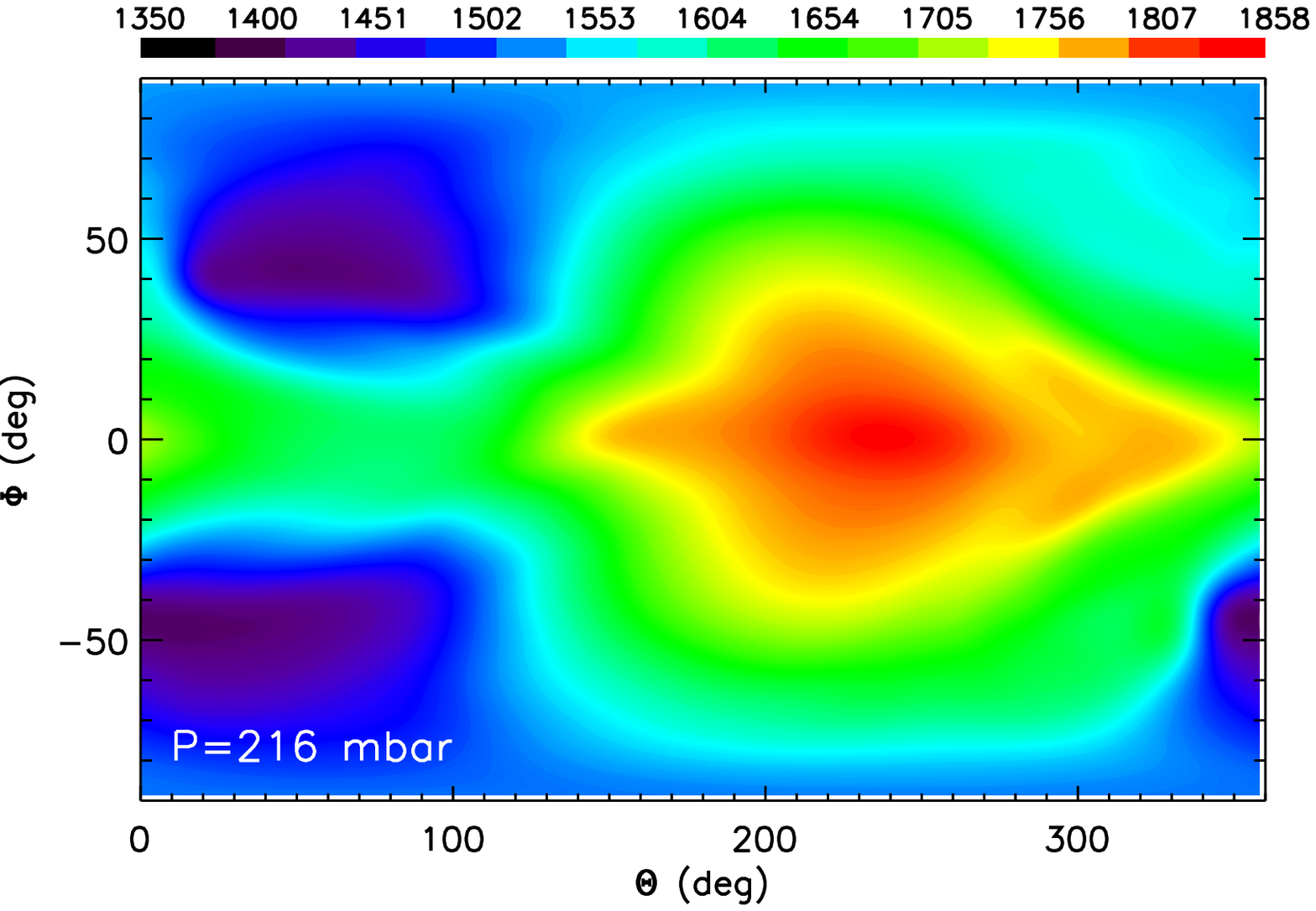}
\includegraphics[width=0.48\columnwidth]{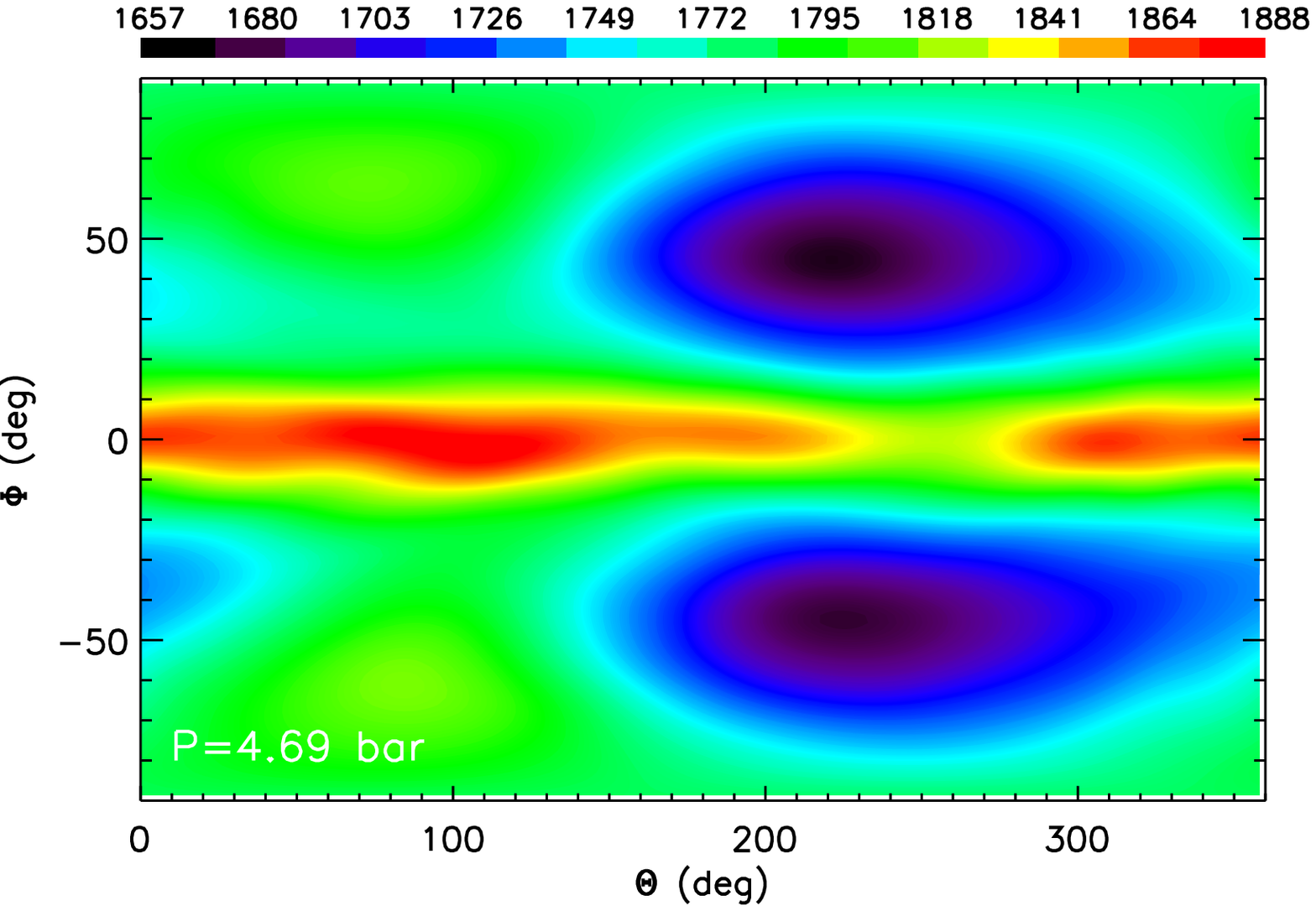}
\includegraphics[width=0.48\columnwidth]{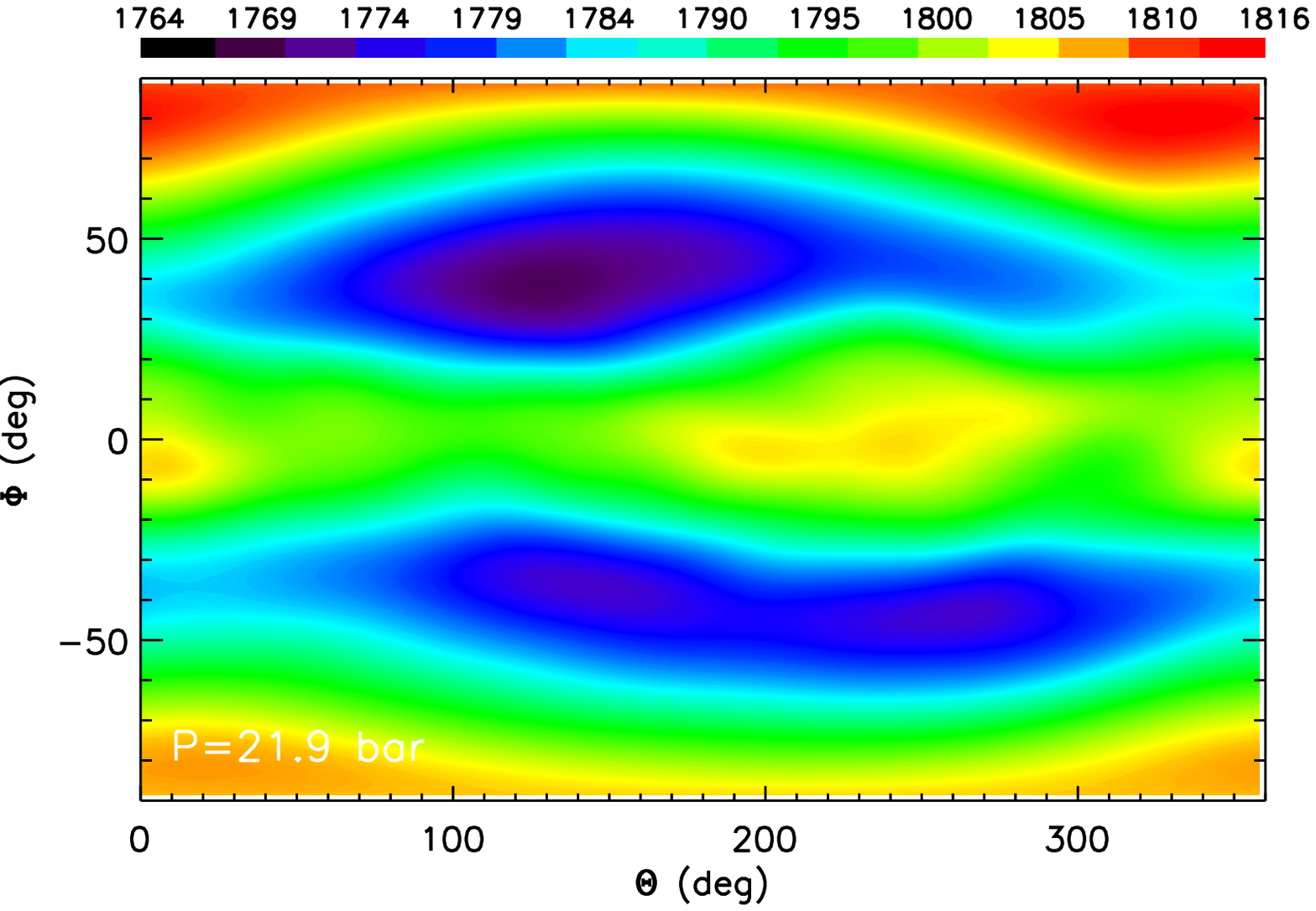}
\end{center}
\vspace{-0.2in}
\caption{Snapshots of the temperature field as functions of latitude ($\Phi$) and longitude ($\Theta$), for the hot Jupiter simulation with $n_{\rm L}=2$ and $\epsilon=2000$, at $P=2.13$ mbar (top left panel), 216 mbar (top right panel), 4.69 bar (bottom left panel) and 21.9 bar (bottom right panel).  The snapshots are taken at 1500 days after the start of the simulation.  Temperatures are given in K.}
\label{fig:fms}
\end{figure}

Figure \ref{fig:fms} shows temperature maps as functions of latitude and longitude at $P=2.13$ mbar, 216 mbar, 4.69 bar and 21.9 bar.  These pressure levels are chosen to match those shown in Figure 8 of \cite{hmp11}.  At $P=2.13$ mbar, the location where the temperature is at its maximum --- which we term the ``hot spot" --- is shifted away from the substellar point, which is somewhat different from the top left panel of Figure 8 of \cite{hmp11} and the second panel (from the top) of Figure 16 of \cite{showman09}.  Otherwise, the characteristic chevron-shaped feature, which is seen in all of the three-dimensional simulations of HD 209458b (e.g., \citealt{showman09,rm10,hmp11}), appears at $P=216$ mbar.  Deeper in the atmosphere, the flow becomes dominated by advection and longitudinal differences in temperature become less apparent.

\subsection{Comparison of different models}
\label{subsect:compare}

\begin{figure}
\begin{center}
\includegraphics[width=0.48\columnwidth]{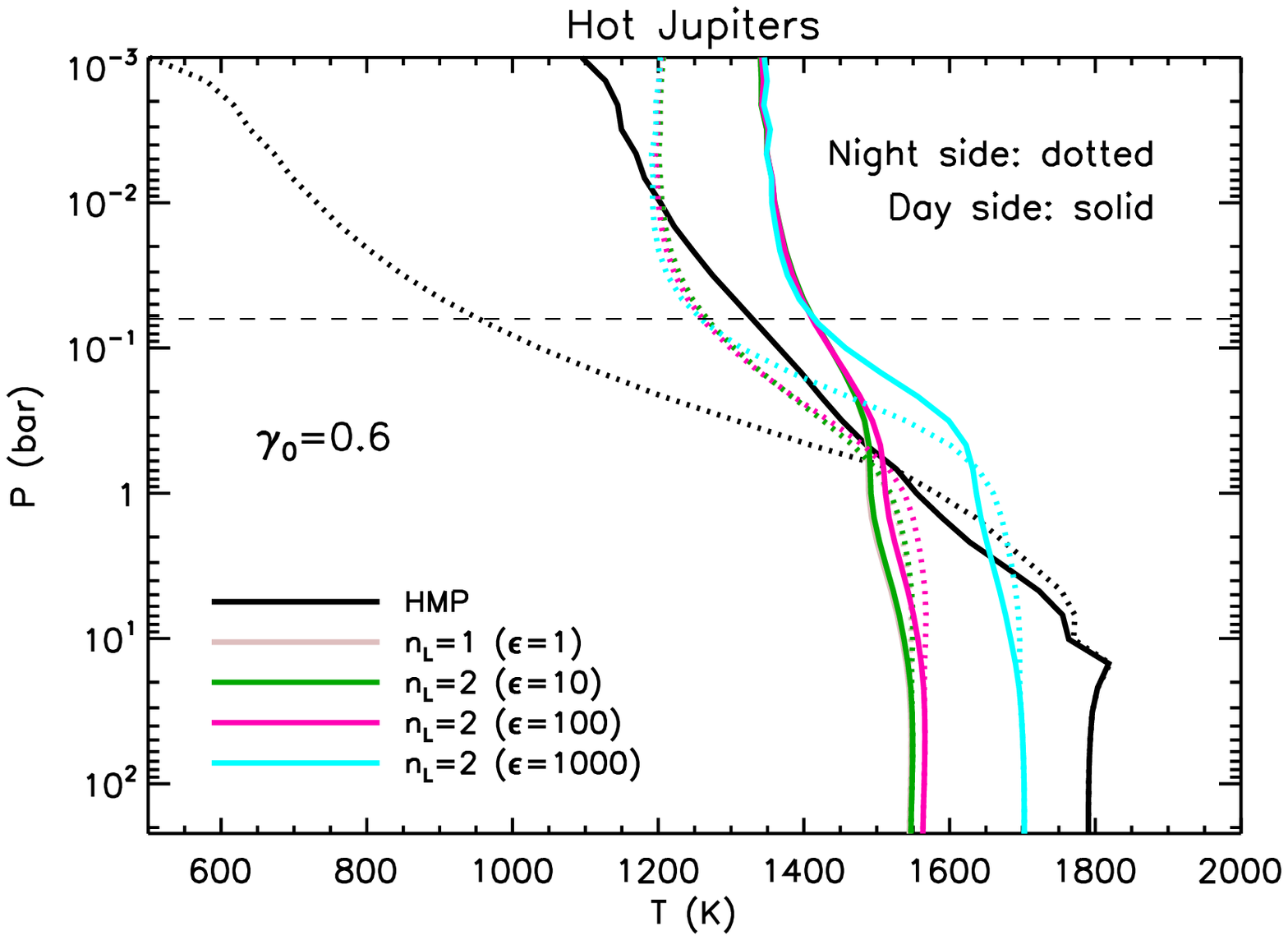}
\includegraphics[width=0.48\columnwidth]{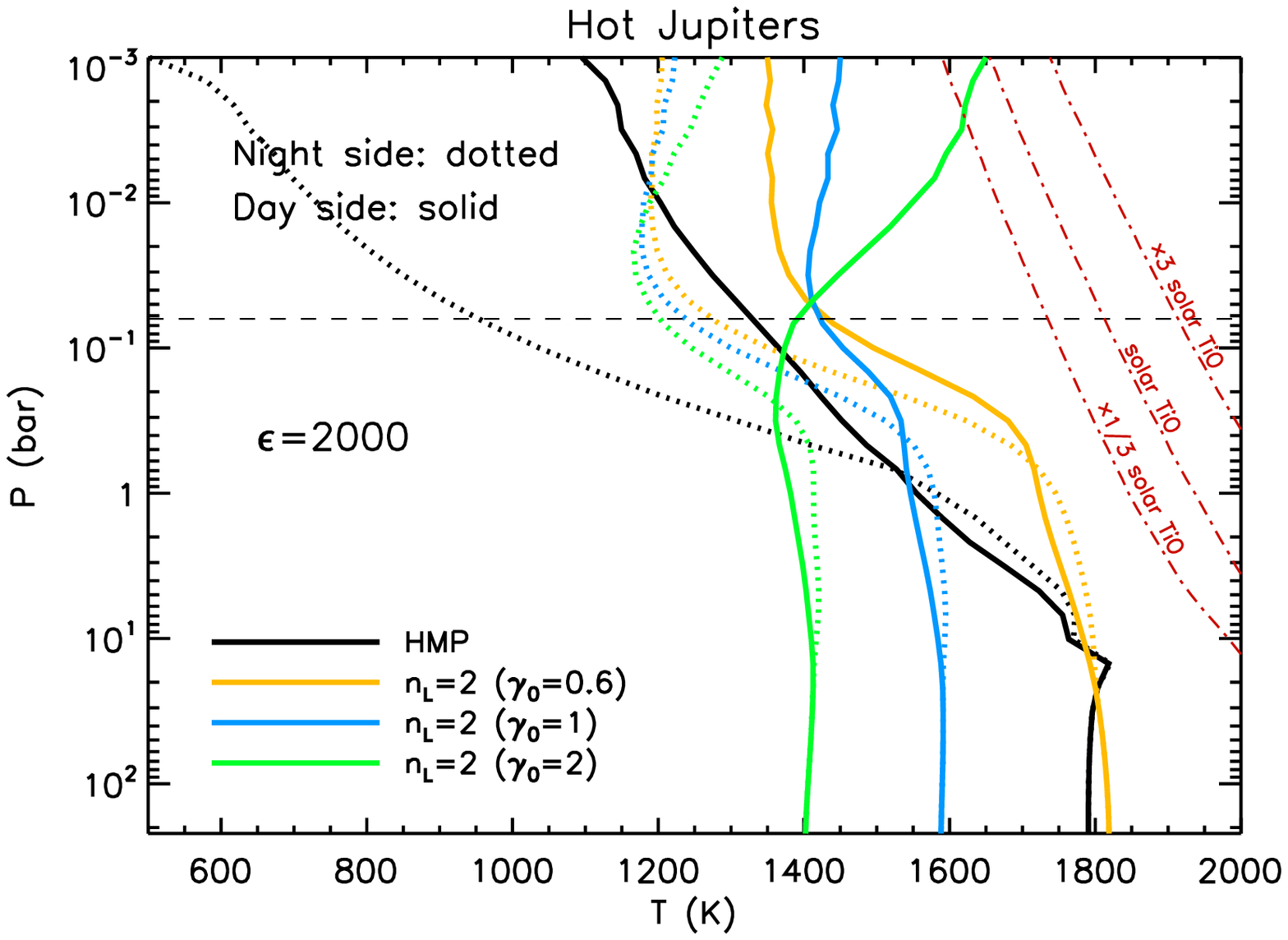}
\end{center}
\vspace{-0.2in}
\caption{Hemispherically-averaged temperature-pressure profiles, separated into the day and night sides, for the hot Jupiter models presented in this study.  Left panel: comparing $n_{\rm L}=1$ and $\epsilon=10, 100$ and 1000 cases.  Note that the $n_{\rm L}=1$ ($\epsilon=1$) and $\epsilon=10$ cases  are almost indistinguishable in this plot.  Right panel: comparing profiles with $\epsilon=2000$ and different values of $\gamma_0$.  The curves labelled by ``HMP" are taken from the HD 209458b model of Heng, Menou \& Phillipps (2011), which only considers atmospheric dynamics and uses Newtonian relaxation to mimic radiative cooling.  The dashed, horizontal line indicates the approximate location of the longwave photosphere.  In the right panel, the condensation curves for TiO from Spiegel, Silverio \& Burrows (2009) are shown.}
\label{fig:tp}
\end{figure}

We next compare hemispherically-averaged temperature-pressure profiles from different models in Figure \ref{fig:tp}, separated into the day and night sides of the simulated hot Jovian atmospheres.    We see that for a fixed value of $\gamma_0=0.6$, the models with $\epsilon=1$--1000 essentially have the same temperature-pressure profiles, both on the day and night sides, at and above the longwave photosphere.  The maximum zonal and meridional wind speeds are virtually unchanged: 5.9 km s$^{-1}$ and from $-0.6$ to $-0.7$ km s$^{-1}$, respectively (at least for the fiducial magnitude of the hyperviscosity adopted in this study).  The day-night temperature contrasts are almost indiscernible, even though the temperatures at depth are modified differently due to the effect of collision-induced absorption.  The day-night temperature contrasts from the purely dynamical model of \cite{hmp11} are somewhat different, but we should keep in mind that this simulation employs Newtonian relaxation via an equilibrium temperature-pressure profile, which in turn contains a parameter $\Delta T_{\rm eq}$ that sets the (initial) day-night temperature contrast as a function of pressure (i.e., $\Delta T_{\rm eq}=530$--1000 K from $P=10$ bar to $P \le 1$ mbar).  In our improved simulations, the day-night temperature contrasts are computed self-consistently.

The model with $\gamma_0=2$ ($\tau_{\rm S_0}=4670, \tau_0=2335$) exhibits temperature inversions on both the day and night sides, consistent with the finding by \cite{hubeny03}, \cite{hansen08} and \cite{guillot10} that $\gamma_0>1$ models should produce temperature inversions.  (See \citealt{bo10} for a review of the observational evidence for temperature inversions in hot Jovian atmospheres.)  Curiously, the temperature at depth is $T_\infty \approx 1400$ K, which is lower than $T_{\rm eq} \approx 1432$ K.  Thus, this model makes a prediction that infrared observations which are able to probe the deep, barotropic layers of a hot Jupiter should infer a equivalent-blackbody temperature which is lower than the equilibrium temperature of the exoplanet.  For the model with $\gamma_0=2$, we note that the Held-Suarez mean flow quantities are qualitatively similar to the baseline hot Jupiter models previously presented (except for the temperature field), despite the differences present in the temperature-pressure profiles.

These differences in the temperature-pressure profiles are relevant to observations, because the longwave photosphere is, in the absence of clouds, located at a pressure of \citep{bl93}
\begin{equation}
P_{\rm L} \approx \frac{2g_p}{3\kappa_{\rm L}} \approx 63 \mbox{ mbar} ~\left( \frac{g_p}{9.42 \mbox{ m s}^{-1}} \right) \left( \frac{\kappa_{\rm L}}{0.01 \mbox{ cm}^2 \mbox{ g}^{-1}} \right)^{-1}.
\label{eq:photosphere}
\end{equation}

We conclude that the quantity $\gamma_0$ controls the presence or absence of a temperature inversion, as well as the day-night temperature contrast.  The key differences from the one- or two-dimensional models is that the redistribution of heat/energy and the depths at which shortwave/longwave absorption occurs are calculated self-consistently once the optical depths are specified.  We also note that while we have shown hemispherically-averaged temperature-pressure profiles for the day and night sides, our calculations generally produce \emph{three-dimensional} $T$-$P$ profiles.

\section{Discussion}
\label{sect:discussion}

\subsection{Observational consequences}

\begin{figure}
\begin{center}
\includegraphics[width=0.48\columnwidth]{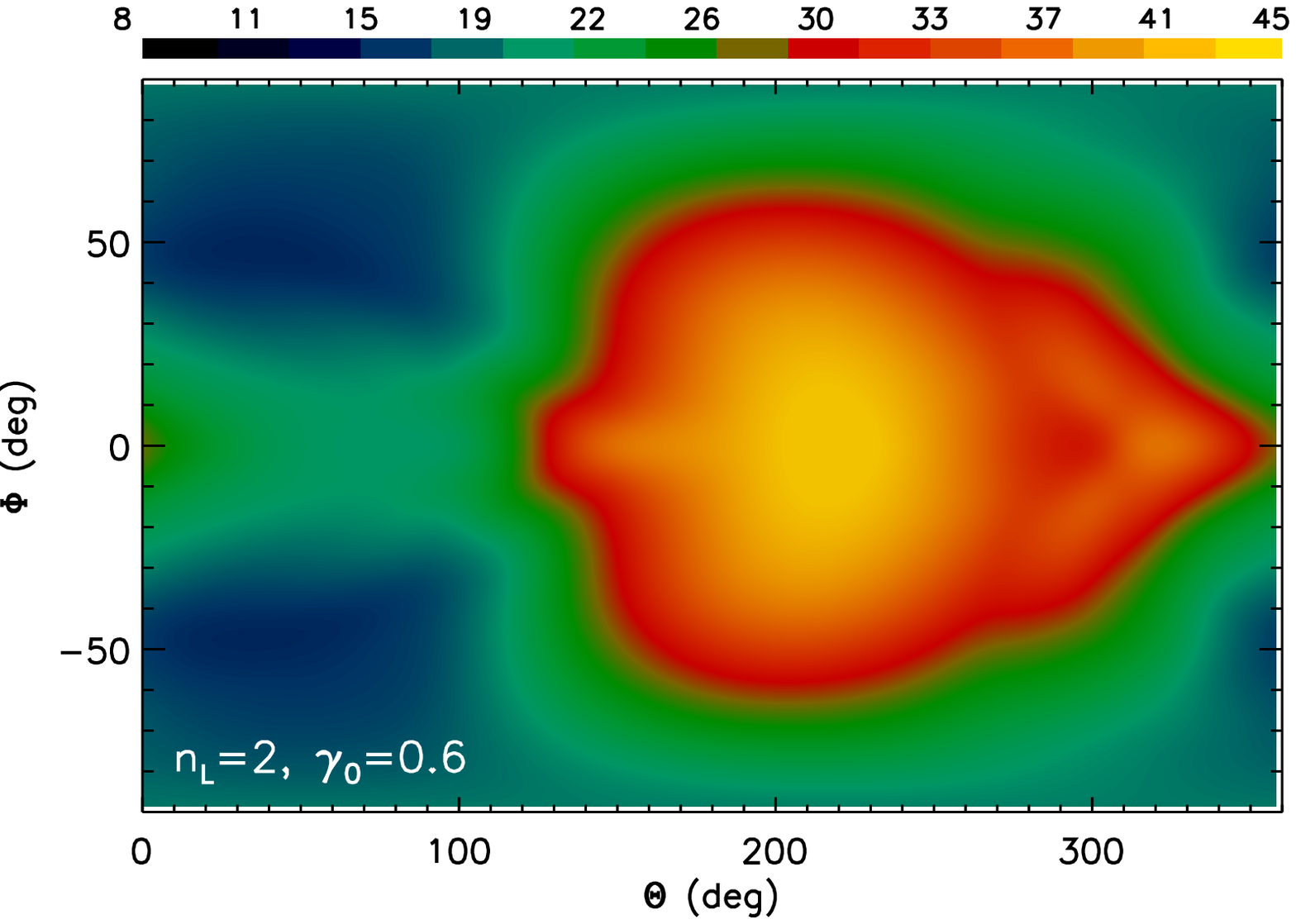}
\includegraphics[width=0.48\columnwidth]{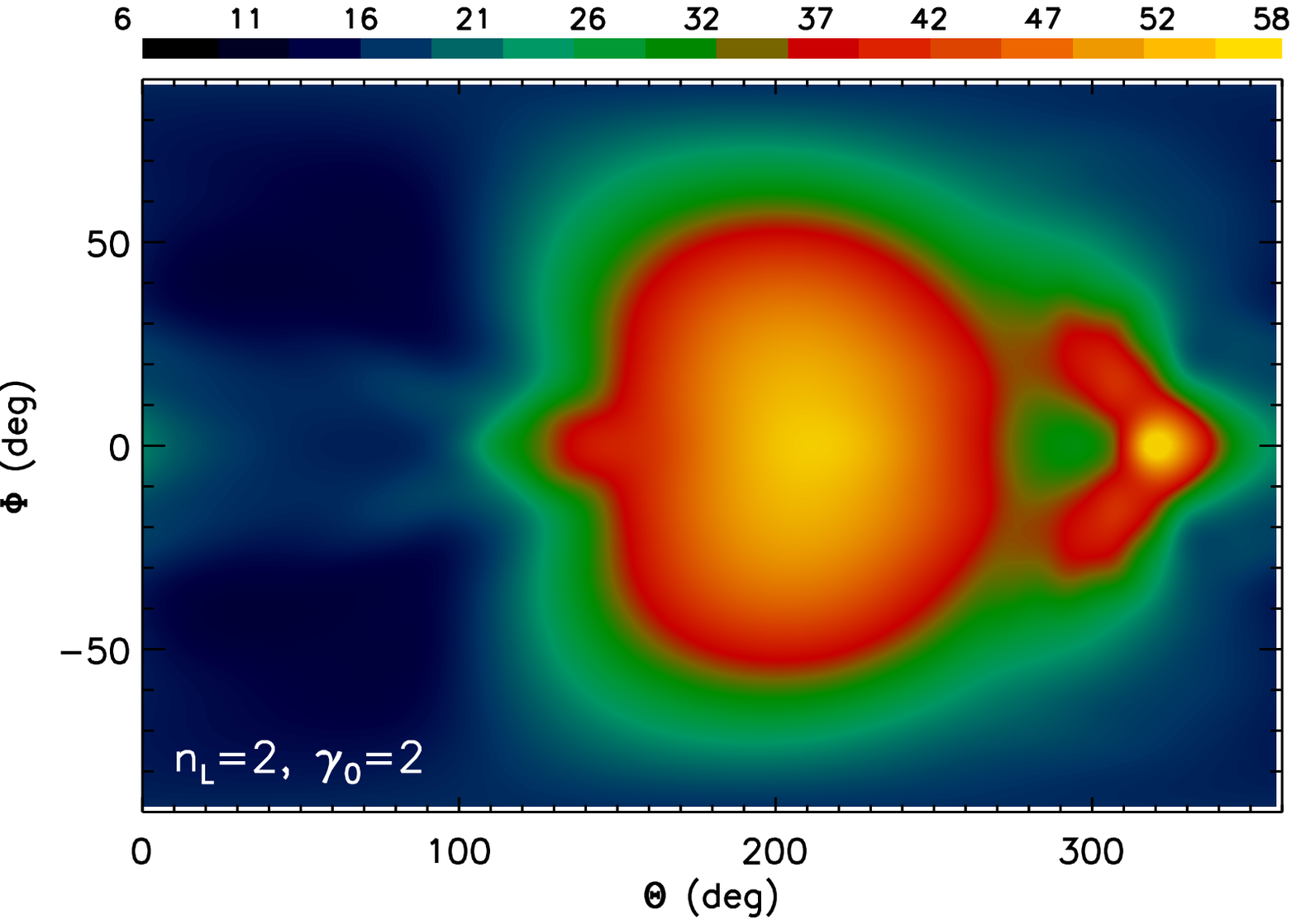}
\end{center}
\vspace{-0.2in}
\caption{Temporally-averaged maps of the flux ${\cal F}_{\rm OLR}$ associated with the outgoing longwave radiation, as functions of latitude ($\Phi$) and longitude ($\Theta$), for two different hot Jupiter models ($n_{\rm L}=2$, $\epsilon=2000$): $\gamma_0=0.6$ (left panel) and 2 (right panel).  Fluxes are given in units of $\times 10^4$ W m$^{-2}$.}
\label{fig:hotspot}
\end{figure}

\begin{figure}
\begin{center}
\includegraphics[width=0.7\columnwidth]{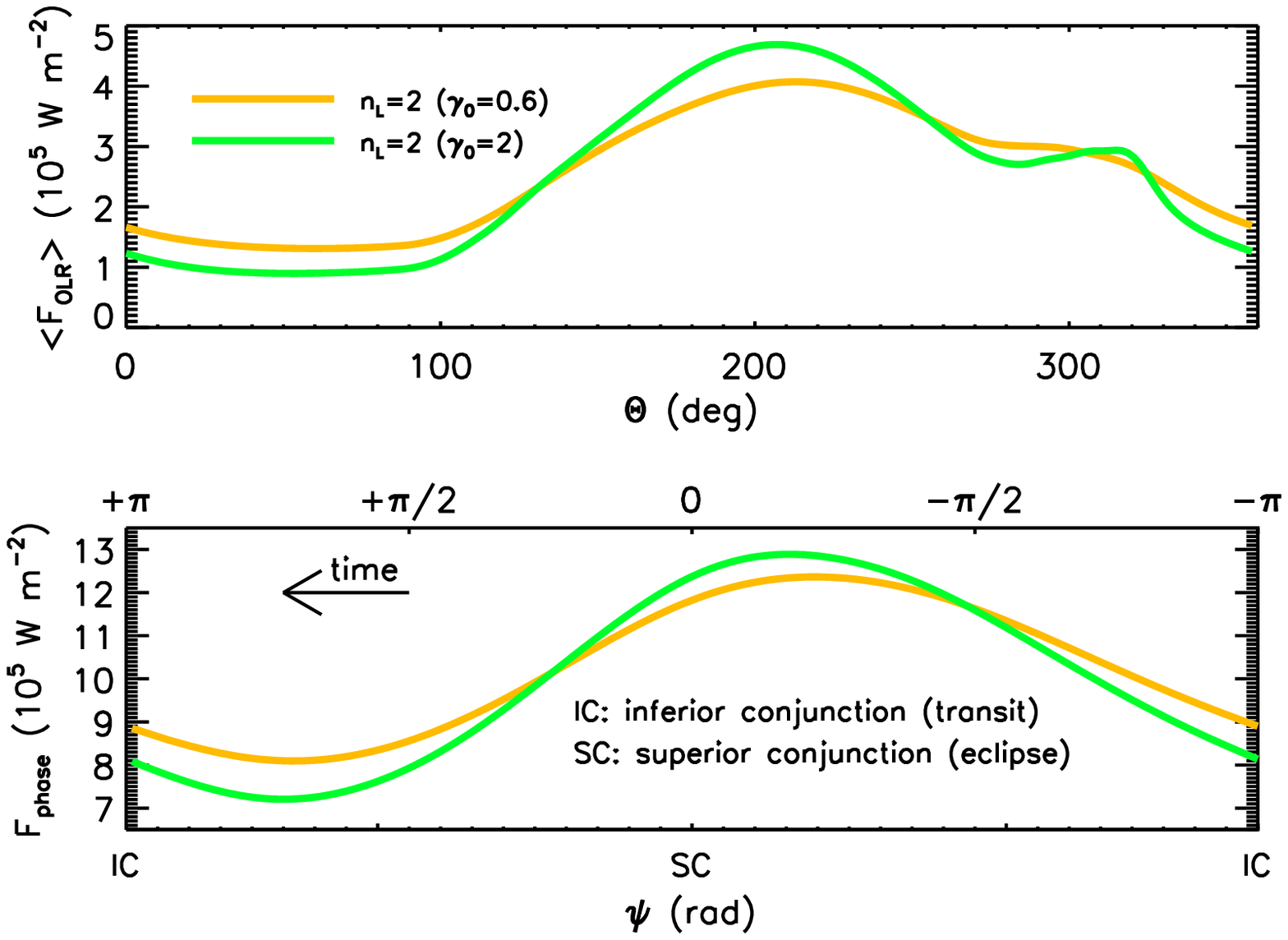}
\end{center}
\vspace{-0.2in}
\caption{Latitudinally-averaged outgoing longwave radiation (OLR; top panel) as a function of the longitude $\Theta$ and thermal phase curve (bottom panel) as a function of the phase angle $\psi$, for the two models shown in Figure \ref{fig:hotspot}.  In the bottom panel, the thermal phase curves peak before the secondary eclipses occur.}
\label{fig:hotspot2}
\end{figure}

A direct consequence of our results concerns the vertical mixing of atmospheric fluid.  As explained in equation \S\ref{subsect:additional}, the strength of atmospheric circulation is quantified by the Eulerian mean streamfunction.  On Jupiter, we have $\Psi \sim 10^{10}$ kg s$^{-1}$ \citep{ls10}.  As shown by our results in Figures \ref{fig:hd209}, \ref{fig:baseline} and \ref{fig:baseline2}, the circulation in hot Jovian atmospheres has a strength of $\Psi \sim 10^{14}$ kg s$^{-1}$, which is $\sim 10^4$ times stronger than on Jupiter.  Since the circulation cells extend from $\sim 1$ mbar to $\sim 10$ bar in our models, the relevant length scale is $\sim 10 H$, where $H = k_{\rm B}T/\bar{m} g_p$ is the pressure scale height.  If we pretend that we may prescribe a ``diffusion coefficient" to the atmospheric circulation, then
\begin{equation}
K_{zz} \sim 10 H ~\int_{\sigma_0}^1 \bar{v}_\Phi\left(\Phi,\tilde{\sigma}_0\right) ~d\tilde{\sigma}_0 \sim \frac{10 H \Psi g_p}{R_p P_0} \sim 10^{10} \mbox{ cm}^2 \mbox{ s}^{-1} ~\left( \frac{T}{1500 \mbox{ K}} \frac{\Psi}{10^{14}\mbox{ kg s}^{-1}} \right) \left( \frac{\bar{m}}{2m_{\rm H}} \frac{R_p}{9.44 \times 10^4 \mbox{ km}} \frac{P_0}{220 \mbox{ bar}} \right)^{-1},
\end{equation}
which is consistent with the required value of $K_{zz} \sim 10^9$--$10^{11}$ cm$^2$ s$^{-1}$, estimated by \cite{spiegel09} for HD 209458b, in order to keep particles associated with TiO, with sizes of 0.1--10 $\mu$m, aloft such that they may act as shortwave absorbers capable of producing a temperature inversion.  Unfortunately, the Eulerian mean streamfunction is not a good representation of the vertical or horizontal mixing of particles embedded in the flow (termed ``tracers" in the atmospheric science community; see \S12 of \citealt{vallis}).  So while the enhanced vertical mixing (relative to Jupiter), via large-scale circulation cells present in hot Jovian atmospheres, may be a tempting and plausible mechanism for TiO to be maintained at high altitudes, a more detailed study of the interaction of tracers with the atmospheric circulation is required.  (See \citealt{ym10} for toy models of vertical, turbulent mixing in hot Jovian atmospheres and their implications for the exoplanetary radii.)

An observational way of distinguishing between different hot Jovian atmospheres is to examine their thermal phase curves \citep{ca08,ca11}.  The flux associated with the outgoing longwave radiation (OLR), denoted by ${\cal F}_{\rm OLR} = {\cal F}_{\rm OLR}(\Theta,\Phi)$, is the emergent flux from the longwave photosphere.  In Figure \ref{fig:hotspot}, we show maps of ${\cal F}_{\rm OLR}$ for two models ($n_{\rm L}=2$, $\epsilon=2000$; $\gamma_0=0.6$ versus 2), where it is apparent that the chevron-shaped feature seen at $P \sim 0.1$ bar resides near the longwave photosphere.  The OLR flux may be used to construct the thermal phase curves, as was done for exo-Earths by \cite{selsis11}.  One has to first average the OLR flux over latitude,
\begin{equation}
\langle {\cal F}_{\rm OLR} \rangle \equiv \frac{1}{\pi} \int^{\pi}_{-\pi} ~{\cal F}_{\rm OLR} ~\cos^2\Phi ~d\Phi.
\end{equation}
The flux associated with the thermal phase curve, ${\cal F}_{\rm phase}$, can then be constructed using equation (7) of \cite{ca08},
\begin{equation}
\begin{split}
&\langle {\cal F}_{\rm OLR} \rangle = A_0 + \sum^{\tilde{n}}_{j=1} ~A_j \cos\left( j\phi \right) + B_j \sin\left(j \phi \right),\\
&{\cal F}_{\rm phase} = 2 A_0 + \sum^{\tilde{n}}_{j=1} ~C_j \cos\left( j\psi \right) + D_j \sin\left(j \psi \right),\\
\end{split}
\label{eq:phase_conversion}
\end{equation}
where $\Theta = \phi + \pi$, the phase angle is denoted by $\psi$ and the summation coefficients are related via the following expressions:
\begin{equation}
C_j = 
\begin{cases}
- \frac{2}{\left( j^2 - 1 \right)} \left( -1 \right)^{j/2} A_j, & j \ge 2, \\
\frac{\pi A_1}{2}, & j=1, \\
\end{cases}
\label{eq:transform}
\end{equation}
and
\begin{equation}
D_j = 
\begin{cases}
\frac{2}{\left( j^2 - 1 \right)} \left( -1 \right)^{j/2} B_j, & j \ge 2, \\
- \frac{\pi B_1}{2}, & j=1. \\
\end{cases}
\label{eq:transform2}
\end{equation}
We perform the fits, using the function described in the first expression in equation (\ref{eq:phase_conversion}), to our computed $\langle {\cal F}_{\rm OLR} \rangle$ curves by considering terms up to $\tilde{n}=13$.  We then obtain ${\cal F}_{\rm phase}$ by transforming the $A_j$ and $B_j$ coefficients, using equations (\ref{eq:transform}) and (\ref{eq:transform2}), to $C_j$ and $D_j$ coefficients.  As pointed out by \cite{ca08}, the odd sinusoidal modes do not contribute to the thermal phase curve: we set $C_j=D_j=0$ when $j$ is odd except for $j=1$.  This implies that information is lost when the latitudinally-averaged OLR flux is converted to the thermal phase curve.  For example, for the $\gamma_0=0.6$ model, we have $[(A_2/A_0)^2 + (B_2/A_0)^2]^{1/2} \approx 0.08$ while $[(A_3/A_0)^2 + (B_3/A_0)^2]^{1/2} \approx 0.04$.  For the $\gamma_0=2$ model, we have $[(A_2/A_0)^2 + (B_2/A_0)^2]^{1/2} \approx 0.15$ while $[(A_3/A_0)^2 + (B_3/A_0)^2]^{1/2} \approx 0.09$.  These estimates suggest that the second lowest odd mode ($j=3$) has a small but non-negligible contribution to the OLR map.  It is worth noting that the odd modes do not contribute to the thermal phase curve only when the following assumptions are valid: the exoplanet rotates edge-on; there is no limb darkening associated with the exoplanet; and there is no temporal variability associated with the OLR.

In Figure \ref{fig:hotspot2}, we calculate $\langle{\cal F}_{\rm OLR}\rangle$ and ${\cal F}_{\rm phase}$ for the $\gamma_0=0.6$ and $\gamma_0=2$ models previously shown in Figure \ref{fig:hotspot}.  For the latitudinally-averaged OLR flux (top panel), the angular offsets of the peaks from the substellar point ($\Theta=180^\circ$) are $33.8^\circ$ and $26.3^\circ$ for the $\gamma_0=0.6$ and $\gamma_0=2$ models, respectively.  As realized by \cite{ca11}, the corresponding offsets in the thermal phase curve (bottom panel) are somewhat larger: $39.4^\circ$ and $31.9^\circ$, which correspond to temporal offsets of 9.3 and 7.5 hours, respectively, easily discernible with current technology.  Notice that the phase angle $\psi$ is defined in such a manner that the peaks of the thermal phase curves occur before superior conjunction (i.e., secondary eclipse).

\begin{figure}
\begin{center}
\includegraphics[width=0.7\columnwidth]{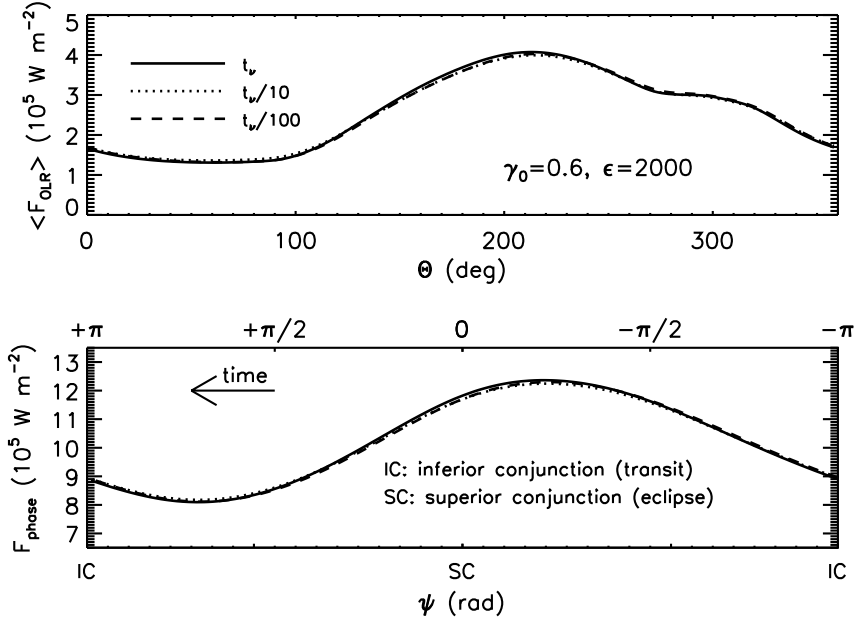}
\end{center}
\vspace{-0.2in}
\caption{Same as Figure \ref{fig:hotspot2}, but for comparison of the $\gamma_0=0.6$ and $\epsilon=2000$ model with hyperviscous time scales of $t_\nu = 10^{-5}, 10^{-6}$ and $10^{-7}$ HD 209458b day.  The hot spot offset appears to be somewhat robust to our ignorance of $t_\nu$.}
\label{fig:hypervis}
\end{figure}
 
We may be concerned that drag within the atmospheres may alter the location of the hot spot, as implied by the study of \cite{sp11}.  Within the context of the simulations, drag may manifest itself in two forms: physical (e.g., Ohmic) versus numerical (i.e., horizontal dissipation; see \S3.3 of \citealt{hmp11}).  We restrict our discussion to the uncertainties associated with numerical drag, which takes the form of hyperviscosity in our spectral simulations.  To illustrate the effects of numerical drag, we execute two more simulations with $\gamma_0=0.6$ and $\epsilon=2000$ (our default model), but with the time scale associated with hyperviscosity ($t_\nu$) decreased by factors of 10 and 100.  The corresponding latitudinally-averaged OLR flux and thermal phase curves are shown in Figure \ref{fig:hypervis}.  As already demonstrated in \cite{hmp11}, changing the hyperviscosity alters the zonal wind speeds: the maximum and minimum speeds of $6.1$ km s$^{-1}$ and $-0.8$ km s$^{-1}$, respectively, from our default model now become $5.8$ km s$^{-1}$ and $-1.2$ km s$^{-1}$ for $t_\nu = 10^{-6}$ HD 209458b day and $5.5$ km s$^{-1}$ and $-1.1$ km s$^{-1}$ for $t_\nu = 10^{-7}$ HD 209458b day.  Since hyperviscosity is a numerical tool and it is difficult to specify its magnitude from first principles, zonal wind speeds in hot Jovian atmospheres are \textit{not} robust predictions of general circulation models, at least at the $\lesssim 10\%$ level.  However, the hot spot offset appears to be a somewhat robust prediction of the models: for $t_\nu = 10^{-5}, 10^{-6}$ and $10^{-7}$ HD 209458b day, the temporal offsets are 9.3, 9.7 and 9.7 hours, respectively.  Therefore, despite our ignorance of $t_\nu$ the hot spot offsets measured from thermal phase curves offer an opportunity to distinguish between hot Jovian atmospheres characterized by different values of $\gamma_0$.

\subsection{Summary}

The salient points of our study can be summarized as follows:
\begin{itemize}

\item We have successfully implemented a simple (dual-band, two-stream) radiative transfer scheme in conjunction with the \texttt{FMS} spectral dynamical core previously described in \cite{hmp11}.  We have tested our computational setup by simulating the atmospheres of Earth-like (exo)planets and hot Jupiters.  Our models possess an intermediate degree of sophistication in a required hierarchy of three-dimensional models of atmospheric circulation.

\item We have generalized the analytical formalism of \cite{guillot10} to calculate temperature-pressure profiles of hot Jovian atmospheres, in radiative equilibrium, which include the effect of collision-induced absorption via a single parameter $\epsilon$.

\item Atmospheric circulation on hot Jupiters is characterized by the presence of large circulation cells extending from the equator to the poles, due to the effect of Coriolis deflection being weakened by slower rotation.  While a transitional layer (``tropopause") may exist between the baroclinic and barotropic components of the atmosphere, the former is generally located at \emph{greater} altitudes while the latter resides deeper in the atmosphere.  This is the opposite from the situation on Earth (and a hypothetical, tidally-locked Earth), where the stratosphere sits on top of the troposphere.

\item Large-scale circulation cells in hot Jovian atmospheres offer a plausible mechanism for maintaining TiO at high altitudes, such that it may act as a shortwave absorber capable of producing a temperature inversion.  However, a more detailed study of the interaction between particles embedded in the flow (``tracers") and the atmospheric circulation is required before any robust conclusions can be drawn.

\item The absence or presence of a temperature inversion in the baroclinic component of a hot Jovian atmosphere, as well as the temperature contrast between the day versus night sides, is determined by the ratio of shortwave to longwave opacity normalizations ($\gamma_0$), at least in a cloud- or haze-free scenario.

\item The angular offset between the hot spot --- associated with the equatorial, super-rotating wind expected to be present in hot Jovian atmospheres --- and the substellar point may help us distinguish between hot Jovian atmospheres characterized by different values of $\gamma_0$, and appears to be robust to our ignorance of hyperviscosity.

\end{itemize}

\section{Acknowledgments}

K.H. acknowledges support from the Zwicky Prize Fellowship and the Star and Planet Formation Group at ETH Z\"{u}rich.  D.M.W.F. acknowledges support from NSF grants ATM-0846641 and ATM-0936059.  K.H. thanks Kristen Menou, Fr\'{e}d\'{e}ric Pont, Nikku Madhusudhan, Hans Martin Schmid, Tim Merlis, Eric Agol, David Sing, Dave Spiegel, Michael Meyer, Emily Rauscher, Tapio Schneider and Paul O'Gorman for useful conversations.  We additionally thank Nick Cowan, Andrew Youdin and Ren\'{e} Heller for useful comments following their reading of an earlier version of the manuscript.  K.H. is especially grateful to Kristen Menou for providing invaluable guidance during the initial stages of this study and Nick Cowan for providing a crash course on thermal phase curves.  We thank the anonymous referee for constructive criticism which improved the quality of the manuscript.  This work benefited from competent and dependable technical support by Olivier Byrde, Eric M\"{u}ller, Adrian Ulrich et al. of the \texttt{Brutus} computing cluster team, as well as steadfast administrative support by Marianne Chiesi of the Institute for Astronomy, at ETH Z\"{u}rich.


\appendix

\section{Atmospheric Boundary Layer}
\label{append:pbl}

There are two main pieces of key physics which need to be included within the atmospheric boundary layer.  Firstly, the turbulence generated within the layer acts as a form of drag between the surface and the atmosphere.  Secondly, since the turbulent eddies generated are expected to be three-dimensional, energy flows towards the smallest length scales (i.e., a forward Kolmogorov cascade) where it is dissipated by molecular viscosity.  The process of turbulent mixing may be described by the diffusion equation.  Therefore, a boundary layer scheme needs to approximately capture the effects of drag and diffusion.

Operationally, drag between the surface and the lowest model layer is expressed in the form of drag coefficients acting on the temperature and velocity fields, which are calculated using the ``Monin-Obukhov" drag laws, considered standard in the atmospheric science community (see \S2c of \citealt{fhz06}):
\begin{equation}
{\cal C}_{\rm MO} = 
\begin{cases}
\kappa_{\rm vK}^2 \zeta^{-2}, & {\cal R}_{i_0} < 0, \\
\kappa_{\rm vK}^2 \zeta^{-2} \left( 1 - \frac{{\cal R}_{i_{\rm 0}}}{{\cal R}_{i,{\rm crit}}} \right)^2, & 0 < {\cal R}_{i_0} < {\cal R}_{i,{\rm crit}}, \\
0 & {\cal R}_{i_0} > {\cal R}_{i,{\rm crit}},\\
\end{cases}
\label{eq:dragcoeff}
\end{equation}
where
\begin{equation}
\zeta \equiv \ln \left(\frac{z_0}{z_{\rm rough}} \right)
\end{equation}
and $z_0$ is the height of the lowest model layer.  It should be noted that equation (\ref{eq:dragcoeff}) is considered a simplified Monin-Obukhov formulation --- more sophisticated functions exist to describe ${\cal C}_{\rm MO}$.  The key physics is captured in three numbers: $z_{\rm rough}$, ${\cal R}_i$ and $\kappa_{\rm vK}$.  The first of these is the roughness length $z_{\rm rough}$, which parametrizes the macroscopic effects of the surface type on the drag.  Within the Monin-Obukhov framework, it is the vertical height at which the horizontal winds go to zero.  For example, it is $z_{\rm rough} \sim 10^{-4}$ m over open water and $\sim 1$ m over urban terrain.  Following \cite{fhz06}, we adopt $z_{\rm rough} = 3.21 \times 10^{-5}$ m for our baseline Earth-like model (\S\ref{subsect:earth}).

Denoting the horizontal wind speed by $v$, the second of these numbers is the bulk Richardson number (e.g., \S3.7.1 of \citealt{wp05}),
\begin{equation}
{\cal R}_i = \frac{g_p}{\theta_{\rm T}} \frac{\partial \theta_{\rm T}}{\partial z} \left( \frac{\partial v}{\partial z} \right)^{-2},
\end{equation}
which quantifies the effects of shear versus vertical stability.  When the bulk Richardson number falls below a critical value ${\cal R}_{i,{\rm crit}} \sim 1$--10, the atmosphere is turbulent.  Above this critical value, the drag coefficients are set to zero.  Again following \cite{fhz06}, we set ${\cal R}_{i,{\rm crit}}=1$ for our Earth-like model.  In equation (\ref{eq:dragcoeff}), the quantity ${\cal R}_{i_0}$ is the bulk Richardson number evaluated at the lowest model layer.  

The third of these numbers is the von K\'{a}rm\'{a}n constant, which from laboratory experiments and terrestrial atmospheric observations is estimated to be $\kappa_{\rm vK} \approx 0.4$ (see \citealt{a06} and references therein).  We thus adopt $\kappa_{\rm vK} = 0.4$ in our Earth-like simulations.

To treat turbulent diffusion, we first determine the boundary layer depth $h$ by calculating the height where ${\cal R}_i = {\cal R}_{i,{\rm crit}}$ \citep{fhz06}.\footnote{At the top of the boundary layer, there often exists a stable layer where the turbulent motions from beneath do not penetrate into, termed the ``capping inversion".}  The turbulent diffusion coefficient is computed as \citep{tm86}
\begin{equation}
{\cal K} = 
\begin{cases}
{\cal K}_0\left(z\right) & z < z^\prime, \\
{\cal K}_0 \left(z^\prime\right) \left(1 - \frac{z - z^\prime}{h -z^\prime} \right)^2 \frac{z}{z^\prime}, & z^\prime < z < h, \\
\end{cases}
\end{equation}
where
\begin{equation}
z^\prime \equiv f_b h.
\end{equation}
The boundary layer fraction $f_b$ defines a ``surface layer" of height $z^\prime$ which is assumed to be in heat and momentum equilibrium with the surface --- this formulation is made in the interest of computational efficiency, so as to avoid an iterative calculation to establish equilibrium.  Following \cite{fhz06}, we adopt $f_b=0.1$.  Above the surface layer, the diffusion coefficient smoothly goes to zero at the top of the boundary layer ($z=h$).  The surface layer diffusion coefficient is
\begin{equation}
{\cal K}\left( z \right) = 
\begin{cases}
\kappa_{\rm vK} ~v_0 z ~{\cal C}_{\rm MO}^{1/2}, & {\cal R}_{i_0} < 0, \\
\kappa_{\rm vK} ~v_0 z ~{\cal C}_{\rm MO}^{1/2} \left[ 1 + \frac{{\cal R}_i ~\ln\left(z/z_{\rm rough}\right)}{{\cal R}_{i,{\rm crit}} - {\cal R}_i} \right]^{-1}, & {\cal R}_{i_0} > 0, \\
\end{cases}
\end{equation}
where $v_0$ represents the horizontal velocity in the lowest model layer.  The diffusion coefficients act on the velocity field and the dry static (specific) energy, the later of which is defined as
\begin{equation}
{\cal E}_{\rm dry} \equiv c_P T + g_p z.
\end{equation}

The Monin-Obukhov framework describes mixing induced via both shear and convection, and allows for a smooth transition between the two regimes.   A key difference between it and the convective adjustment scheme (\S\ref{subsect:convection}) is that the former mixes momentum while the latter does not.  An obstacle to applying the Monin-Obukhov scheme to the atmospheres of hot Jupiters is that the four parameters are considered to be free.

\label{lastpage}

\end{document}